\RequirePackage{lhchiggs}
\documentclass{cernyrep}
\usepackage{graphicx}
\usepackage{rotating}
\usepackage{color}
\usepackage{hhline}
\usepackage{subfigure}
\usepackage{lscape}
\usepackage{floatpag}
\floatpagestyle{plain}
\usepackage{axodraw}
\usepackage{lineno}
\usepackage{lhchiggs,heppennames2,cernunits,hepparticles}
\usepackage[abs]{overpic}
\usepackage[overlay,absolute]{textpos}

\providecommand{\HERWIG}{{\sc HERWIG}}
\providecommand{\PYTHIA}{{\sc PYTHIA}}

\DeclareRobustCommand{\PA}{\HepParticle{A}{}{}\Xspace}
\DeclareRobustCommand{\PV}{\HepParticle{V}{}{}\Xspace}
\DeclareRobustCommand{\PX}{\HepParticle{X}{}{}\Xspace}
\DeclareRobustCommand{\Pf}{\HepParticle{f}{}{}\Xspace}
\DeclareRobustCommand{\PAf}{\HepAntiParticle{\Pf}{}{}\Xspace}

\newcommand{\bqas}{\begin{eqnarray*}}
\newcommand{\eqas}{\end{eqnarray*}}

\newcommand{\bq}{\begin{equation}}                    
\newcommand{\eq}{\end{equation}}
\newcommand{\bqa}{\arraycolsep 0.14em\begin{eqnarray}}
\newcommand{\eqa}{\end{eqnarray}}
\newcommand{\ba}[1]{\begin{array}{#1}}
\newcommand{\ea}{\end{array}}
\newcommand{\ben}{\begin{enumerate}}
\newcommand{\een}{\end{enumerate}}
\newcommand{\bei}{\begin{itemize}}
\newcommand{\eei}{\end{itemize}}
\newcommand{\eqn}[1]{Eq.(\ref{#1})}

\newcommand{\fig}[1]{Fig.~\ref{#1}}

\newcommand{\mh}{\mathswitch {M_{\PH}}}
\newcommand{\mw}{\mathswitch {M_{\PW}}}
\newcommand{\mz}{\mathswitch {M_{\PZ}}}
\newcommand{\mt}{\mathswitch {M_{\PQt}}}

\providecommand\POWHEG{{\sc POWHEG}}

\providecommand{\MCatNLO}{M\scalebox{0.8}{C}@N\scalebox{0.8}{LO}\xspace}
\providecommand{\Sherpa}{S\scalebox{0.8}{HERPA}\xspace}

\newcommand{\GoSam}{G\protect{O}S\protect{AM}\xspace}

\newcommand{\smartpap}{p\hskip-7pt\hbox{$^{^{(\!-\!)}}$}}
\newcommand{\smallw}{{\scriptscriptstyle W}}
\newcommand{\mwav}{m_\smallw} 
\newcommand{\mwc}{m_{\smallw 0}} 
\newcommand{\mwi}{m_{\smallw,i }} 

\providecommand{\sla}[1]{\ifmmode%
  \setbox0=\hbox{$#1$}%
  \setbox1=\hbox to\wd0{\hss$/$\hss}\else%
  \setbox0=\hbox{#1}%
  \setbox1=\hbox to\wd0{\hss/\hss}\fi%
  #1\hskip-\wd0\box1 }

\newcommand\Herwigpp{{\tt Herwig++}}
\newcommand{\POWHEGBOX}{{\tt POWHEG BOX}}
\newcommand{\MadGraph}{{\tt MadGraph}}
\newcommand{\ALPGEN}{{\tt ALPGEN}}
\newcommand{\Hgg}{{\tt H}}
\newcommand{\HggJ}{{\tt HJ}}
\newcommand{\HggJJ}{{\tt HJJ}}

\begin{document}

  

\thispagestyle{empty}
{
\setlength{\unitlength}{1mm}
\begin{picture}(0.001,0.001)
\put(0,-50){\huge\bf INFN}
\put(18,-100){\LARGE\bfseries
            The Standard Model from LHC to future colliders}
\put(117,-185){\Large Editors:}
\put(135,-185){\Large S.~Forte}
\put(135,-191){\Large A.~Nisati}
\put(135,-197){\Large G.~Passarino}
\put(135,-203){\Large R.~Tenchini}

\end{picture}
}


\pagenumbering{roman}




\newpage

\newpage
\vspace*{10cm}
\begin{center} 
 {\bf Abstract}
\end{center}
\vspace{0.5cm}
This review summarizes the results of the activities which have taken
place in  $2014$ within the
 Standard 
Model Working Group of the ``What Next'' Workshop organized by INFN, Italy. 
We present a framework, general questions, and some indications of possible 
answers on the main issue for Standard Model physics in the LHC era and in view of possible 
future accelerators.

\newpage
\mbox{}

\newpage
\begin{flushleft}

 S.~Forte$^{1}$,\,
 A.~Nisati$^{2}$,\,
 G.~Passarino$^{3}$\,and\,
 R.~Tenchini$^{4}$\,(eds.);\\
 C.M.~Carloni~Calame$^{5}$,\,
 M.~Chiesa$^{6}$,\,
 M.~Cobal$^{7}$,\,
 G.~Corcella$^{8}$,\,
 G.~Degrassi$^{9}$,\,
 G.~Ferrera$^{1}$,\,
 L.~Magnea$^{3}$,\,
 F.~Maltoni$^{10}$,\,
 G.~Montagna$^{5,6}$,\,
 P.~Nason$^{11}$,\,
 O.~Nicrosini$^{6}$,\,
 C.~Oleari$^{11}$,\,
 F.~Piccinini$^{6}$,\,
 F.~Riva$^{12}$,\,
 A.~Vicini$^{1}$

\end{flushleft}
 
\begin{itemize}

\item[$^{1}$] Dipartimento di Fisica, Universit\`a di Milano and \\ 
         INFN,      Sezione di Milano, Via Celoria 16, I-20133 Milano, Italy

\item[$^{2}$] INFN, Sezione di Roma \\
              Piazzale Aldo Moro 2, I-00185 Roma, Italy

\item[$^{3}$]  Dipartimento di Fisica, Universit\`a di Torino and\\
               INFN, Sezione di Torino, Via P. Giuria 1, I-10125 Torino, Italy

\item[$^{4}$] INFN, Sezione di Pisa \\
              Largo B. Pontecorvo 3, I-56127 Pisa, Italy

\item[$^{5}$] Dipartimento di Fisica, Universit\`a di Pavia, \\
              via Bassi 6, I-27100 Pavia, Italy

\item[$^{6}$] INFN, Sezione di Pavia, via Bassi 6, I-27100 Pavia, Italy

\item[$^{7}$] Dipartimento di Chimica, Fisica e Ambiente,
 Universit\`a di Udine and\\INFN, Gruppo Collegato di Udine, 
   Via delle Scienze, 206 - I-33100 Udine, Italy

\item[$^{8}$] INFN,  Laboratori Nazionali di
              Frascati  \\
              Via E. Fermi 40, I-00044 Frascati, Italy

\item[$^{9}$] Dipartimento di Matematica e Fisica, Universit\`a' Roma Tre and \\
              INFN, sezione di Roma Tre, 
              Via della Vasca Navale 84, I-00146 Roma, Italy

\item[$^{10}$] Centre for Cosmology, Particle Physics and Phenomenology (CP3),\\  
               Universit\'e Catholique de Louvain, B-1348 Louvain-la-Neuve, Belgium

\item[$^{11}$] Dipartimento di Fisica, Universit\`a di Milano-Bicocca and\\ INFN, Sezione di Milano-Bicocca,
Piazza della Scienza 3, I-20126 Milano, Italy

\item[$^{12}$] Institut de Th\'eorie des Ph\'enom\'enes Physiques,\\
  \'Ecole Polytechnique F\'ed\'erale de Lausanne, 1015 Lausanne,
                Switzerland

\end{itemize}

\newpage
\tableofcontents
\newpage
\mbox{}

\newpage
\pagenumbering{arabic}
\setcounter{footnote}{0}
\section{Synopsis \footnote{S.~Forte, A.~Nisati, G.~Passarino, R.~Tenchini}}

 The goal of this review is to develop the community's long-term physics 
 aspirations. Its narrative aims to communicate the opportunities for
 discovery in physics with high-energy colliders:
this area includes experiments 
 on the Higgs boson, the top quark, the electroweak and strong interactions. It also 
 encompasses direct searches for new particles and interactions at high energy.
 To address these questions a debate in the community is necessary, 
 having in mind that several topics have overlapping boundaries.
This document summarizes some aspects of our  
current, preliminary understanding, expectations, and recommendations.
 
Inevitably, the starting point is the need for a better understanding of  the current
``We haven't seen anything (yet)'' theoretical  environment. Is the Standard Model (SM) with a 
$125\UGeV$ Higgs the Final
 Theory, or indeed could it be? The associated problems are known. Some
 of them (neutrino masses, strong CP, gauge coupling unification,
 cosmological constant, hierarchy problem) in principle could not be problems at all, but
 just a theoretical prejudice. Others (\eg dark matter) seem rather harder 
to put aside. 
Indeed, while some of these questions point to particular energy
scales and types of experiments,  there is no scientific reason to justify the 
 belief that all the big problems have solutions, let alone ones we
 humans can find.

 Since exploration of the TeV scale is still in a preliminary stage it would be 
 advisable to keep options open, and avoid investing all resources on a
 single option, be it increasing precision of theory predictions (see 
Ref.~\cite{xsectcomp} for an extensive compilation)
 and experimental results, or the search  
 for new models (and the resolution of the issue of the relevance of naturalness), see 
 \Brefs{Giudice:2013nak,Dittmaier:2011ti}.

 In order to set up a framework for addressing these issues, in this
 introduction we draw a very rough roadmap for future scenarios.
 The bulk of this document will then be devoted to a summary
 of what we believe to be the key
 measurements and some of the main tools which will be necessary in
 order to be able to interpret their results with the goal of answering some of the broader
 questions. 
 
 \subsection {Scenarios}
 It is useful to discuss possible scenarios separating three different
 timescales: a short one, which more or less coincides with the
 lifetime of LHC and its luminosity upgrade (HL-LHC), a medium one, related to an 
 immediate successor of the LHC (such as the ILC), and a long timescale for future machines such as
 future circular colliders (such as a FCC).
 
 \subsubsection{Scenarios for LHC physics} 
In the short run, a possible scenario  is that  
nothing but the Standard Model is seen at LHC energies, with no detection 
of dark matter: an uncomfortable situation, in view of the fact that
dark matter is at least ten percent of the mass 
density of the  Universe~\cite{Agashe:2014kda}. 
A minimal approach to this situation could be to simply ignore some of the problems (hierarchy, 
gauge coupling unification,  strong CP, cosmological constant), and extend the SM in the minimal 
way that accommodates cosmological data. For instance, introduce real scalar dark matter, 
two right-handed neutrinos, and a real scalar inflaton. 
With the risk, however, of ending up with a Ptolemaic theory, in which new pieces of data 
require ever new epicycles.

A more agreeable scenario (obviously, a subjective point of view)
is one in which nonstandard physics is detected in the Higgs sector
(or, possibly less likely, in the flavor sector). 
This could possibly occur while looking at the Higgs width (possibly through interferometry ~\cite{Dixon:2003yb,Kauer:2012hd,Caola:2013yja,Campbell:2013una,Passarino:2013bha,Campbell:2015vwa}
beyond the narrow width approximation~\cite{Kauer:2012hd}), decays
(including vector meson~\cite{Bodwin:2013gca} and rare Dalitz~\cite{Passarino:2013nka}), and 
more generally anything that would use the Higgs as a probe for new physics (Higgs, top-Higgs
anomalous production modes, with new loop contributions, associate productions, trilinear couplings).

It is likely that, if such a discovery (\ie a discovery connected to Higgs interactions)
will happen, it will be through the
combination of electroweak precision data with Higgs physics~\cite{Baak:2014ora}. This
means, on the one hand, controlling the general features of
electroweak symmetry breaking (EWSB), specifically through the determination
of SM parameters  ($\mt, \mw, \alphas$, \etc) from global fits but
also through the study of processes which are directly sensitive to
EWSB, such as  $\PV\PV\,$-scattering~\cite{Degrande:2013yda}. On the
other hand, it means developing predictions and tools~\cite{Artoisenet:2013puc,Cranmer:2013hia}
to constrain the space of couplings of the 
Effective Field Theory (EFT).
 
The most powerful strategy for looking for deviations from the
SM~\cite{LHCHiggsCrossSectionWorkingGroup:2012nn} will require the
determination of the Wilson coefficients for the most general set of
$\mathrm{dim}=6$ operators (see Ref.~\cite{Henning:2014wua}). With enough statistics these could 
be determined, and they will then be found to be either close to zero (their SM value), or 
not. In the former case, we would conclude that both next-to-leading (NLO) corrections (and the 
residual theoretical uncertainty at NNLO level) and the coefficients are small: so the SM is 
actually a minimum in our Lagrangian space or very close to it. This would be disappointing, 
but internally consistent, at least up to the Planck scale~\cite{Giudice:2013nak}. 

The latter case would be more interesting, but also problematic.
Indeed, operators whose coefficients are found to be large would have
to be treated beyond leading-order (LO). This means that we should move in the
Lagrangian space and adopt a new renormalizable Lagrangian in which the Wilson 
coefficients become small; local operators would then be redefined with respect to the new  
Lagrangian. 
Of course there will be more  Lagrangians projecting into the same set of operators but still we
could see how our new choice handles the rest of the data.
 In principle, there will be a blurred arrow in our space of Lagrangians, and we should simply 
focus the arrow. This is the so-called inverse problem~\cite{ArkaniHamed:2005px}: if LHC finds 
evidence for physics beyond the SM, how can one determine the underlying theory?
It is worth noting that we will always have problems at the interpretation level of the 
results.
 
The main goal in the near future will be to identify the structure of the effective Lagrangian 
and to derive qualitative  information on new physics; the question of the ultraviolet 
completion~\cite{Adams:2006sv,Dvali:2010jz} cannot be answered, or at least not in a unique way, 
unless there is sensitivity to  operators with dimension greater than
6. 
The current goals are therefore rather less ambitious that
the ultimate goa;l of  understanding if the effective theory can be 
UV completed~\cite{Contino:2013kra,Passarino:2012cb,Ghezzi:2014qpa,Low:2009di}.
 
 What might actually be needed is an overall roadmap to Higgs precision measurements.
 From the experiments we have some projections on which experimental
 precision is reachable in  different channels in the next few years.
The logical next step would be to determine what kind of theoretical precision we
 need for each channel to match this experimental precision, as a function
 of time. Based on this, we can then define the precision we need for each parameter 
measurements using EFT. This then determines in 
 a very general way what kind of work is needed from the theory community. 

 \subsubsection{Physics at the ILC}
As a  next step, ILC~\cite{Asner:2013psa} 
plans to provide the next significant step in the 
 precision study of Higgs boson properties~\cite{Peskin:2013xra}. LHC precision measurements 
 in the $5{-}10\%$ range should be brought down to the level of  $1\%$. But this 
 means that the strategy  discussed
 above~\cite{LHCHiggsCrossSectionWorkingGroup:2012nn} must be
 upgraded by the inclusion of higher order electroweak corrections.

This is not precision for precision's sake, rather, the realization
that precision measurements and accurate theory predictions represent
a fundamental approach in the search for new physics beyond the SM.
For instance,
while a machine  with limited precision may only claim a discovery of
a SM-like Higgs boson, once greater precision is achieved,  it may be
possible to rule out the SM nature of the  Higgs boson through the accurate
determination of its couplings. A tantalizing example of such a
situation is provided by the current status of the vacuum stability problem:
the vacuum is at the  verge or being stable or metastable, and a sub-percent change of 
$\sim 1\UGeV$ in either  $\mt$ or $\mh$ is all it takes to tip the scales~\cite{Degrassi:2012ry,Buttazzo:2013uya}.

This, however, raises new challenges.
For example, the ILC plans to measure $\sigma_{\PZ\PH}$. Of course, this is a
 pseudo-observable: there are  neither $\PZ$ nor $\PH$ particles in a detector.
Precision physics thus raises the issue of how  ``unobservable'' 
 theoretical entities are defined, which is a very practical issue, 
given that an unobservable quantity is not uniquely defined
 (what is the up quark mass? or even the top quark mass?).

It is important however to understand that naturalness, which has been
perhaps so far the main guiding principle, has largely lost this 
role~\cite{Dine:2015xga,Giudice:2013yca}. 
It is still well possible that naturalness can be relaxed to a
sufficient extent that it still holds in some plausible sense ---
after all, the SM is a renormalizable theory, up to Landau poles
it is completely fine and predictive, and it can thus stretched at
will~\cite{Altarelli:2013lla}. 
It is plausible to assume that Nature has a way, still hidden to us, to realize 
a deeper form of naturalness at a more fundamental level, but this
gives no guidance on the relevant scale: we then have no alternative
to looking for the smallest possible deviations.
 
 \subsubsection{The far future}
 Given that sufficiently precise measurements of the Higgs properties and
 the EWSB parameters are ideal probes for the new physics scale, a
 future circular collider (FCC) 
 could be the best complementary machine to LHC. This includes the,
 partly complementary, 
 $\Pe\Pe$, $\Pe\Pp$ and $\Pp\Pp$ options. At $\sqrt{s} = 500\UGeV$ the
 luminosity of a FCC-$ee$~\cite{Gomez-Ceballos:2013zzn} and ILC would be
 comparable; additional  
 luminosity would improve the precision of Higgs couplings of only a factor $\sqrt{2}$. However, 
 the opening of $\Pep\Pem \to \PAQt\PQt\PH$ process allows the
 $\PAQt\PQt\PH$ coupling to 
 be measured, with a global fit, with a precision of $10\%$ at the FCC-$ee$. The potential
 of the $\Pe\Pp$  and $\Pp\Pp$ options  has just started being explored.

 \subsubsection{A new frontier?}
As we already mentioned several times, many of the currently
outstanding problems --- naturalness, the UV behavior of the Higgs sector ---
point to the possibility that electroweak symmetry breaking may be
linked to the vacuum stability problem: is the Higgs potential at $M_{\mathrm{plank}}$ 
flat~\cite{Degrassi:2012ry,Buttazzo:2013uya}, and if so, why?  This
then raises the question whether perhaps EWSB might be
determined by Planck-scale physics, which, in turn,  begs the
question of the matching of the SM to gravity. 
Of course, BSM physics (needed for dark matter) could change the picture, 
by making the Higgs instability problem worse, or by curing it.

But the fact remains that we do not have a renormalizable quantum field theory of gravity. The
ultimate theoretical frontier then would be understanding how to move beyond quantum field theory. 
 
 \subsection{Measurements and tools}
 
 In order to address the issues outlined in the previous section, a number of crucial measurements 
 are necessary. Some of these have a clear time frame: for example, Higgs couplings will surely be
 extensively measured at the LHC, while double Higgs production (and trilinear
Higgs couplings) will only be accurately measured 
 at future accelerators. Other measurements will be performed with increasing precision on 
 different  timescales. Extracting from these measurements the information that
 we are after will in turn require the development of a set of analysis tools. 

The main purpose of this note is to summarize the status and prospects
for what we believe to be some of the most important directions of progress,
both in terms of measurement and tools. 

First, we will discuss crucial measurements. Specifically, we will address
gauge boson mass measurement, that provide perhaps the most precise of
SM standard candles. We will then discuss the mass of the top quark,
which, being the heaviest fundamental field of the Standard Model
Lagrangian provides a natural door to new physics. We will finally
address the Higgs sector: on the one hand, by analyzing our current
and expected future knowledge of the effective Lagrangian for the
Higgs sector (up to dimension six 
operators), and
then by discussing the implications for the stability of the
electroweak vacuum, which is ultimately related to the way the
Standard Model may open up to new physics.

We will then turn to some selected methods and tools: resummation
techniques, which are expected to considerably improve the accuracy
and widen the domain of applicability of perturbative QCD
computations, and then the Monte Carlo tools which provide 
the backbone of data analysis, both for the strong and the
electroweak interactions.

\hfill\eject
\section{The $\PW$ and $\PZ$ mass and electroweak precision physics\footnote{A.~Vicini}}
\label{sec:WZ}
\subsection{Introduction}

\subsubsection{Relevance of a high-precision $\mwav$ measurement}

The $\PW$ boson mass has been very precisely measured at the Tevatron 
CDF ($\mwav=80.387\pm 0.019\UGeV$) \cite{Aaltonen:2013vwa} and D0 ($\mwav=80.375\pm 0.023\UGeV$) \cite{D0:2013jba} experiments,
with a world average now equal to $\mwav=80.385\pm 0.015\UGeV$ \cite{Beringer:1900zz}.
There are prospects of a further reduction of the total error
by the LHC experiments,
and a value of $15$ or even $10\UMeV$ is presently discussed \cite{Buge:2006dv,Besson:2008zs}.
These results offer the possibility of a high precision test of the gauge sector of the Standard Model (SM).
The current best prediction in the SM is 
($\mwav=80.361\pm 0.010\UGeV$)  from muon decay \cite{Ferroglia:2012ir};
it has been computed including the full 2-loop electroweak (EW) corrections
to the muon decay amplitude \cite{Awramik:2003rn}
and partial three-loop (${\mathcal O}(\alpha\alpha_{s}^{2})$, ${\mathcal O}(\alpha_{t}^{2}\alpha_{s})$, ${\mathcal O}(\alpha_{t}^{3})$) and four-loop QCD corrections ${\mathcal O}(\alpha_{t}\alpha_{s}^{3})$, where $\alpha_{t}=\alpha m_{t}^{2}$ \cite{Schroder:2005db,Chetyrkin:2006bj,Boughezal:2006xk}.
Alternatively, the value $\mwav=80.358\pm 0.008\UGeV$ is obtained 
from a global EW fit of the SM \cite{Baak:2014ora}.
The error on this evaluation is mostly due to parametric uncertainties of the inputs of the calculation, the top mass value, the hadronic contribution to the running of the electromagnetic coupling, and also to missing higher-order corrections.

The comparison of an accurate experimental value with the predictions of different models might provide an indirect signal of physics beyond the SM \cite{Baak:2013fwa}.
The value $\mwav^i$ computed in model $i$ follows form the relation
$\frac{G_\mu}{\sqrt{2}} = \frac{g^2}{8(\mwav^i)^2} (1+\Delta r^i)$
where the radiative corrections to the muon decay amplitude 
are represented by the parameter 
$\Delta r^i=\Delta r^i(\mwav^i,M_{\mathrm{SM}},M_{\mathrm{BSM}})$ 
and possibly offer
sensitivity to new particles present in the considered extension $i$ of the SM,
whose mass scale is generically indicated with $M_{\mathrm{BSM}}$.

\subsubsection{Physical observables}
At hadron colliders, the $\PW$ boson mass is extracted from the study of the charged-current (CC) Drell-Yan (DY) process,
$p\smartpap\to \Pl^+\PGn_l+X$ (and also $p\smartpap\to \Pl^-\PAGn_l+X$).
In the leptonic final state the neutrino is not measured, so that the invariant mass of the lepton pair can not be reconstructed.
The value of $\mwav$ is determined from the study of the lepton transverse momentum, of the missing transverse energy and of the lepton-pair transverse mass distributions.
These observables have an enhanced sensitivity to $\mwav$ because of their jacobian peak at the $\PW$ resonance\cite{Rujula:2011qn}.
More precisely, it is the study of their shape, rather than the study of their absolute value,
which provides informations about $\mwav$
These observables are defined in terms of the components of the lepton momenta in the transverse plane.
The main experimental uncertainties are related to the determination of the charged lepton energy or momentum
on one side, and, on the other side, to the reconstruction of the missing transverse energy distribution, so that the neutrino transverse momentum can be inferred in an accurate way. 
The modeling of the lepton-pair transverse momentum distribution also plays a major role in the determination of the neutrino components.
A systematic description of the size of the experimental uncertainties affecting the measurement and of their impact on the $\mwav$ measurement can be found in \cite{Aaltonen:2013vwa,D0:2013jba,Besson:2008zs,Buge:2006dv}.

\subsubsection{Sensitivity to $\mwav$ of different observables}
\begin{figure}[!h]
\begin{center}
\includegraphics[width=70mm,angle=0]{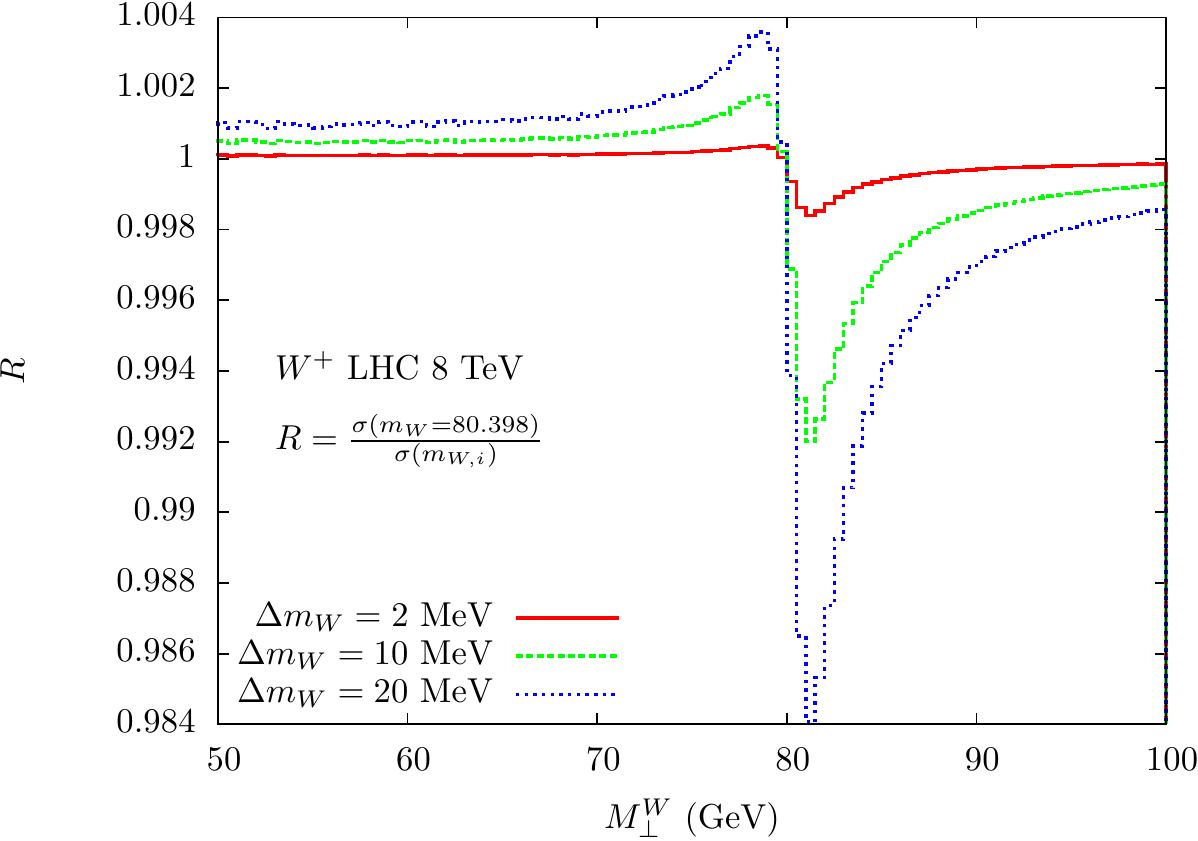}
\includegraphics[width=70mm,angle=0]{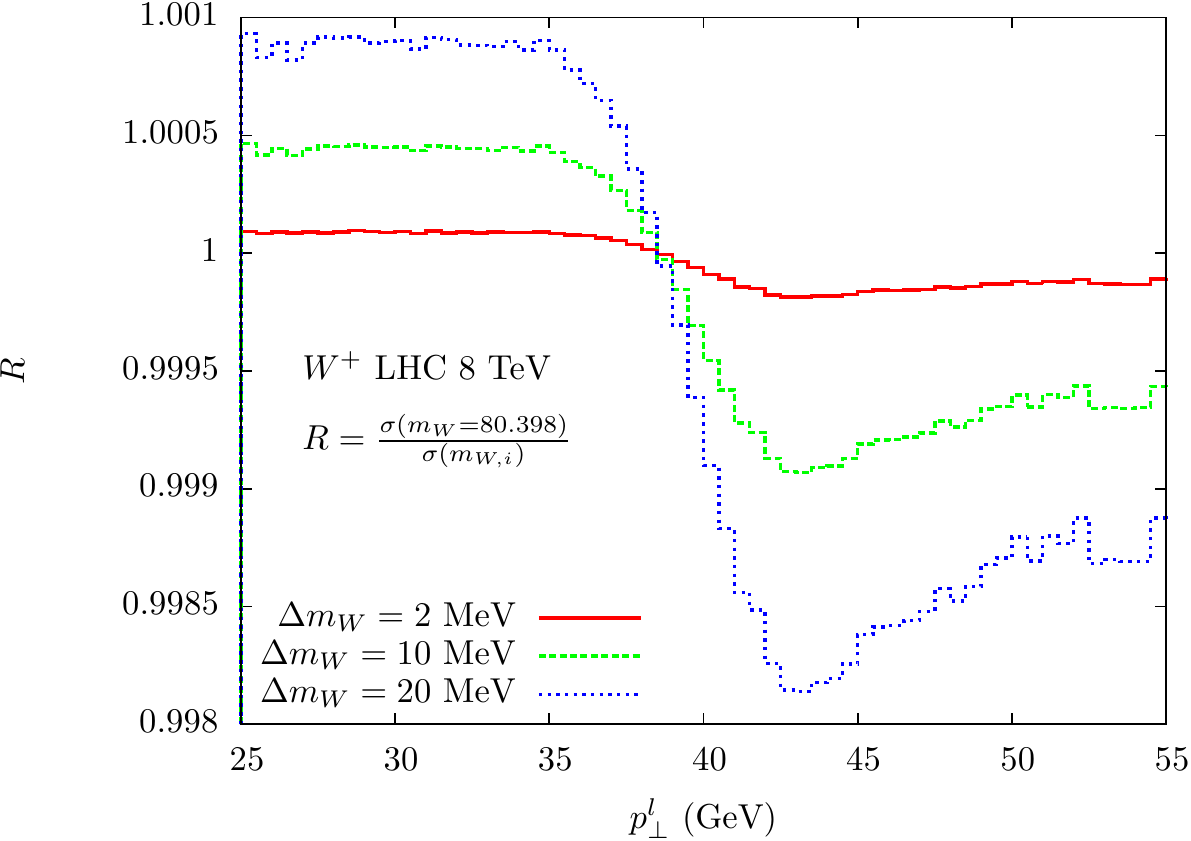}
\caption{ 
Ratio of lepton-pair transverse mass (left) and lepton transverse momentum (right) distributions
which have been generated with different $\PW$ boson masses.
\label{fig:sensitivity}
}
\end{center}
\end{figure}

The sensitivity of the observables to the precise $\mwav$ value
can be assessed with a numerical study of their variation under a given shift of this input parameter.
In \Fref{fig:sensitivity} we show the ratio of two distributions obtained with $\mwc=80.398\UGeV$ and shifted, $\mwi=\mwc+\Delta\mwav$. The distortion of the shapes amounts to one to few parts per mill, depending if one considers the lepton transverse momentum or the lepton-pair transverse mass.
We can rephrase this remark by saying that a measurement of $\mwav$ at the $10\UMeV$ level requires the control of the shape of the relevant distributions at the per mill level.
The codes used to derive the results in \Fref{fig:sensitivity} do not include the detector simulation; the conclusions about the sensitivity to $\mwav$
should be considered as an upper limit, which can be reduced by additional experimental smearing effects. 

The $\PW$ boson mass is measured by means of a template fit approach:
the distributions are computed with Montecarlo simulation codes for different values of $\mwav$ and are compared with the corresponding data; the value which maximizes the agreement is chosen as the preferred value.
The templates are theoretical objects, computed with some assumptions about input parameters, proton PDF choices and
perturbative accuracy.
The uncertainties affecting the templates, missing higher orders, PDF and input parameters uncertainties,
have an impact on the result of the fit and should be treated as a theoretical systematic error.

\subsection{Available tools and sources of uncertainty}
The DY reaction in LO is a purely EW process, which receives perturbative corrections due to the EW and to the QCD interactions; in higher orders also mixed QCD{-}EW contributions appear and are of phenomenological relevance.
The observables under study have a different behaviour with respect to the perturbative corrections,
so that in some cases a fixed-order prediction is not sufficient and the resummation to all orders of logarithmically-enhanced contributions becomes necessary. 
With the resummation, three different kinds of entangled ambiguities appear in the preparation of the templates:
1) missing higher-order logarithmically-enhanced terms in the resummed expression, 
2) ambiguities of the matching between fixed-order and all-order results, 
3) the interplay, in the region of low lepton-pair transverse momenta, of perturbative and non-perturbative QCD corrections.
This latter source of uncertainty is also related to the non-perturbative effects parametrized, 
in the collinear limit, in the proton PDFs.

The $\mwav$ value follows from the precise study of the shape of the observables; 
for this reason, the use of distributions normalized to their respective integrated cross sections removes
an important class of uncertainties associated to the DY total rate determination.

\subsubsection{EW radiative corrections}
EW radiative corrections to CC and neutral-current (NC) DY are available with NLO-EW accuracy and are implemented in several public codes: 
{\sc WZGRAD} \cite{Baur:2004ig,Baur:2001ze}, 
{\sc RADY} \cite{Dittmaier:2001ay}, 
{\sc SANC} \cite{Arbuzov:2005dd}, 
{\sc HORACE} \cite{CarloniCalame:2006zq,CarloniCalame:2007cd}.
The effect of multiple photon emissions is accounted for in {\sc HORACE}
by a QED Parton Shower (PS), properly matched with the fixed-order calculation; higher-order universal effects, that can be reabsorbed in a redefinition of the tree-level couplings, 
are also available in the above codes and play an important role in the description of the NC invariant mass distribution.

Real-photon emissions from the final state leptons greatly modify the value of the measured lepton energies and momenta.
The distortion of the jacobian peak is at the level of 5-18\%, depending on the observable, on the kind of lepton and on the procedure that recombines QED radiation that surrounds the lepton into an effective calorimetric object.
The impact at ${\mathcal O}(\alpha)$ of this radiation can be estimated to yield a 
shift of $\mwav$ of ${\mathcal O}(150 \,\UMeV)$ \cite{Aaltonen:2013vwa}. Additional radiation induces a further change in the result of ${\mathcal O}(10\%)$ of the ${\mathcal O}(\alpha)$ effect.

Subleading terms, i.e. not enhanced by a final state lepton mass logarithm,
are exactly available as part of the ${\mathcal O}(\alpha)$ calculation
and are partially available at ${\mathcal O}(\alpha^2)$ thanks to the matching procedure between QED PS and exact ${\mathcal O}(\alpha)$ matrix elements. Their impact amounts to a few contributions, each yielding a shift of ${\mathcal O}(5 \,\UMeV)$.
The residual uncertainty due to missing higher orders has been estimated to be smaller than $5\UMeV$, in the framework of a purely EW analysis; it should be however kept in mind that the interplay of EW and QCD corrections leads, for some observables like e.g. the lepton tranvserse momentum distribution, 
to an increase of the purely EW estimate.

\subsubsection{QCD radiative corrections}
QCD corrections to lepton-pair production are available at fully differential level through ${\mathcal O}(\alphas^2)$ and are implemented in the Montecarlo integrators 
{\sc FEWZ} \cite{Gavin:2012sy},
{\sc DYNNLO} \cite{Catani:2009sm} and 
{\sc SHERPA} \cite{Hoeche:2014aia}.
The gauge boson transverse momentum distribution is known 
with NNLL+NLO accuracy (and with NNLO accuracy on the total cross section)
and is implemented in the Montecarlo integrator {\sc DYqT} \cite{Bozzi:2008bb}, 
without the description of the decay into leptons
\footnote{The $\beta$-version of the code that includes the gauge boson decay is available from the authors of the code.}.
The NNLL resummation, without the NNLO accuracy on the total cross section, is available in the integrator {\sc ResBos} \cite{Balazs:1997xd,Berge:2004nt}.
The effects on the total cross section and on the gauge boson rapidity distribution of the logarithmic threshold corrections have been included up to N$^3$LO+NNLL accuracy \cite{Ahmed:2014cla,Catani:2014uta}.
Standard tools for the experimental analyses are the Shower Montecarlo (SMC) event generators with NLO-QCD accuracy,
like {\sc MC@NLO} \cite{Frixione:2002ik} or {\sc POWHEG} \cite{Alioli:2008gx} (more recently {\sc HERWIG} \cite{Frixione:2010ra} or {\sc SHERPA} \cite{Hoche:2010pf}).
They have NLO-QCD accuracy on the total cross section, but only LO-QCD accuracy in the description of the lepton-pair transverse momentum.
Recently, progresses have been made in the direction of a merging of NNLO-QCD matrix elements with a QCD PS, 
in {\sc SHERPA} \cite{Hoeche:2014aia} or in {\sc NNLOPS} \cite{Karlberg:2014qua} or in {\sc GENEVA} \cite{Alioli:2012fc}.

The QCD corrections have important effects on the DY observables in terms of absolute normalization and in terms of shapes. The former can be mitigated by considering normalized distributions, while the latter are the most critical ingredient in the theoretical framework.
Among the observables relevant for the $\mwav$ measurement,
the lepton transverse momentum distribution is a paradigmatic example: 
its prediction in fixed order is affected by the very large logarithmic corrections for small lepton-pair transverse momenta and only after their resummation a sensible description becomes possible.
In this case, the evaluation of the QCD uncertainty on $\mwav$ is possible
with a joint systematic study of matching ambiguities, renormalization/factorization scale variations, of the effect of subleading logarithmic terms and of the modeling of the non-perturbative effects at very low transverse momenta \cite{Bozzi:2008bb,Guzzi:2013aja}.
A very naive estimate of the combination of all these effects, in a simplified setup, might be translated into a shift of the measured $\mwav$ by ${\mathcal O}(50-100)\UMeV$,
which would clearly be a dramatic conclusion of the uncertainty analysis.
It has been proposed in \cite{Giele:1998uh} 
to consider ratios of $\PW$ and $\PZ$ observables, with an evident reduction of the scale uncertainties both in size and in shape.
A study of the residual uncertainty on $\mwav$ in this approach is in progress
\cite{ferrera-talk}.
The published Tevatron results \cite{Aaltonen:2013vwa,D0:2013jba}
do not quote a comprehensive QCD uncertainty that includes perturbative effects; they rather use the generator {\sc ResBos} with a fixed choice of the perturbative scales and of the proton PDF
to describe the $\PZ$ boson transverse momentum distribution;
this analysis allows to fit the parameters of a model describing the non-perturbative low-transverse-momentum components of QCD radiation, which are then used to simulate the CC DY process;
this approach assumes universality of these parameters and their independence on the process energy scale.
In the Tevatron analyses the error assigned to the $p_\perp^{\PW}$ modeling is only due to a variation of the non-perturbative parameters in the range allowed by the fit of the $\PZ$ boson data.

The impact of the different QCD uncertainties mentioned above 
is milder in the case of the lepton-pair transverse mass, 
because this observable is more stable with respect to the inclusion of higher-order QCD corrections. 
The shape distorsion observed when comparing its NLO- and NNLO-QCD determinations is minimal;
the scale variations do not significantly modify the shape around the jacobian peak,
and so the impact on the $\mwav$ determination is limited.

\subsubsection{Proton PDF uncertainty}
The proton PDFs enter in the $\mwav$ determination because they are needed to compute the templates used in the fit of the data.
Different PDF set choices, or different replica choices within the same set, imply a change of the templates shape and in turn of the preferred $\mwav$ value.
The propagation of the PDF error is computed according to the prescription of each PDF collaboration, and eventually the different results can be combined with the PDF4LHC prescription \cite{Botje:2011sn,Forte:2013wc}.

Neglecting all detector effects, which have an important impact on the acceptance determination,
the PDF uncertainty on the $\mwav$ extracted from the study of the normalized lepton-pair transverse mass distribution remains below the $10\UMeV$ level \cite{Bozzi:2011ww,Rojo:2013nia},
whereas the spread in the case of the lepton transverse momentum distribution, 
again estimated at generator level,
ranges between $6$ and $18\UMeV$, depending on the chosen PDF set, collider energy and final state \cite{bcv}.
A crucial role is played by the acceptance cuts, on the leptons but also an the lepton pair.
At higher collider energies, the PDF uncertainty associated to the lepton-pair transverse mass remains stable,
whereas the one on $\mwav$ extracted from the lepton transverse momentum distribution increases 
for proton-proton collider energies between $8$ and $100\UTeV$ (cfr. table \ref{tab:mwunc}); the application of a cut $p_\perp^{\PW}<15\UGeV$ on the lepton-pair transverse momentum keeps the estimated uncertainty below the 1$5\UMeV$ level \cite{bcv}.

\subsubsection{Mixed QCD{-}EW radiative corrections}
QCD corrections, via initial state radiation, modify the kinematics of the DY events, whereas the leading EW effects are due to a variation of the lepton spectra due to final state radiation.
The interplay between these two groups of corrections is not trivial
and strongly depends on the observable under study.
The first perturbative order at which these mixed corrections appear is
${\mathcal O}(\alpha\alphas)$, but no exact calculation is available, so that one has to rely on some approximations.
The NLO-QCD and NLO-EW exact matrix elements have been implemented in {\sc POWHEG}
and have been consistently matched with both QCD-PS and QED-PS for CC \cite{Bernaciak:2012hj,Barze:2012tt} and NC \cite{Barze':2013yca} DY.
In this approach
all the QCD-LL (initial state collinear logarithms) and all the QED-LL (final state mass logarithms) corrections, in all possible combinations,  are taken into account, including the leading ${\mathcal O}(\alpha\alphas)$ terms.
The first terms that are beyond the accuracy of the code are of
${\mathcal O}(\alpha\alphas)$ and subleading in the expansion with respect to the EW logarithms.
The role of the mixed corrections is particularly relevant in the prediction of the lepton transverse momentum distribution \cite{Barze:2012tt,Barze':2013yca}.
For this quantity, as discussed in \cite{Dittmaier:2014qza,Dittmaier:2014koa}, a naive factorization recipe to combine QCD and EW corrections, fails.
The {\sc POWHEG} implementation of the QCD{-}EW combination misses, on one hand, subleading effects of ${\mathcal O}(\alpha\alphas)$; it provides, on the other hand, an exact treatment of the kinematics of each radiated parton 
and thus gives the correct convolution of QCD and EW corrections including those effect that break the factorization {\it ansatz}.
The study of the impact of different combinations of QCD and EW effects, with and without NLO accuracy, is in progress \cite{mwew}.

\subsection{Prospects of improvement}

Let us briefly discuss the prospects for a high-precision measurement of $\mwav$ at a high- energy/luminosity proton-proton collider in the next 10-20 years, 
under the assumption that progresses that today can be wished, or expected in the long term, will be available.

\subsubsection{Montecarlo generators}
\begin{enumerate}
\item
Definition of a matching procedure that allows a Montecarlo event generator
to reach NNLO-QCD accuracy on the DY total cross section and NNLL-QCD accuracy
in the resummation of the logarithms of the lepton-pair transverse momentum
(partial results are already available, by different groups).
\item
 Evaluation of the N$^3$LO-QCD corrections to the DY processes, 
   as the first step towards the construction of an integrator code
   that reaches N$^3$LO accuracy on the total cross section and N$^3$LL accuracy in the resummation
of the logarithms of the lepton-pair transverse momentum
(the results in the soft limit are already available, by different groups).\\
The formulation of an integrator with this accuracy on the lepton-pair transverse momentum
is intertwined with the consistent definition of the non-perturbative contributions to the same observable.
With a similar  tool, and with the event generator of item 1), the evaluation of the ratio of $\PW$ to $\PZ$ observables should be sufficiently stable from the QCD point of view and
the residual corresponding uncertainty on $\mwav$ could fall down to the $5\UMeV$ level; this estimate is, at the moment, a guess that can become more sound after the estimate with the presently available tools
of the QCD uncertainty on $\mwav$ extracted from ratios of $\PW$ over $\PZ$ observables.

\item
 Completion of the full calculation of the corrections at ${\mathcal O}(\alpha\alphas)$ to the DY processes,
   to fix the ambiguity affecting the combination of QCD{-}EW corrections at the first non trivial order
(partial results in the $\PW$ pole approximation are already available, matrix elements for different subprocesses that contribute at this order are available).
The analysis of the purely EW effects on the $\mwav$ determination indicates a residual uncertainty at the $5\UMeV$ level,
but suffers from being a LO-QCD study; the inclusion of the ${\mathcal O}(\alpha\alphas)$ corrections will make the conclusion more stable against QCD-scale variations.

\item
Determination of proton PDFs which can be consistently matched with an
${\mathcal O}(\alpha\alphas)$ calculation (NLO accuracy mixed QCD{-}EW).

\item
Completion of the calculation of the full set of ${\mathcal O}(\alpha^2)$ corrections, 
   to reduce the uncertainties in the calibration phase ($\PZ$ mass determination
   and precise understanding of the absolute lepton energy scale).
\end{enumerate}

\subsubsection{ Uncertainty reduction with higher energy/luminosity}
We compare the perspective at future colliders for a measurement of $\mwav$
from the lepton transverse momentum and from the lepton-pair transverse mass.
With the high luminosity projected at a high-energy (13, 33 or 100 TeV)
hadron collider,
and in particular with the high-luminosity programs planned at 13 TeV,
the number of events useful for an accurate $\mwav$ measurement
will be extremely large, making the uncertainty of statistical nature negligible,
compared to those of systematic origin (theoretical and experimental).

{\bf Higher energy and PDFs}.
The energy scale of the DY processes, relevant for the $\PW$ mass measurement, is given by the masses of the $\PW$ and $\PZ$ gauge bosons.
An increase of the center-of-mass energy of a hadron collider
reduces the values of the partonic-$x$,
the fraction of the hadron momenta carried by the colliding partons,
relevant to produce a final state of given invariant mass,
and modifies the so called parton-parton luminosity, 
i.e. the effective number of colliding partons, and eventually the cross section.
The change of collider energies has thus an impact on the PDF uncertainty, because of the different partonic-$x$ range probed. 
The PDF uncertainty on $\mwav$ measured from the lepton-pair transverse mass distribution is already today at the $10\UMeV$ level
and is improving as long as LHC data become available, with some realistic chances that a contribution to the uncertainty on $\mwav$ will become soon of the order of $5\UMeV$ \cite{Rojo:2013nia,Baak:2013fwa}. 
A preliminary estimate, at generator level, of the PDF uncertainty 
associated to the lepton transverse momentum distribution, using only the PDF set {\sc NNPDF3.0},
can be found in Table \ref{tab:mwunc}. 
These results assume the possibility of a cut on the lepton-pair transverse momentum; in the case that such an assumption could not be verified, a steeper growth of the uncertainty, up to ${\mathcal O}(25)\UMeV$ at 100 TeV, would be observed.

It will require a global effort to reduce the present ${\mathcal O}(20)\UMeV$
uncertainty down below the $10\UMeV$ level, 
because of the contribution to the uncertainty
of all the parton densities in a wide range of partonic $x$.
The use of ratios of $\PW$ over $\PZ$ observables
should partially reduce the PDF uncertainty, especially the one associated to  
gluon-induced subprocesses.
\begin{table}[!h]
\begin{center}
\begin{tabular}{|c|c|c|c|c|}
\hline
   \multicolumn{5}{|c|}{normalized distribution, additional cut $p_\perp^{\PW}<15\UGeV$} \\
\hline
& 8 \UTeV & 13 \UTeV & 33 \UTeV & 100 \UTeV \\
\hline
$W^+$ & $80.395 \pm 0.009$ & $80.400 \pm 0.010 $ & $80.402 \pm 0.010 $  & $80.404 \pm 0.013 $ \\
\hline
$W^-$ & $80.398 \pm 0.007$  & $80.391 \pm 0.006 $ & $80.385 \pm 0.007 $ & $80.398 \pm 0.011 $ \\
\hline
\end{tabular}
\caption{
Estimate of the central values and of the PDF uncertainty on $\mwav$, extracted from the normalized lepton transverse momentum distributions simulated with the {\sc NNPDF3.0\_nlo\_as\_0118} PDF set and with the {\sc POWHEG} NLO-QCD event generator matched with the {\sc PYTHIA 6.4.26} QCD Parton Shower. The fit interval is $p_\perp^l \in [29,49]\UGeV$. The templates used in the fit have been prepared with {\sc NNPDF2.3\_nlo\_as\_0118}
\label{tab:mwunc}
}
\end{center}
\end{table}


{\bf Higher luminosity and neutrino momentum determination.}
The very large number of collisions occuring at each bunch crossing in the collider
will make the so-called pile-up phenomenon more and more pronounced
with higher collider luminosity:
the latter increases the hadronic activity in the transverse plane,
making the reconstruction of the missing transverse momentum (and eventually of the neutrino transverse momentum) problematic.
As a consequence, the uncertainty on the shape of the lepton-pair transverse mass will limit the possibility of a high-precision measurement.

\subsection{Conclusions}
\begin{itemize}
\item
The progress in the calculation of higher order QCD and EW corrections
seems to offer some chances that adequate theoretical tools will become available
to perform a $\mwav$ measurement at the $10\UMeV$ level.

\item
The lepton transverse momentum distribution has a very clean experimental definition
and does not suffer from the pile-up problems that show-up with high-luminosity conditions, provided that appropriate lepton isolation criteria are validated and applied.
On the other hand it is extremely sensitive to any detail of the QCD description,
both in the perturbative regime and for what concerns the PDF uncertainties, which could jeopardise any hope of measuring $\mwav$ at the $10\UMeV$ level.
The definition of $\PW$ over $\PZ$ ratios could be the clue to significantly reduce all common theoretical systematics, as demonstrated in \cite{Giele:1998uh}; this same approach could also help to mitigate the PDF uncertainty.
The availability of predictions with N$^3$LO+N$^3$LL accuracy should make it possible
to reduce the QCD systematic error below the $10\UMeV$ level.

\item
The lepton-pair transverse mass distribution
has a very mild dependence on the details of QCD corrections, 
so that it should be possible to make its theoretical prediction
accurate enough,
to contribute with a systematic error at the $10\UMeV$ level. 
The PDF uncertainty on this observable is moderate
and will benefit of the inclusion of more LHC data in the global PDF fit.
On the other hand,
the accuracy of the measurement will deteriorate in presence of higher luminosity conditions, mostly because of 
increasing pile-up effects that disturb the identification of the hard scattering process.

\end{itemize}

\hfill\eject
\section{Top quark physics\footnote{M.~Cobal and G.~Corcella}}

\subsection{Introduction}

The top quark, discovered in 1995 \cite{Abe:1995hr,Abachi:1995iq}, 
is nowadays the heaviest among the known elementary particles.
It plays a crucial role in the Standard Model phenomenology and the 
electroweak symmetry breaking:  thanks to its large mass, it exhibits the
largest Yukawa coupling with the Higgs boson and therefore it is very important in
the precision tests of the electroweak interactions. 
The top-quark mass $m_{\PQt}$ is a fundamental parameter of the
Standard Model: even before the Higgs boson discovery 
\cite{Aad:2012tfa,Chatrchyan:2012ufa}, it was used, together with the
$\PW$ boson mass, to constrain the Higgs boson mass in the global fits.
With few exceptions which will be discussed in the following,
all measurements for both $\PQt\PAQt$ and single-top production
are in agreement with the Standard Model expectations. 
Nevertheless, top quark phenomenology will remain one of the main fields
of investigation in both theoretical and experimental particle physics,
at any present and future facility, \ie both lepton and hadron colliders, as well
as linear and circular accelerators. 
Hereafter, we shall discuss the future perspectives regarding
the measurement of the top-quark properties, taking
particular care about its mass, couplings and final-state
kinematic distributions.

\subsection{Top quark mass}
The mass of the top quark ($m_{\PQt}$) is a fundamental physical
quantity and its current world average is
$m_{\PQt}=173.34\pm 0.27$ (stat) $\pm 0.71$ (syst) \UGeV\cite{ATLAS:2014wva}.
Besides its role in the precision tests, it was 
found 
that, using updated values for Higgs and top masses
and assuming that possible new physics interactions at the Planck scale
do not affect the stability phase diagram
and the electroweak vacuum lifetime
(see, \eg \cite{Branchina:2014usa} 
for an alternative treatment of this point),
the Standard Model vacuum lies on the border
between stability and metastability regions \cite{Degrassi:2012ry}.
This result implies that, 
if the central value of $m_{\PQt}$ had to shift or the uncertainty got
reduced or enhanced, the vacuum may still sit on the border between
stability and metastability zones, or be 
located completely inside one of them. 
Therefore, it is mandatory to measure the top mass with the highest 
possible precision and having all sources of errors under control.
Moreover, a crucial assumption employed by the authors of \cite{Degrassi:2012ry}
is that the measured mass corresponds to the top-quark pole mass. 
Nevertheless, 
as will be clarified later on, the connection 
between the top mass measured in current analyses of experimental 
data and the pole mass is not
straightforward and, although the two values should be reasonably close, 
any effort to clarify the top
mass interpretation is important in order to validate or modify the outcome of electroweak fits or 
the study in Ref.~\cite{Degrassi:2012ry}.

Furthermore, the top mass plays a role in 
inflationary universe theories and in the open issue regarding whether the 
inflaton can be the Higgs field or not. As discussed, \eg
in \cite{DeSimone:2008ei}, in inflationary theories 
the running of the couplings is important and, once the 
the Yukawa coupling is determined from the top mass,
the spectral index crucially depends on both the top and
Higgs masses. 


The standard methods to measure the top mass at hadron colliders, where $\PQt\PAQt$
pairs are produced in $\PQq\PAQq$ (dominant at the Tevatron) or $\Pg\Pg$ 
(dominant at the LHC) annihilation, are based on the the investigation
of the properties of the final states in top decays ($\PQt\to \PQb\PW$), which, 
according to the $\PW$ decay mode, are classified
as all leptons, leptons+jets or all jets. 
In all cases, there are two b-tagged jets, whereas the $\PW$ decay
products are reconstructed as isolated leptons (muons or electrons) or as jets 
(for $\PW\to \PQq\PAQq'$ processes).
After requiring energy-momentum conservation and constraining the $\PW$ mass, 
the final-state invariant-mass distribution exhibits a peak, 
which is interpreted as the production of a top quark.

The conventional likelihood-type techniques to reconstruct the top mass are 
the matrix-element and template methods.
The matrix-element method compares the measured quantities with predictions
obtained by convoluting the LO $\PQt\PAQt$ cross section with the detector
response. The template method is based on investigating
several distributions of observables depending on $m_{\PQt}$, under the assumption that
the final state is $\PW\PQb\PW\PQb$ and the $\PW$ mass is known; the data are then
confronted with Monte Carlo templates and $m_{\PQt}$ is the value which minimizes the
$\chi^2$. Matrix-element and template methods are those used in the world average
determination, based on the updated measurements from D0, CDF, ATLAS and
CMS Collaborations.
The projections for the LHC run at $14\UTeV$, with a $\PQt\PAQt$ cross section
about 951 pb, according to the template/matrix-element
methods \cite{Juste:2013dsa}, are quoted in Table~\ref{tab:tempme}.
\begin{table}[!ht]
\begin{center}
\begin{tabular}{|c|c|c|}
\hline
${\cal L}$ & $\delta m_{\rm stat}$ & 
$\delta m_{\rm sys}$ \\
\hline
$100$\ifb & $40\UMeV$ & $700\UMeV$\\
$300$\ifb & $30\UMeV$ & $600\UMeV$\\
\hline
\end{tabular}
\caption{
Estimated statistical and systematic uncertainties on the
top mass measurement at the LHC, 
by using template and matrix-element methods, at 14~TeV and
100 and 300\ifb integrated luminosity.
\label{tab:tempme}
}
\end{center}
\end{table}
Other strategies which have been proposed are the so-called endpoint 
\cite{Chatrchyan:2013boa}
and $J/\psi$ 
\cite{Kharchilava:1999yj,Chierici:2006dh}
methods.
In fact, in the dilepton channel, 
the endpoint of distributions like the $b$-jet+$\ell$ invariant mass $m_{b\ell}$
or the $\mu_{\PQb\PQb}$ and $\mu_{\ell\ell}$ variables, related to the $\PQb\PQb$ and 
$\ell\ell$ invariant masses as discussed in \cite{Burns:2008cp}, 
after costraining the $\PW$ and neutrino masses, are directly
comparable with $m_{\PQt}$. 
\Bref{CMS:2013wfa} presents the projections for statistical and systematic
errors on the top mass reconstruction by means of the endpoint method,
as reported in Table~\ref{tab:endp}.
\begin{table}[!h]
\begin{center}
\begin{tabular}{|c|c|c|c|}
\hline
$\sqrt{s}$ & ${\cal L}$ & $\delta m_{\rm stat}$ & 
$\delta m_{\rm sys}$ \\
\hline
$13\UTeV$ & $30$\ifb & $400\UMeV$ & $1\UGeV$\\
$14\UTeV$ & $300$\ifb & $100\UMeV$ & $600\UMeV$\\
$14\UTeV$ & $3000$\ifb & $40\UMeV$ & $500\UMeV$\\
\hline
\end{tabular}
\caption{Projections of the expected uncertainties on the
top mass, by using the endpoint metod.
\label{tab:endp}
}
\end{center}
\end{table}
\begin{table}[!h]
\begin{center}
\begin{tabular}{|c|c|c|c|}
\hline
$\sqrt{s}$ & ${\cal L}$ & $\delta m_{\rm stat}$ & 
$\delta m_{\rm sys}$ \\
\hline
$13\UTeV$ & $30$\ifb & $1\UGeV$ & $1.5\UGeV$\\
$14\UTeV$ & $300$\ifb & $300\UMeV$ & $800\UMeV$\\
$14\UTeV$ & $3000$\ifb & $100\UMeV$ & $600\UMeV$\\
\hline
\end{tabular}
\caption{As in Table~\ref{tab:endp}, but using the $J/\psi$ method.}
\label{tab:jpsi}
\end{center}
\end{table}
In all cases, the dominant uncertainties are the ones
due to hadronization and jet energy scale.

The $J/\psi$ method relies instead on the fact that, although the $B\to J/\psi$ decay
is a rare one, in the dilepton channel and exploiting the 
$J/\psi\to \PGmp\PGmm$ mode, the three-lepton invariant mass $m_{3\ell}$ as well
as the $m_{J/\psi\ell}$ spectra allow a reliable fit of $m_{\PQt}$ at the LHC, especially
in scenarios with both high energy and high luminosity.
Table~\ref{tab:jpsi} contains the expectations for statistical and systematic
uncertainties at $13\UTeV$ (${\mathcal L}=30$\ifb)
and at $14\UTeV$ (${\mathcal L}=300$ and $3000$\ifb), as presented
in \Bref{CMS:2013wfa}.
Given such numbers, calculating the overall
uncertainty on $m_{\PQt}$ is straightforward.
In all cases, the dominant source of theory error is the treatment
of bottom-quark fragmentation in top decays, discussed in
\cite{Corcella:2009rs} in the framework of parton shower generators and in
\cite{Cacciari:2002re,Biswas:2010sa} by using NLO QCD calculations.
As far as possible future runs at $33\UTeV$ and $100\UTeV$ are concerned, 
the total error on the recostruction of the top mass based on the $J/\psi$  method
is predicted to be $1\UGeV$ and $600\UGeV$, respectively
\cite{Juste:2013dsa}.

Generally speaking, in most analyses the experimental results are compared with simulations
based on Monte Carlo generators (an exception is the endpoint method) and, 
strictly speaking, the reconstructed
top mass cannot be precisely identified with theoretical definitions like, \eg the pole mass.
In fact, programs like \HERWIG 
\cite{Corcella:2000bw} or
\PYTHIA \cite{Sjostrand:2006za} are equivalent to LO QCD calculations, 
with the resummation of all leading
(LL) and some next-to-leading soft/collinear logarithms (NLL) 
\cite{Catani:1990rr}. 
In order to fix a renormalization scheme and
get the pole or $\overline{\rm MS}$ mass, 
one would need at least a 
complete NLO computation, while
parton showers only contain the soft/collinear part of
the NLO corrections.
Furthermore, any observable yielded by such codes depends on parameters which are to  
be tuned to experimental data, in particular non-perturbative quantities, such as
the shower cutoff or the parameters entering in the hadronization models,
namely the cluster \cite{Webber:1983if} (\HERWIG) or string (\PYTHIA) 
\cite{Andersson:1983ia} models.
In fact, in 
the non-perturbative phase of the event simulation, the $\PQb$ quark from top decay hadronizes, \eg 
in a meson B$^{\pm,0}$, by combining 
with a light (anti) quark $\PAQq$, which may come from final- as well as initial-state
radiation. Since the b quark likely radiates gluons before hadronizing, the initial colour and part
of the four-momentum of the top quark may well be transferred to light-flavored hadrons, rather than
only B-hadrons. As a result, there is no unique way to 
assign the final-state particles to the initial (anti)
top quark and this leads to another contribution to 
the uncertainty (about 300 MeV in the world average) on the top mass, when reconstructed
from the invariant mass of the top-decay products.

Also, parton shower algorithms neglect the top width, $\Gamma_{\PQt}\simeq (2.0\pm 0.5)\UGeV$, 
\cite{Agashe:2014kda}) and top-production and decay phases
are assumed to factorize. 
But $\Gamma_{\PQt} /m_{\PQt}\sim {\cal O}(10^{-2})$ and therefore, for a precise mass definition with an uncertainty 
below $1\UGeV$, even width effects should be taken into account.  
Therefore, one often refers to the measured mass as a `Monte Carlo mass', which must
be related to a given theoretical definition.
Since the top mass is extracted from final-state top-decay observables,
relying on the on-shell kinematics of its decay products (leptons and jets), one should
reasonably expect the measured mass to be close to the pole mass, which is a
definition working well for an on-shell heavy quark.

In fact, calculations based
on Soft Collinear Effective Theories (SCET) \cite{Hoang:2008xm}
have proved that, assuming that the Monte Carlo mass is the
SCET jet mass evaluated at a scale of the order of
the shower cutoff, \ie $Q_0\sim {\mathcal O}(1\UGeV)$, it
differs from the pole mass by an amount
$\sim{\mathcal O}(\alphas\Gamma)\sim 200\UMeV$.
A foreseen investigation, which may help to shed light on this
issue, is based on the simulation of fictitious top-flavoured hadrons,
\eg $T^{\pm,0}$ mesons \cite{Corcella:2014rya}.
It is well known how to relate the mass of a meson to a
quark mass in any renormalization scheme. Therefore, a comparison
of final-state quantities with the top quark decaying before or
afer hadronization, and the subsequent extraction of the top mass
from their Mellin moments, can be a useful benchmark
to address the nature of the reconstructed $m_{\PQt}$ and the uncertainty
due to non-perturbative effects, such as colour reconnection.
In standard top-quark events the top quark gets its colour from an
initial-state quark or gluon and, after decaying, gives it to the bottom quark;
on the contrary, if it forms $\PQt$-hadrons, it is forced to 
create a colour-singlet.  

More recently, in order to weaken the dependence on 
the shower algorithms and non-perturbative corrections, 
other methods have been proposed to measure the top mass
at the LHC. 
One can use the total $\PQt\PAQt$ cross section, recently computed to
NNLO+NNLL accuracy \cite{Czakon:2013goa}, and extract a quantity
consistent with a theoretical mass definition, such as the pole 
mass \cite{Chatrchyan:2013haa}.
However, this analysis, though theoretically well
defined, still relies on the assumption that the mass in the Monte Carlo codes,
used to determine the experimental acceptance, is the pole mass. 
Moreover, since the total cross section exhibits
a quite weak dependence on the top mass, the resulting uncertainty is too large
for this strategy to be really competitive.
Nevertheless, the very fact that the mass determined from the cross section is 
in agreement with the value yielded by the template and matrix-element
techniques, confirms the hint that
the extracted top mass mimics the pole mass.

Another possible strategy consists of using the $\PQt\PAQt$ invariant mass 
in events with a hard jet ($j$), since it is an observable more sensitive to the
top mass than the inclusive cross section \cite{Alioli:2013mxa}.
The claim of the authors is that the unknown higher-order corrections to
the $\PQt\PAQt j$ rate should contribute less than $1\UGeV$ to the uncertainty
on $m_{\PQt}$
and that the detector effects account for ${\mathcal O}(100~{\rm MeV})$.
The ATLAS Collaboration has recently performed an analysis on the top mass
extraction by using the $\PQt\PAQt j$ rate \cite{Aad:2015waa}, along
the lines of  \cite{Alioli:2013mxa}, where the calculation of the
$\PQt\PAQt j$ cross section is performed at NLO, by using the pole 
top-quark mass. The result  $m_t^{\rm pole}=[173.7\pm 1.5 ({\rm stat})
\pm 1.4 ({\rm syst})^{+1.0}_{-0.5}({\rm theory})]\UGeV$ is presently the
most precise extraction of the pole mass.

One can also reconstruct the top mass by using the Mellin moments
of lepton ($\ell^\pm$) observables
in the dilepton channel, such as $p_{\PQt}(\ell^\pm)$, $p_{\PQt}(\ell^+\ell^-)$, $m_{\ell^+\ell^-}$, 
$E(\ell^+)+E(\ell^-)$ and $p_{\PQt}(\ell^+)+p_{\PQt}(\ell^-)$, which are typically linear
functions of the top mass \cite{Frixione:2014ala}. 
The advantage is that such
observables exhibit very little dependence on showers and non-perturbative
effects and do not require the reconstruction of the top quarks. 
The current estimate, relying on aMC@NLO \cite{Alwall:2014hca}
and MadSpin \cite{Artoisenet:2012st} for
the top-quark spin correlations, is that $m_{\PQt}$ can be reconstructed 
with an error around $800\UMeV$.

Future lepton facilities
will be an excellent environment to measure the top mass,
as it will be easier to identify top quark events than at hadron colliders.
At the moment, we have several 
proposals for lepton colliders, mainly $\Pep\Pem$ machines: the International
Linear Collider (ILC), the Compact Linear Collider (CLIC)
as well as circular colliders (TLEP). 
The potential for top-quark physics at ILC and CLIC has been
recently revisited \cite{Seidel:2013sqa}, 
with simulations of the luminosity spectra and detector response. 
Top-quark analyses at both CLIC and ILC are affected by the background due
to $\PGg\PGg$ annihilation into hadrons,
which has to be reduced.

At $\Pep\Pem$ colliders, top-pair production near threshold 
is an interesting process, where two main contrasting effects play a role:
because of the strong interaction, the $\PQt$ and the $\PAQt$ can form
a Coulomb bound state, whereas the electroweak interaction smears out
the peak of the cross section. 
The resonant cross section, computed up to NNLO accuracy
\cite{Hoang:2000yr}
by using Non Relativistic QCD, is peaked at $\sqrt{s}\simeq 2m_{\PQt}$
and behaves like $\sigma_{\rm res}\sim \alphas^3/(m_{\PQt}\Gamma_{\PQt})$; the NNNLO calculation 
is nowadays among the main
challenges in perturbative QCD.
The top mass can thus be reconstructed through
a so-called threshold scan. Besides pole and 
$\overline{\rm MS}$ masses, a particularly suitable mass definition
at threshold is the 1S mass
\cite{Hoang:1999zc} $m_{\PQt}^{1S}$, a short-distance
mass defined as half the mass of a fictitious $^3S_1$ 
toponium ground state for stable top quarks.


In order to estimate the 
uncertainty on the measurement of the top mass at a lepton collider,
a simulation scanning the range $346\UGeV <\sqrt{s}< 354\UGeV$ in steps of $1\UGeV$,
by using the TOPPIK program \cite{Hoang:1999zc}
and assuming an integrated luminosity
${\mathcal L}=300$\ifb was carried out in \cite{Martinez:2002st}.
The overall uncertainty is gauged to be about $100\UMeV$, after summing in quadrature the uncertainties
due to statistics ($30\UMeV$), luminosity ($50\UMeV$), beam energy ($35\UMeV$) 
and on the functional form of  $f(\sqrt{s_{\mathrm{res}}},m_t)$ ($80\UMeV$).
The luminosity spectrum of the machine affects the (statistical) uncertainty of the measurement:
passing from CLIC to ILC the uncertainty on the mass should improve by $10{-}20\%$.
The theoretical error, due to missing higher orders and uncertainties on the
quantities entering in the calculation, such as $\Gamma_{\PQt}$ and
$\alphas$, is predicted to be $3\%$ of the full uncertainty. 
Furthermore, a 2D template fit to the cross section can be performed as well, measuring simultaneously 
$m_{\PQt}$ and $\alphas$. Through this method, 
one can reach an uncertainty on the pole $m_{\PQt}$ of $60\UMeV$ and on the $1\mathrm{S}$ mass of 
$30\UMeV$.
Above threshold, the top mass can still be determined by using final-state distributions,
in the same manner as at hadron colliders: with $\sqrt{s}= 500\UGeV$ and 
${\mathcal L}=500$\ifb, current estimates foresee an uncertainty of
$80\UMeV$ \cite{Seidel:2013sqa}. 

\subsection{Top quark couplings}

The determination of the coupling of the top quarks to
$\PW$, $\PZ$ and Higgs bosons, as well as to photons and gluons,
is certainly a challenge in top-quark phenomenology.
In particular, possible direct measurements of the Yukawa coupling 
will be a crucial test of the Standard Model and will help
to shed light on some new physics models. 

The strong coupling constant $\alphas$ can be extracted from the measurement
of the $\PQt\PAQt$ and $\PQt\PAQt j$ cross sections. 
\Bref{Chatrchyan:2013haa} compared the NNLO calculation \cite{Czakon:2013goa}
with the measured $\PQt\PAQt$ cross section in terms of $m_{\PQt}$
and $\alphas(m_{\PZ})$. Once the top pole mass in the computation
is fixed to the world average, one can extract the strong coupling constant from
the comparison, obtaining the value $\alphas(m_{\PZ})= 0.1151^{+0.0033}_{-0.0032}$,
which is at present the first $\alphas$ determination in top-quark events
and within a NNLO analysis. The experimental (about $3.5\%$) 
and theory (about $5\%$) uncertainties are of similar order of magnitude
and are not expected to change dramatically in the future LHC operation,
namely centre-of-mass energy $13\UTeV$ and luminosity 300 fb$^{-1}$.
In future perspectives, at a linear collider, through a threshold
scan of the total cross section, it will be possible to extract
$\alphas$ with an uncertainty smaller than $1\%$ and the width
$\Gamma_{\PQt}$ with an accuracy of a few percent \cite{Martinez:2002st}.

The coupling of the top quarks to $\PW$ bosons can be measured through 
top decays and single-top production.
The helicity fractions of $\PW$ bosons in top decays have been calculated
to NNLO accuracy in \cite{Czarnecki:2010gb}, and therefore
the theory uncertainty is by far smaller than the experimental one.
A higher level of precision of the measurement of such helicities,
by exploiting the leptonic angular distributions, is thus mandatory
in the next LHC operations, 
in order to test the Standard Model in the top-decay sector as well.
As for single-top production, the LHC cross sections in the 
$s$- and $\PQt$-channel, as well
as in the $\PW\PQt$ associated-production mode, are in agreement with the Standard
Model expectations, but are affected by rather large uncertainties
(see, \eg \Brefs{Khachatryan:2014iya,Aad:2014fwa}
for the $\PQt$-channel case), with the systematic ones being even above $10\%$. 
Increasing the energy and
the luminosity of the LHC will not improve too much the accuracy
of this measurement, but nevertheless a precision of $5\%$ in the
determination of the single-top cross section and of $2.5\%$ in
the measurement of the CKM matrix element $V_{\PQt\PQb}$ is foreseen
\cite{Schoenrock:2013jka}.

Future $\Pep\Pem$ colliders will be able to measure the $\PQt\PW\PQb$ coupling with an
accuracy about $2\%$, by scanning the centre-of-mass energy between $m_{\PQt}$ and
$2m_{\PQt}$ \cite{Juste:2006sv}. Furthermore, a $\gamma e$ collider is predicted
to have a precision reach for the $\PQt\PW\PQb$ coupling between $10^{-1}$ and $10^{-2}$
\cite{Boos:2001sj},
while an $ep$ accelerator using the LHC facility at $1.3\UTeV$ 
may aim at a sensitivity within  $10^{-2}$ and $10^{-3}$
\cite{Dutta:2013mva}.

As for the top coupling to photons, although measurements of the top
charge \cite{Aad:2013uza} and of the inclusive $\PQt\PAQt\PGg$
cross section \cite{ATLAS:2011nka} are available, with the results
being in agreement with the Standard Model predictions,
it would be desirable determining the $\PQt\PAQt\gamma$ coupling with
a higher level of precision. In fact, this process suffers from 
large QCD backgrounds, and it is therefore necessary to set strong cuts to
suppress them; the NLO calculation for $\PQt\PAQt\gamma$ production
\cite{Melnikov:2011ta} will help an improved measurement 
at the LHC. 
At $14\UTeV$, with a luminosity of $300$\ifb, the coupling
to photons is expected to be measured with a precision of $4\%$, whereas at
$3000$\ifb\  the expected accuracy is expected to be about $1\%$.
As for $\PQt\PAQt\PZ$, improving the cross section measurement
as well as detecting single tops in association 
with a $\PZ$ are important challenges for the next LHC run.
At $300$\ifb\  the $\PQt\PAQt\PZ$
axial coupling can be measured
with an uncertainty of about $6\%$, but the vector one only with an
accuracy of $50\%$; increasing the luminosity to $3000$\ifb
should allow a determination of the vector coupling with an uncertainty
of $17\%$ \cite{Adelman:2013jga}.

A linear collider will certainly be an ideal environment to test the
coupling of top quarks with $\PGg$ and $\PZ$ bosons. As the 
$\Pep\Pem \to \PQt\PAQt$ process mixes photon and $\PZ$ exchanges, having
polarized beams will be fundamental to measure independently such
couplings. \Bref{Baer:2013cma} studied the reach of the linear
colliders ILC and CLIC, with polarizations of electrons and positrons
equal to $80\%$ and $30\%$, respectively, and $\sqrt{s}=500\UGeV$, finding 
that the expected precision is at the level of permille,
namely $2\times 10^{-3}$ for the coupling to photons and 
between $3\times 10^{-3}$ and $5\times 10^{-3}$ for $\PQt\PAQt\PZ$.
FCC-$ee$ should be able to permit such measurements with an even better
sensitivity, thanks to a higher luminosity; however, the absence of
polarization will not allow to disentagle of the $\PGg$ and
$\PZ$ contributions.

The determination of the Yukawa coupling of top quarks is clearly
a crucial one, since the top-Higgs coupling provides the largest 
corrections to the Higgs mass at one loop,
leading to the well known
naturalness problem (see the discussion on the
naturalness issue in Section 1).
In order to extract the Yukawa coupling, one
would need to measure the cross section of the process
$pp\to \PQt\PAQt \PH$: the LHC analyses at $7$ and
$8\UTeV$ yielded  upper limits on the
$\PQt\PAQt \PH$ cross section slightly above the
Standard Model expectations \cite{CMS:2013sea,Aad:2015iha}.
Measurements foreseen at 13 and $14\UTeV$ should
shed light on the observed excess: 
the expected accuracy on the $\PQt\PAQt \PH$ cross section
is $15\%$ at $300$\ifb and 
$10\%$ at $3000$\ifb \cite{Dawson:2013bba}.

Even better measurements of the Yukawa coupling are among the goals of 
lepton colliders: for an ILC of 1000~fb$^{-1}$, the foreseen accuracies are
10\% at $\sqrt{s}=500\UGeV$ and $4\%$ at $1\UTeV$, under the assumption that 
the polarization rates are $80\%$ for electrons and $30\%$ for positrons.
As for CLIC, the note \cite{CLICdp-Note-2014-001} investigates
the potential for a direct measurement of the top Yukawa coupling.
The relative uncertainty 
scales like $0.53\times \Delta\sigma/\sigma$, $\sigma$ being the
cross section for $\PQt\PAQt\PH$ production, so that,  
for $\Pem\Pep$ annihilation at $1.4\UTeV$,  
a precision of $4.27\%$ can be achieved without beam polarization.
At FCC-$ee$, the only possible strategy to extract the Yukawa coupling is 
a threshold scan of the $\Pep\Pem \to \PQt\PAQt$ cross section, in order to
be sensitive to Higgs exchange, besides the $\PZ$ and photon contributions.
The projections are about $30\%$, thus worse than
the expectations of ILC and CLIC \cite{Agashe:2013hma}.

\subsection{Final-state kinematics}

Studying kinematic distributions relying on top production and decay 
does provide important tests of the Standard Model and allows one
to investigate several new physics scenarios.
The complete differential process 
$\Pp\Pp\to \PQt\PAQt\to \PW^+\PQb\PW^-\PAQb$ has been computed
to NLO accuracy, with \cite{Denner:2012yc,Bevilacqua:2010qb}
and without \cite{Melnikov:2009dn} including top width effects.

Among the observables which have been investigated, the top transverse
momentum spectrum has been calculated by means of resummed calculations,
carried out using standard techniques \cite{Kidonakis:2010dk}
and in the framework of Soft Collinear Effective Theories  \cite{Ahrens:2011mw},
wherein even the $\PQt\PAQt$ invariant mass $m_{\PQt\PAQt}$ has been computed.
Although such computations generally agree with the experimental data,
it was found \cite{Jung:2013vpa}, 
by using the NLO MCFM program \cite{Campbell:2010ff}, that 
the uncertainty on the $p_{\PQt}$ spectrum in the boosted regime, \ie the top decay
products clustered into a single jet, is about twice larger than in the unboosted case.
Such a result clearly calls for a full NNLO calculation in that regime.

An important final-state observable is the forward-backward asymmetry,
which has represented for some time an open issue, since it
exhibited a $2\sigma$ deviation at the Tevatron
\cite{Aaltonen:2014eva}, when compared
with NLO QCD predictions. However, the recent calculation 
\cite{Czakon:2014xsa} of the full NNLO corrections to the
asymmetry, which is also the first differential NNLO computation
for $2\to 2$ QCD processes, has shown agreement
with the D0 data \cite{Abazov:2014cca}, whereas the
disagreement with CDF \cite{Aaltonen:2014eva}
is reduced to $1.5$ standard deviations.
 At the LHC, such a measurement,
which is straightforward for $\PQq\PAQq $ initial states, 
is more difficult: in a $\Pp\Pp$ collider  $\PQt\PAQt$ production
is mostly driven by $\Pg\Pg$ annihilation.
In fact, ATLAS and CMS performed measurements of the asymmetry,
in agreement with the Standard Model, but 
affected by large errors \cite{Chatrchyan:2011hk,Abazov:2014cca}.
Enhancing the energy to $14\UTeV$ will increase the production of
$\PQt\PAQt$ pairs through $\Pg\Pg$ annihilation, which does not produce any
forward-backward asymmetry. However, as discussed in \cite{Jung:2013vpa}, 
the uncertainties due to background modelling and lepton identification scale with
the luminosity as $1/\sqrt{{\mathcal L}}$ and therefore, after setting appropriate
cuts on the $\PQt\PAQt$ invariant mass and centre-of-mass rapidity, 
the fraction of $\PQq\PAQq $ annihilation can be enhanced, thus allowing
an improved measurement of the asymmetry.
Two alternatives to the standard forward-backward asymmetry have been
proposed in \cite{Berge:2013xsa} in events with $\PQt\PAQt$+jet: 
they are the energy and incline asymmetries, expressed in terms of 
the energy difference between the $\PQt$ and the 
$\PAQt$ and of the rapidity of the
$\PQt\PAQt j$ system. After setting suitable cuts, 
the incline-asymmetry distribution, evaluated at NLO in QCD in \cite{Berge:2013xsa},
can reach the value of $-4\%$ at $14\UTeV$ LHC,
and can be observed  with a significance of 5 standard deviations
at a luminosity of 100\ifb.
As for the energy-asymmetry distribution, its maximum value at $14\UTeV$ is $-12\%$
and it can be measured at ${\cal L}=100$\ifb with
a significance of 3~$\sigma$.

At a linear collider, the main kinematic properties which are foreseen
to be measured are the top production angle $\theta_{\PQt}$ and the helicity
angle $\theta_h$. In this way, one will be able to determine the
forward-backward asymmetry and the slope of the helicity angle $\lambda_{\PQt}$
with an accuracy of 2\% in semileptonic events,
as obtained in the simulations at $\sqrt{s}=500\UGeV$
carried out in \cite{Amjad:2013tlv}. 
In the $\PQt\PAQt$ threshold regime, where a number of measurements at the
linear collider is planned, at present only the total cross section
has been computed at NNLO, whereas the calculation of NNLO differential
distributions is highly desirable, in order to take full advantage of
such a machine.

\hfill\eject
\newcommand\sss{\scriptscriptstyle}
\newcommand{\mW}{m_{\sss W}}
\newcommand{\sW}{s_{\sss W}}
\newcommand{\cW}{c_{\sss W}}
\newcommand{\hc}{+\,\mathrm{h.c.}}
\def\bsp#1\esp{\begin{split}#1\end{split}}
\def\bpm{\begin{pmatrix}}
\def\epm{\end{pmatrix}}
\def\ie{{\it i.e.}}
\def\eg{{\it e.g.}}

\section{Effective Field Theories for the Higgs sector\footnote{F.~Maltoni and F.~Riva}}
\label{sec:EFT}
\subsection{Introduction}

The discovery by  the ATLAS and CMS\cite{Aad:2012tfa,Chatrchyan:2012ufa} 
collaborations of a scalar boson with mass $m_{\PH} \simeq 125\UGeV$, has
prompted unprecedented theoretical and experimental activities to accurately determine its properties, especially the strength and the structure of the coupling to the other Standard Model (SM) particles. Even though the present measurements point to  production cross section and decay rates compatible with those predicted for the Higgs boson of the SM, the uncertainties are still quite large, \ie,   at the level of $10-20\%$. One of the  aims of the next LHC runs and possibly of future linear or circular colliders,  is therefore to bring down these uncertainties to  the percent level~\cite{Dawson:2013bba}.

This program highlights the need for a framework to systematically organize precision tests of the SM and to parametrize its plausible deformations. Here we argue that a SM Effective Field Theory (EFT) provides such a framework.

The essence of a bottom-up EFT approach is  that, since no new physics has been observed at the LHC, we can assume that it is much heavier than the energy accessible to our experiments and expand the Lagrangian in powers of energy (derivatives) over the New Physics scale, $D_\mu/\Lambda$ and in powers of (SM) fields.\footnote{The scope of the EFT approach assumes that no other new state  of mass $m<\Lambda$ exists. While scenarios with light states are obviously interesting and worth investigating, a model-independent approach is not really suitable there. Interactions between SM particles could still be affected by loops where light new physics states could be propagating, leading to fully model-dependent dynamical features. Were this the case, a model dependent (possibly simplified) approach should be employed. 
}
 In this way one builds an effective description by systematically adding to the Lagrangian of the SM all possible higher-dimensional operators  compatible with the SM $SU(3) \times SU(2) \times U(1)$ symmetries and containing only SM fields, see e.g.~\cite{Burges:1983zg,
Leung:1984ni,Buchmuller:1985jz,Hagiwara:1992eh,Hagiwara:1993ck,Hagiwara:1993qt,GonzalezGarcia:1999fq,Eboli:1999pt,Giudice:2007fh,Grzadkowski:2010es,Bonnet:2011yx,Biswal:2012mp,Contino:2013kra,Elias-Miro:2013mua,Gainer:2013rxa,Chen:2013waa}. It is important to notice that the expansion in fields lies on a  different footing w.r.t. the expansion in derivatives. In fact, while the former is necessarily associated with inverse powers of the mass-scale $\Lambda$, the latter must also involve a coupling, which we generically call $g_*$ (this is easily seen by restoring powers of $\hbar$ in the Lagrangian: since couplings scale $[g_*]\sim\hbar^{-1/2}$ while fields scale as $[H]\sim\hbar^{1/2}$, the genuine dimensionless building block of the Lagrangian is, e.g., $g_*H/\Lambda$). For this reason the  effective description  is valid as long as new physics  states appear at a scale $\Lambda \gg E$ much larger than the scale at which experiments are performed (e.g  $\Lambda \gg m_{\PH}$), \emph{and} as long as $\Lambda\gg g_*v$. It is worth noting that there are contexts where the former expansion is good, while this latter expansion fails, corresponding to scenarios where the BSM is directly responsible for EWSB, as in Technicolor models, and a description in terms non-linearly realized EW symmetry becomes more appropriate. In these contexts the leading-order (dimension-4) Lagrangian does not coincide with the SM, since the 125 GeV scalar has no relation with the Higgs boson and, considering that all observations made by the LHC experiments so far are in good agreement with SM predictions, it is natural to consider this option as disfavored.
Moreover, if the underlying theory respects custodial symmetry and one considers only observables which involve the same number of Higgs particles, the effective description is the same, independently on whether or not we perform the expansion in powers of the Higgs field.  For these reasons we will assume the validity of the expansion in fields or, equivalently, we assume that the observed Higgs scalar is part of an $SU(2)_{\mathrm L}$ doublet together with the longitudinal polarizations of the  $W$, $Z$ bosons.
 Under this assumption the effective Lagrangian can be expanded into a sum of operators with increasing dimensionality, with only one operator of dimension five (the one associated to the Majorana mass of the neutrinos) and a set of 76 operators  at dimension six (for one fermion family, counting real independent parameters in the effective Lagrangian)~\cite{Grzadkowski:2010es}.

The use of an effective field theory approach brings significant advantages, above all with respect to alternative parametrizations, such as those based on generic anomalous couplings~\cite{Bolognesi:2012mm}. 
First of all, EFTs represent a consistent and flexible framework to perform precision tests of the SM, where radiative corrections can be rigorously incorporated, different assumptions (e.g. custodial symmetry, flavor symmetries,...) can be independently tested, and it is easily and systematically improvable.
Secondly, expressing precision SM tests in terms of EFTs allows us to interpret the results in terms of physics Beyond the SM (BSM) in a generalization of the popular $S,T$ parameters. This relation represents a channel to compare  precision searches with direct searches and also provides one simple but important motivation for performing precision tests: if new physics resides at a scale $\Lambda$ and couples to the Higgs field with strength $g_*$, at low energy it might only induce a relative change of order $\sim  (g_* v/\Lambda)^2$ to some couplings, where $v=246\UGeV$ is the Higgs vacuum expectation value. For maximally strongly coupled theories with $g_*\lesssim 4\pi$, a  10\% (1\%)  deviation from the SM would correspond to a new physics scale $\Lambda\sim10 \, (40) \UTeV$, unreachable with direct searches. Finally, another motivation for EFTs is that they provide an educated principle to organize deformations from the SM into a leading/next-to-leading hierarchy (corresponding to increasingly higher orders in the EFT expansion parameters) and, moreover, such a hierarchical scheme reflects in an (almost) generic way the low-energy behavior of large classes of BSM scenarios. This model-independence  represents one further advantage of precision tests (in the form of precision searches) over direct searches for New Physics, that typically require a concrete model to extract the most out of them. Furthermore, the breakdown of the EFT description at energies $E\simeq\Lambda$ provides an important self-consistency check: issues such as unitarity violation (which is a major problem of any anomalous coupling description), are automatically taken into account by the EFT~\cite{Degrande:2012wf,Biekoetter:2014jwa}.
can be clearly identified and analysed in the context of an EFT~\cite{Degrande:2012wf,Biekoetter:2014jwa}.

The Lagrangian of the SM+higher-dimensional operators is renormalizable \`a la Wilson. In other words, order by order in $\Lambda$, higher order corrections in the couplings can be consistently computed. 
Moreover, in principle, the inclusion of higher-order $E^n/\Lambda^n$ effects to a given observable (measured at energies $E<\Lambda$), allows to consistently incorporate BSM effects with higher and higher accuracy. All this is essential in the extraction of information from cross section measurements at hadron colliders where higher-order QCD effects are always relevant and at $\Pep\Pem$-colliders, where the precision is so high that SM EW corrections become important. An example of the utility of such a parametrization, and the importance of being able to include EW corrections, is given by the popular S,T precision parameters \cite{Peskin:1990zt}, that represent a subset of the EFT parametrization suitable for universal (i.e. where the new physics couples only to gauge bosons) BSM theories \cite{Barbieri:2004qk}. 

The more general EFT contains operators that affect EW precision observables as well as operators that affect Higgs physics, and, since in the SM the Higgs excitation is always accompanied by the Higgs vev, $v+h$, some operators contribute to both Higgs and EW physics. The latter can therefore be strongly constrained independently of Higgs physics. So even though the number of free parameters in the EFT seems  quite large at face value, it is possible that  by identifying a suitable set of observables to constrain all of them {\it at the same time}, by performing a global fit. Work in this direction has already started~\cite{Pomarol:2013zra,deBlas:2014ula,Ellis:2014jta,Gupta:2014rxa,Falkowski:2014tna}, but unexplored avenues remain, in particular in the relation between flavor observables in Higgs and non-Higgs processes.

In summary, the EFT provides a consistent and systematically improvable framework to quantify and interpret deviations from the Standard Model predictions due to physics residing at higher scales, $\Lambda$, not only in Higgs physics but for all SM particles and interactions. The key questions that we would like to address  are: What are the prospects to precisely determine the Higgs couplings and parametrise possible deviations in terms of an EFT in the coming LHC runs and possibly beyond? What are the current and foreseeable theoretical and experimental challenges in 
pursuing a precise determination of all the parameters entering dim=6 SM Lagragian, in particular for the part concerning the Higgs?  

The plan of this contribution is as follows. In the following section, the basic features and properties of the Higgs EFT's  reviewed, with the aim of clarifying the main points and presenting the state-of-the-art. In Section~\ref{ssec:snowmass} the results of the Snowmass study are summarised. In Section~\ref{ssec:EFTWN} the main directions of theoretical and experimental activity where significant work is expected to meet the required accuracy and precision are highlighted. 

\subsection{The dim=6 Standard Model Lagrangian}
\label{ssec:EFTintro}

We start from the  $SU(3)_c\times SU(2)_{\mathrm L}\times U(1)_Y$ gauge symmetry of the SM. 
The gauge vector fields lie in the adjoint representation of the relevant gauge subgroup,
\begin{equation}\label{eq:gaugecontent}
  SU(3)_c \to G^a_\mu  = ({{\bf 8}},{{\bf 1}},0) \ ,\quad 
  SU(2)_{\mathrm L} \to \PW^k_\mu  = ({{\bf 1}},{{\bf 3}},0) \ ,\quad 
  U(1)_{\mathrm Y}  \to \PB_\mu    = ({{\bf 1}},{{\bf 1}},0) \ ,
\end{equation}
The chiral matter content of the theory is organized in three generations of
left-handed and right-handed quark ($\PQq_{\mathrm L}$,
$\PQu_{\mathrm R}$ and $\PQd_{\mathrm R}$) and lepton ($L_{\mathrm L}$ and $e_{\mathrm R}$) fields (we ignore neutrino masses in this context),
\begin{equation}\bsp
  &\PQq_{\mathrm L} = \bpm \PQu_{\mathrm L}\\\PQd_{\mathrm L} \epm =
      \big({\bf 3}, {\bf 2},\frac16\big) \ , 
 \quad 
  \PQu_{\mathrm R} = \big({\bf 3}, {\bf 1}, \frac23\big) \ , \quad
  \PQd_{\mathrm R} = \big({\bf 3}, {\bf 1},-\frac13\big) \ ,\\
  &L_{\mathrm L} = \bpm \nu_{\mathrm L} \\ \ell_{\mathrm L} \epm =
     \big({\bf 1}, {\bf 2},-\frac12\big)\ ,
  \quad 
   e_{\mathrm R} = \big({\bf 1}, {\bf 1},-1\big)\ ,
\esp \end{equation}
Finally, the scalar sector contains a single $SU(2)_{\mathrm L}$ doublet of fields,
\begin{equation}
  \Phi = \bpm -i G^+ \\ \frac{1}{\sqrt{2}} \Big[ v + h + i G^0\Big] \epm = 
      \big({\bf 1}, {\bf 2},\frac12\big) \ .
\end{equation}
With the first equality, we show the component fields of the doublet after shifting the neutral field $h$
by its vacuum expectation value $v$. Moreover, 
we have included the Goldstone bosons $G^{+,0}$ to be absorbed by the weak bosons to
get their longitudinal degree of freedom.

In the effective field theory approach that we adopt, the SM is defined as the leading part (including relevant and marginal operators) of an expansion in fields and derivatives, while new interactions possibly due to 
non-observed heavy states, at a scale of order $\Lambda\gtrsim m_{\PW}$, are parametrized by operators of higher dimension.
Ignoring  interactions of dimension 5, that lead to  Majorana masses for the neutrinos, the next-to-leading  terms in this expansion come from operators of dimension six. Here we focus on operators that contain the Higgs doublet, so that they can potentially be relevant for Higgs physics. In a convenient basis of independent operators ${\mathcal O}_i$~\cite{Giudice:2007fh,Grzadkowski:2010es,Elias-Miro:2013mua,Pomarol:2013zra,Gupta:2014rxa,Masso:2014xra} these can be written 
as
\begin{equation}\label{eq:effL}
  {\mathcal L} = {\mathcal L}_{\mathrm{SM}}  + \sum_i \bar c_i {\mathcal O}_i =
    {\mathcal L}_{\mathrm{SM}} + {\mathcal L}_{\rm H-only} + {\mathcal L}_{\mathrm{EW}}+ 
   {\mathcal L}_{\mathrm{CP}}
    + {\mathcal L}_{\mathrm{dip}}  +  {\mathcal L}_{\rm no-Higgs}\ ,
\end{equation}
assuming baryon and lepton number conservation (at the scales relevant for Higgs physics). 

${\mathcal L}_{\rm H-only}$  corresponds to the set of $CP$-conserving
operators that contain the Higgs doublet appearing as $\Phi^\dagger\Phi$:
\begin{equation}
\label{eq:silh}
\vphantom{00}\hspace{-.7cm}
\begin{split}
\Delta {\mathcal L}_{\rm H-only} =
\, & \frac{\bar c_\Phi}{2v^2}\, \partial^\mu\!\left( \Phi^\dagger \Phi \right) \partial_\mu \!\left( \Phi^\dagger \Phi \right) 
- \frac{\bar c_6\, \lambda}{v^2}\left( \Phi^\dagger \Phi \right)^3 \\[0.2cm]
& + \left[ \left( \frac{\bar c_u}{v^2}\,  y_{u}\, \Phi^\dagger \Phi\,   {\PAQq}_{\mathrm L} \Phi^c \PQu_{\mathrm R} +  \frac{\bar c_d}{v^2}\,  y_{d}\, \Phi^\dagger \Phi\,   {\PAQq}_{\mathrm L} \Phi \PQd_{\mathrm R} 
+ \frac{\bar c_l}{v^2}\,  y_{l}\, \Phi^\dagger \Phi\,   {\bar L}_{\mathrm L} \Phi l_{\mathrm R} \right)  + {\it h.c.} \right]
\\[0.2cm]
&+\frac{\bar c_{BB}\,  {g'}^2}{4m_{\PW}^2}\, \Phi^\dagger \Phi \PB_{\mu\nu}\PB^{\mu\nu}  +\frac{\bar c_{WW} \, g^2}{4m_{\PW}^2}\, \Phi^\dagger \Phi \PW_{\mu\nu}^a \PW^{a\mu\nu}
   +\frac{\bar c_{GG} \, g_{\mathrm S}^2}{4m_{\PW}^2}\, \Phi^\dagger \Phi G_{\mu\nu}^a G^{a\mu\nu}
\end{split}
\end{equation}
where the Wilson coefficients $\bar c$ are real free parameters, $\lambda$ stands for the Higgs quartic coupling
and $y_u$, $y_d$ and $y_\ell$ are the $3\times 3$ Yukawa coupling matrices in flavor space
(all flavor indices are understood for clarity). In this expression, we also denote
the $U(1)_{\mathrm Y}$, $SU(2)_{\mathrm L}$ and $SU(3)_c$ coupling constants by $g'$, $g$ and $g_s$, and $\PQq_{\mathrm L}\cdot\Phi = \epsilon_{ij}\ \PQq_{\mathrm L}^i\ \Phi^j
  \quad\text{and}\quad  \Phi^\dag\cdot \PAQq_{\mathrm L} = \epsilon^{ij}\ \Phi^\dag_i\ \PAQq_{Lj} $  and $\Phi^c\equiv \epsilon^{ij} \Phi^{*}_j$, with the rank-two antisymmetric tensors being defined by $\epsilon_{12}=1$ and $\epsilon^{12}=-1$.
 Finally,
our conventions for the gauge-covariant derivatives and the gauge field strength tensors are
\begin{equation}\bsp
  \PB_{\mu\nu} =&\ \partial_\mu \PB_\nu - \partial_\nu \PB_\mu \ ,\\
  \PW^k_{\mu\nu} =&\ \partial_\mu \PW^k_\nu - \partial_\nu \PW^k_\mu + g \epsilon_{ij}{}^k \ \PW^i_\mu \PW^j_\nu\ ,\\
  G^a_{\mu\nu} =&\ \partial_\mu G^a_\nu - \partial_\nu G^a_\mu + g_s f_{bc}{}^a\ G^b_\mu G^c_\nu\ ,\\
  D_\rho \PW^k_{\mu\nu} = &\ \partial_\mu\partial_\rho \PW^k_\nu - \partial_\nu\partial_\rho \PW^k_\mu +
    g \epsilon_{ij}{}^k \partial_\rho\big[\PW_\mu^i \PW_\nu^j\big] +
    g \epsilon_{ij}{}^k \PW_\rho^i \big[\partial_\mu \PW_\nu^j - \partial_\nu \PW_\mu^j\big]\\ &\quad + 
    g^2  \PW_{\rho i}\big[\PW_\nu^i \PW_\mu^k - \PW_\mu^i \PW_\nu^k\big] \ , \\
  D_\mu\Phi =&\ \partial_\mu \Phi - \frac12 i g' \PB_\mu \Phi -  i g T_{2k} \PW_\mu^k \Phi \ , 
\esp\label{eq:covder}\end{equation}
$\epsilon_{ij}{}^k$ and $f_{ab}{}^c$ being the structure constants of $SU(2)$ and
$SU(3)$. Notice that we have normalized the Wilson coefficients using SM scales and couplings (following the discussion above), which is equivalent to absorbing powers of $m_{\PW}/\Lambda$ or $g_*v/\Lambda$ inside the Wilson coefficient: in this way (since the relevant experiments are performed at energies $E\sim m_{\PW}$) we can easily keep track of the validity of the perturbative expansion by requiring that $1\gg c_i$.

The interesting feature about  Eq.~\eqref{eq:silh} is that it contains effects that can be studied only in physics that involves the physical Higgs particle $h$. In fact, in the vacuum $\Phi\to v$, the effect of Eq.~\eqref{eq:silh} can absorbed into a redefinition of the SM parameters~\cite{Elias-Miro:2013mua,Pomarol:2013zra}. The Wilson coefficients $\bar c$ of Eq.~\eqref{eq:silh} can, at leading order,  be mapped one-to-one with observables in the context of Higgs physics, in particular
\begin{equation}\label{kappas}
{\upkappa_{\PQu},\upkappa_{\PQd},\upkappa_{\Pl},\upkappa_{\PV},\upkappa_{\PGg},\upkappa_{\Pg},\upkappa_{\PZ\PGg}}\, ,
\end{equation}
which have already been the subject of LHC Run1 experiments, and the Higgs self coupling $\upkappa_{h^3}$, which will be measured during the next Run; (we denote $\upkappa_{\PV}=\upkappa_{\PZ}=
\upkappa_{\PW}$). We discuss these couplings in the next section.

Contrary to Eq.~\eqref{eq:silh}, other operators involving the Higgs field also affect EW observables and are therefore already constrained by other experiments. In particular
\begin{equation}
\label{eq:EW}
\vphantom{00}\hspace{-.7cm}
\begin{split}
\Delta {\mathcal L}_{\mathrm{EW}} =
& \frac{i\bar c_{\PW}\, g}{2m_{\PW}^2}\left( \Phi^\dagger  \sigma^i \overleftrightarrow {D^\mu} \Phi \right )( D^\nu  \PW_{\mu \nu})^i
+\frac{i\bar c_B\, g'}{2m_{\PW}^2}\left( \Phi^\dagger  \overleftrightarrow {D^\mu} \Phi \right )( \partial^\nu  \PB_{\mu \nu})   
\\[0.2cm]
\, & \frac{\bar c_T}{2v^2}\left (\Phi^\dagger {\overleftrightarrow { D^\mu}} \Phi \right) \!\left(   \Phi^\dagger{\overleftrightarrow D}_\mu \Phi\right)  +\frac{\bar c_{WB}g g^\prime}{4 m_{\PW}^2} \, \Phi^\dagger \sigma^i \Phi \, \PW_{\mu\nu}^i \PB^{\mu\nu} 
\\[0.2cm]
\, & \frac{i \bar c_{Hq}}{v^2}  \left(\PAQq_{\mathrm L} \gamma^\mu \PQq_{\mathrm L}\right)  \big( \Phi^\dagger{\overleftrightarrow D}_\mu \Phi\big)
+ \frac{i \bar c^\prime_{Hq}}{v^2}  \left(\PAQq_{\mathrm L} \gamma^\mu \sigma^i \PQq_{\mathrm L}\right)  \big(\Phi^\dagger\sigma^i {\overleftrightarrow D}_\mu \Phi\big) 
\\[0.2cm]
& + \frac{i \bar c_{Hu}}{v^2}  \left(\PAQu_{\mathrm R} \gamma^\mu \PQu_{\mathrm R}\right)  \big( \Phi^\dagger{\overleftrightarrow D}_\mu \Phi\big)
+ \frac{i \bar c_{Hd}}{v^2}  \left(\PAQd_{\mathrm R} \gamma^\mu \PQd_{\mathrm R}\right)  \big( \Phi^\dagger{\overleftrightarrow D}_\mu \Phi\big) \\[0.2cm]
& +\left(  \frac{i \bar c_{Hud}}{v^2}  \left(\PAQu_{\mathrm R} \gamma^\mu \PQd_{\mathrm R}\right)  \big( \Phi^{c\, \dagger} {\overleftrightarrow D}_\mu \Phi\big) +{\it h.c.} \right)
 \\[0.2cm]
& + \frac{i \bar c_{HL}}{v^2}  \left(\bar L_{\mathrm L} \gamma^\mu L_{\mathrm L}\right)  \big( \Phi^\dagger{\overleftrightarrow D}_\mu \Phi\big)
+ \frac{i \bar c^\prime_{HL}}{v^2}  \left(\bar L_{\mathrm L} \gamma^\mu \sigma^i L_{\mathrm L}\right)  \big(\Phi^\dagger\sigma^i {\overleftrightarrow D}_\mu \Phi\big)
 \\[0.2cm]
& + \frac{i \bar c_{Hl}}{v^2}  \left(\bar l_{\mathrm R} \gamma^\mu l_{\mathrm R}\right)  \big( \Phi^\dagger{\overleftrightarrow D}_\mu \Phi\big) \, ,
\end{split}
\end{equation}
beside modifying Higgs physics, they also contribute to precision observables, such as those measured at LEP; here  $\sigma_k$ are the Pauli matrices and we have introduced
the Hermitian derivative operators ${\overleftrightarrow D}_\mu$ defined as
\begin{equation}
  \Phi^\dag {\overleftrightarrow D}_\mu \Phi = 
    \Phi^\dag D^\mu \Phi - D_\mu\Phi^\dag \Phi \ .
\end{equation} 
Notice that two of the  operators that have been introduced are redundant and can be removed through \cite{Grojean:2006nn}
\begin{equation}\label{eq:redund}\bsp
  {\mathcal O}_{\sss W} =&\ -2 {\mathcal O}_{\sss H} + \frac{4}{v^2} \Phi^\dag \Phi D^\mu\Phi^\dag D_\mu\Phi + {\mathcal O}'_{\sss HQ} + 
    {\mathcal O}'_{\sss HL} \ , \\
  {\mathcal O}_{\sss B} =&\ 2 \tan^2\theta_{\sss W} \Big[ \sum_\psi Y_\psi {\mathcal O}_{\sss H\psi} - {\mathcal O}_{\sss T} \Big] \ ,
\esp\end{equation}
where we sum over the whole chiral content of the theory and $\theta_{\sss W}$ stands for the weak mixing angle. Once this redundancy is accounted for, all Wilson coefficients entering Eq.~\eqref{eq:EW} (at least the flavor-diagonal component) can be constrained using data from $Z$-pole observables at LEP1 or information from $\Pep\Pem\to \PW^+\PW^-$ at LEP2, \cite{Pomarol:2013zra,Gupta:2014rxa,Ellis:2014jta,Falkowski:2014tna}.

Similarly the operators in $ {\mathcal L}_{\mathrm{dip}} $ are measured both in Higgs physics and electric dipole moments (EDMs),
\begin{equation}
\label{eq:silh3}
\begin{split}
\Delta {\mathcal L}_{\mathrm{dip}} =
\, & \frac{\bar c_{uB}\, g'}{m_{\PW}^2}\,  y_u \,  {\PAQq}_{\mathrm L} \Phi^c \sigma^{\mu\nu} \PQu_{\mathrm R} \,  \PB_{\mu\nu}
 + \frac{\bar c_{uW}\, g}{m_{\PW}^2}\,  y_u \, {\PAQq}_{\mathrm L}  \sigma^i \Phi^c \sigma^{\mu\nu} \PQu_{\mathrm R}  \, \PW_{\mu\nu}^i
 + \frac{\bar c_{uG}\, g_{\mathrm S}}{m_{\PW}^2}\, y_u \,  {\PAQq}_{\mathrm L} \Phi^c \sigma^{\mu\nu} \lambda^a \PQu_{\mathrm R}  \, G_{\mu\nu}^a 
\\[0.2cm]
\, &  +  \frac{\bar c_{dB}\, g'}{m_{\PW}^2}\, y_d \, {\PAQq}_{\mathrm L} \Phi \sigma^{\mu\nu} \PQd_{\mathrm R} \, \PB_{\mu\nu}
+  \frac{\bar c_{dW}\, g}{m_{\PW}^2}\,   y_d \, {\PAQq}_{\mathrm L} \sigma^i \Phi \sigma^{\mu\nu} \PQd_{\mathrm R} \, \PW_{\mu\nu}^i
+  \frac{\bar c_{dG}\, g_{\mathrm S}}{m_{\PW}^2}\,  y_d \,  {\PAQq}_{\mathrm L} \Phi \sigma^{\mu\nu}  \lambda^a \PQd_{\mathrm R} \, G_{\mu\nu}^a
\\[0.2cm]
\, & + \frac{\bar c_{lB}\, g'}{m_{\PW}^2}\,  y_l \,   {\bar L}_{\mathrm L} \Phi  \sigma^{\mu\nu}  l_{\mathrm R} \, \PB_{\mu\nu}
+ \frac{\bar c_{lW}\, g}{m_{\PW}^2}\,  y_l \,  {\bar L}_{\mathrm L} \sigma^i \Phi  \sigma^{\mu\nu}  l_{\mathrm R} \, \PW_{\mu\nu}^i
+ {\it h.c.}  
\end{split}
\end{equation}
with complex coefficients (where real and imaginary part correspond respectively to CP-even and CP-odd effects). Other contributions from CP-violating physics BSM involving the Higgs are captured by
\begin{equation}\label{eq:silhCPodd}\bsp
  {\mathcal L}_{\mathrm{CP}} = 
&  \left[\left(i \frac{\tilde c_{ u}}{v^2}\,  y_{u}\, \Phi^\dagger \Phi\,   {\tilde q}_{\mathrm L} \Phi^c \PQu_{\mathrm R} + i \frac{\tilde c_d}{v^2}\,  y_{d}\, \Phi^\dagger \Phi\,   {\PAQq}_{\mathrm L} \Phi \PQd_{\mathrm R} 
+ i\frac{\bar c_l}{v^2}\,  y_{l}\, \Phi^\dagger \Phi\,   {\bar L}_{\mathrm L} \Phi l_{\mathrm R} \right)  + {\it h.c.} \right] \\
  &\
    +\frac{\tilde c_{WB}g g^\prime}{4 m_{\PW}^2} \, \Phi^\dagger \sigma^i \Phi \, \PW_{\mu\nu}^i \widetilde \PB^{\mu\nu}  + \frac{g'^2\  \tilde c_{\sss BB}}{4m_{\PW}^2} \Phi^\dag \Phi \PB_{\mu\nu} {\widetilde B}^{\mu\nu}\\
 &\
  +\!  \frac{g_s^2\ \tilde c'_{\sss g}}{4m_{\PW}^2}      \Phi^\dag \Phi G_{\mu\nu}^a {\widetilde G}^{\mu\nu}_a+\frac{g^2\ \tilde c'_{\sss g}}{4m_{\PW}^2}      \Phi^\dag \Phi \PW_{\mu\nu}^a {\widetilde W}^{\mu\nu}_a\, ,
\esp\end{equation}
where the dual field strength tensors are defined by
\begin{equation}
  \widetilde \PB_{\mu\nu} = \frac12 \epsilon_{\mu\nu\rho\sigma} \PB^{\rho\sigma} \ , \quad
  \widetilde \PW_{\mu\nu}^k = \frac12 \epsilon_{\mu\nu\rho\sigma} \PW^{\rho\sigma k} \ , \quad
  \widetilde G_{\mu\nu}^a = \frac12 \epsilon_{\mu\nu\rho\sigma} G^{\rho\sigma a} \ .
\end{equation}
and the coefficients are real. The assumption of one-family in flavor space can easily be abandoned by promoting the Wilson coefficients of the fermionic operators to tensors in flavor space.

So, contrary to the operators in $ {\mathcal L}_{\rm H-only}$, the ones of ${\mathcal L}_{\mathrm{EW}}+ {\mathcal L}_{\mathrm{CP}}
    + {\mathcal L}_{\mathrm{dip}} $ have not yet received attention in the context of Higgs physics. Although they are already constrained by other experiments, it is not clear whether, in some cases, Higgs physics could lead to a higher sensitivity (see e.g. \Bref{Biekoetter:2014jwa} for an example).

Finally, the complete dimension-6 EFT Lagrangian also includes many operators that do not contain the Higgs, such as four-fermion interactions and operators involving three field strengths. Although these do not contain the Higgs field, some of them might interfere in the extraction of constraints for the operators mentioned here (in particular the operator $ \frac{g^3\ \bar c'_{\sss 3W}}{\Lambda^2} \epsilon_{ijk} \PW_{\mu\nu}^i \PW^\nu{}^j_\rho \PW^{\rho\mu k}$ enters measurements of triple gauge couplings and it might reduce the sensitivity to operators involving the Higgs \cite{Falkowski:2014tna}).

It is important to  stress that the basis we proposed is not unique, as highlighted by Eq.~\eqref{eq:redund}. Field redefinitions  proportional to the leading-order equations of motion, and integration by parts can be used to express some of the operators above in terms of others, whose physical interpretation might be slightly different. 
This redundancy is a feature of higher-dimensional operators that is unfamiliar from the Standard Model. Because of this redundancy, there is a great deal of flexibility in which set of operators to use. In principle, any set of  independent operators constitutes a good basis and there is no physically preferred basis, as long as all operators are included in the analysis. However a particular experimental measurement generally depends on only a few of the operators (at any finite order in perturbation theory) in a given basis and it might be the case that the relation between operators and observables might be simpler in one basis than in others. In this sense the basis of \emph{BSM primaries} \cite{Gupta:2014rxa,Masso:2014xra} was designed to minimize the theoretical correlation between operators and provides an almost one-to-one correspondence between operators and observables. On the other hand the \emph{SILH} basis of Refs.~\cite{Giudice:2007fh,Contino:2013kra,Elias-Miro:2013mua,Pomarol:2013zra} is more BSM oriented and is easily matched to universal microscopic models (including SUSY and Composithe Higgs models), while the basis of \Bref{Grzadkowski:2010es} might be more suitable to describe UV models where the BSM couples to fermions.
Now, in several instances in the literature, bounds on the coefficient of a particular operator have been put by assuming that all the other operators in that basis have vanishing coefficients: this is an ad hoc and meaningless assumption, since typically no BSM scenario gives rise to a single operator (in this sense the analyses of. e.g. Refs~\cite{Giudice:2007fh,Low:2009di} provide an educated guess of how certain patterns of Wilson coefficients might arise from general classes of BSM models). In the absence of an underlying theory, one should always include every dimension-six operator that contributes to the calculation of a physical process and each experimental measurement will generally bound a set of dimension-six operators. Constraining the Wilson coefficients implies adopting a global approach where a sufficiently comprehensive set of observables is mapped onto the full set of operators in the EFT. 


\subsection{Expected precision on the couplings strength: the Snowmass study}
\label{ssec:snowmass}

A first useful starting point to assess the reach of the next LHC runs and possibly at future accelerators in the determination of  the Higgs couplings is studying the precision in searching for deviations in the strength of the couplings~\cite{Heinemeyer:2013tqa}, which corresponds to the operators of Eq.~\eqref{eq:silh} through the parameters mentioned in Eq.~\eqref{kappas}. Such a study has been completed in the Snowmass workshop in summer 2013~\cite{Dawson:2013bba}.

It is important to recall the simplified working assumptions when extrapolating in luminosity. First, the structure of the coupling is the same as of the SM and only the normalisations are let free to float and determined by a global fit on the observed rates that depend on production cross sections and 
branching ratios. Within this approach, shapes and distributions are unchanged with respect to the SM and are used to select signal vs background only. It is important to stress that this methodology allows to test the SM hypothesis, but not to interpret possible deviations.  The experimental efficiencies are assumed to be the same as those of Run I at the LHC. The  evolution of the theoretical uncertainties is treated differently by ATLAS and CMS. For ATLAS the current uncertainties where either included or not, while CMS considered two scenarios, one with the current ones and one with the theoretical uncertainties reduced by a factor of two.

The results, summarised in Tables~\ref{tab:LHCRateProjections} and \ref{tab:SnowmassLHC}, taken 
from \cite{Dawson:2013bba}, show that 
an expected relative precisions better than $10\%$ may be achieved within the next run and possibly improved by a factor two 
in the HL-LHC. 

\begin{table}[t]
\caption{Estimations by ATLAS and CMS of the expected relative precisions on the signal strengths in different Higgs decay final states. In the last column the $95\%$ CL upper limit on the Higgs branching ratio to the invisible decay from the $\PZ\PH$ search is given. ATLAS and CMS ranges are not directly comparable due to the different treatment of the expected theoretical uncertainties. Table taken from Ref.~\cite{Dawson:2013bba}. }
\small
\begin{center}
\begin{tabular}{ccccccccc}\hline\hline
$\int{\mathcal L}dt$ & \multicolumn{8}{c}{Higgs decay final state} \\ \cline{2-9}
 (fb$^{-1}$) & $\PGg\PGg$ & $\PW\PW^*$ & $\PZ\PZ^*$ &  $\PQb\PAQb$ & $\tau\tau$ & $\mu\mu$ & $\PZ\PGg$ & BR$_{\rm inv}$ \\ \hline
\multicolumn{9}{c}{ATLAS} \\ \hline
 300         &  $9-14\%$   &  $8-13\%$  &  $6-12\%$  &  N/A        &  $16-22\%$ & $38-39\%$ & $145-147\%$ & $<23-32\%$     \\
3000         &  $4-10\%$   &  $5-9\%$   &  $4-10\%$  &  N/A        &  $12-19\%$ & $12-15\%$ & $54-57\%$   & $<8-16\%$     \\ \hline
\multicolumn{9}{c}{CMS} \\ \hline
 300         &  $6-12\%$  &  $6-11\%$  &  $7-11\%$  &  $11-14\%$  &  $8-14\%$  & $40-42\%$ & $62-62\%$  & $<17-28\%$ \\
3000         &  $4-8\%$   &  $4-7\%$   &  $4-7\%$   &  $5-7\%$    &  $5-8\%$   & $14-20\%$ & $20-24\%$  & $<6-17\%$ \\ \hline\hline 
\end{tabular}
\end{center}
\label{tab:LHCRateProjections}
\end{table}

\begin{table}[htb!]
\caption{Precision of Higgs boson couplings as expected by CMS and ATLAS with
300~fb$^{-1}$ and 3000~fb$^{-1}$ integrated luminosity at the LHC. The main assumption in 
the fit is that $\upkappa_{\PQu}\equiv\upkappa_{\PQt}=\upkappa_{\PQc}$, 
$\upkappa_{\PQd}\equiv \upkappa_{\PQb}=\upkappa_{\PQs}$ and $\upkappa_\ell\equiv\upkappa_\tau=\upkappa_\mu$.  The range represents spread from two different assumptions
of systematic uncertainties. Table taken from \Bref{Dawson:2013bba}. }
\begin{center}
\begin{tabular}{cccc}\hline\hline
Luminosity  & 300 fb$^{-1}$   && 3000 fb$^{-1}$ \\ \hline
Coupling parameter  & \multicolumn{3}{c}{7-parameter fit} \\ \hline
$\upkappa_{\PGg}$     & $5-7\%$    &&  $2-5\%$ \\
$\upkappa_{\Pg}$          & $6-8\%$    &&  $3-5\%$ \\
$\upkappa_{\PW}$          & $4-6\%$    &&  $2-5\%$ \\ 
$\upkappa_{\PZ}$          & $4-6\%$    &&  $2-4\%$ \\
$\upkappa_{\PQu}$          & $14-15\%$  &&  $7-10\%$ \\
$\upkappa_{\PQd}$          & $10-13\%$  &&  $4-7\%$  \\ 
$\upkappa_\ell$       & $6-8\%$    &&  $2-5\%$ \\ \hline 
$\Gamma_{\PH}$          & $12-15\%$    &&  $5-8\%$ \\ \\
$\upkappa_{\PZ\PGg}$  & $41-41\%$  && $10-12\%$ \\
$\upkappa_\PGm$        & $23-23\%$  && $8-8\%$ \\ 
BR$_{\rm BSM}$          & $<14-18\%$ && $<7-11\%$ \\ \hline\hline
\end{tabular}
\end{center}
\label{tab:SnowmassLHC}
\end{table}

A summary of the final conclusions of the Snowmass study that are relevant for this discussion is:

\begin{itemize}

\item  Higgs boson  phenomenology will be studied at the LHC in the next decade.  Higgs couplings to fermions and vector bosons, assuming only SM decay modes, can be determined with an 
estimated precision of $4-15\%$ for $300$~fb$^{-1}$ at $14\UTeV$, 
going to $2-10$\% in the high-luminosity run (3000~fb$^{-1}$). 

\item Full exploitation of the LHC and HL-LHC Higgs measurements will require important improvements
in precision of theoretical calculations for production as well as for branching ratios in the SM. 

\item  At an $\Pep\Pem$ collider with sufficient integrated luminosity, SM decays and a wide range of rare 
Higgs boson decays, including invisible or exotic final states, will be accessible 
in the $\PZ\PH$ production through the model-independent recoil mass technique.

\item  Performing precision determinations  of Higgs boson couplings to the one-percent level
will require complementary collider programs, such as  Higgs factories at linear or circular $\Pep\Pem$ colliders or even a muon collider. Only  a multi-prong strategy will allow to constrain many of the couplings in a model-independent way.

\item The determination of the  $\PQt\PQt\PH$ coupling can be done at LHC and with sufficient collision energy also at  ILC.  At the HL-LHC a precision of $7{-}10\%$ per experiment is expected, improving to $\sim 2{-}3\%$ at ILC with luminosity upgrade. 

\item  The Higgs self-coupling is among the most interesting couplings still to be determined and it will remain very challenging in the coming years.  At the HL-LHC a $50\%$ measurement per experiment could
be achieved, while  a linear $\Pep\Pem$ collider  at  $1\UTeV$ could reach $13\%$.  Further improvement would need higher collision energies, with CLIC  and FCC-hh possibly going below the $10\%$ level.

\item $C\!P$-odd couplings to vector bosons (loop induced) and to fermions will be accessible already at the LHC to a few percent precision, with further improvements from VBF production and from  fermions $h\to\tau\tau$ decay and $\PAQt\PQt\PH$ production.

\item  
HL-LHC provides unique capabilities to measure rare statistically-limited SM decay modes such as $\PGmp\PGmm$, $\PGg\PGg$, and $\PZ\PGg$ and make the first measurements of the Higgs self-coupling.

\end{itemize}

Several comments are in order. First, as mentioned above, the Snowmass study is limited to the strength of the couplings and it is therefore suitable only for exploring the operators of Eq.~\eqref{eq:silh}. Second, results have been obtained by extrapolating measurements being performed at the time of writing. Several other opportunities for determining coupling strengths will open up with higher-energy and luminosity runs, such as constraining the Higgs-charm couplings through exclusive $\PH \to J/\psi \PGg$ decays~\cite{Bodwin:2013gca}, searching for exotic Higgs decays~\cite{Curtin:2013fra} and improving  measurements of processes where the Higgs contributes off its mass shell, such as $\Pg\Pg \to \PZ\PZ, \PW\PW$~\cite{Kauer:2012hd}.

\subsection{Towards precision EFT: the road ahead}
\label{ssec:EFTWN}

Measuring Higgs coupling strengths can be thought as the deployment of an exploration strategy. If no large discrepancies are found, such as the case now,  the next logical step is to employ a model independent strategy and set limits on the new physics scale $\Lambda$ through precise determinations of the Wilson coefficients of the dim=6 SM Lagrangian. Several are the challenges, both experimental and theoretical, that such a program faces in the short- and mid-term horizon. In the following a few among the most important issues are presented, mostly related to the theoretical needs. 

\begin{itemize}
\item {\bf Constraints from the UV} 
\\
From a UV perspective, one of the most important features of the EFT approach is its model independence: any new physics theory at high scale will generate  interactions that can be fully parametrized at a given accuracy in $E/\Lambda$ by the corresponding set of operators. Given a UV theory or a general class thereof, however, one can propagate  information down to lower scales and predict operator coefficients a priori. In this case RGE's of the operators should be considered to correctly match the full theory to the effective one. Such information is now fully available at one-loop for the dim=6 SM Lagrangian~\cite{Elias-Miro:2013eta,Elias-Miro:2013mua,Elias-Miro:2013gya,Jenkins:2013zja,Jenkins:2013wua,Alonso:2013hga}. 

Still, it is possible to conceive large classes of UV models and imagine what structure these models could imprint in the coefficients of the EFT. We have already mentioned the power-counting in terms of couplings $g_*$, that provides a hint of what kind of effects could be enhanced, e.g., in strongly coupled theories ~\cite{Giudice:2007fh,Low:2009di}. Another example: if the underlying theory is a  renormalizable gauge theory, it is possible to classify dimension-six operators as being potentially generated at tree level or at one loop~\cite{Arzt:1994gp,Giudice:2007fh,Einhorn:2013kja}. Both the  basis of \Bref{Grzadkowski:2010es} and \Bref{Giudice:2007fh} allow this classification, while the  basis of Ref.~\cite{Hagiwara:1993ck} does not and therefore does not offer a transparent framework to describe this type of UV models. 

An EFT can be thought as building a bridge to consistently collect information  about deviations of the SM in order to constrain possible new physics theories lying at higher scales. Until indications become clear and if enough experimental information is available, one should keep an "agnostic" standpoint and fit the data with dimension-six operators regardless any UV-inspired classification.  On the other hand, having general arguments on where to expect deviations for given classes or specific UV theories and explicit maps UV-EFT predicting the values of the most relevant Wilson coefficients will be extremely useful and necessary to characterise the information available at any given time.

Furthermore, in an initial phase of the program when the limited amount of data imposes limitations to the precision and sensitivity of this types of searches, it will be important to have a hierarchical organization within the EFT description that selects some subsets of Wilson coefficients and prioritizes them.
\\
\item {\bf Precision observables in EFT at NLO accuracy in QCD and EW}\\
Constraining new physics via an EFT implies the ability of controlling uncertainties and therefore performing calculations at one-loop or at NLO  for a wide set of observables in the EFT, including precision observables at LEP. Technical as well as conceptual challenges arise in such computations. First, the complete structure of UV operator mixing and renormalisation in both QCD and EW perturbative expansions needs to be known. In addition, suitable regularization and renormalisation schemes need to be properly defined. Another key point is that  comparison with data will require the inclusion of QCD and EW corrections for the SM predictions.  Understanding the pattern of higher-order QCD corrections  for the dim=6 contributions will also be certainly necessary. On the other hand the inclusion  of EW corrections for predictions featuring dim=6  operators might also turn out to be relevant. These issues have just began now to be explored~\cite{Passarino:2012cb}.
\\
\item {\bf The flavour structure of the dim=6 SM Lagrangian}\\
The counting of the number of operators in the dim=6 SM Lagrangian strongly depends on the assumed flavoured structure. The 76 operators  corresponding to one generation, become 2499 for three families. Many of these operators are four-fermion operators, that are not directly related to Higgs physics, yet can enter in precision measurements
(one simple example being the muon decay width through which $G_F$ is defined). Current global studies, see e.g. \cite{Pomarol:2013zra,Ellis:2014jta,Falkowski:2014tna} assume no special structure in flavor (flavour blindness), while \Bref{Pomarol:2013zra} extends the analysis to the first order in Minimal Flavour Violation~\cite{D'Ambrosio:2002ex} scenarios; it would be interesting to explore other patterns of flavor symmetry breaking and to include direct constraints on Flavor Changing Neutral Current (FCNC) operators. 
\\
\item {\bf Precise predictions for production and decay processes: dim=4 SM Lagrangian}\\
Searching for deviations from the SM predictions can only be done if sufficiently accurate and precise predictions from the dim=4 SM Lagrangian can be obtained for the relevant observables, both for signal as well as for backgrounds. The state of the art of the predictions for SM processes in Higgs physics is often summarised as follows:  NNLO in QCD  and NLO in EW are known for total cross sections as well for distributions of all main Higgs production channels (barring $\PAQt\PQt\PH$ which is known only at NLO in QCD). There are, however, important exceptions and notable improvements that will be needed in order to bring the precision EFT program to a success and that will keep the theoretical community busy during the LHC Run II and possibly beyond.  The first important set of improvements concerns Higgs production in  gluon fusion.  A significant reduction of the theoretical uncertainties for the total rates is expected   from the computation of the full N3LO QCD corrections on one side and from better PDF determination on the other. This latter point will drastically rely on our ability to perform an accurate an precise measurement of the gluon PDF using LHC data using NNLO predictions and the corresponding data on inclusive jet and top pair production. A full NNLO computation of the $H$+jet rates, and improvement, through resummation, of the exclusive jet rates is also expected in the coming years. Both these improvements are required already at the level of the precision on the Higgs boson signal strengths only. For the EFT program, information on the effective Higgs-gluon couplings can be gained via measurements at high $p_T(H)$, a region where top-mass effects from the loops are important and currently known only at LO. This is just one example of a rather important class of computations at NLO that will be needed, i.e., those involving loop induced processes at the Born level. Other notable examples belonging to this class are the $\Pg\Pg \to BB$ processes, with $BB=\PH\PH, \PZ\PH, \PZ\PZ, \PW^+\PW^-$ which are presently known only at 1-loop level (Born). $\Pg\Pg \to \PH\PH$ represents the dominant Higgs pair production channel which can provide information first on the trilinear Higgs coupling, but also on the top-Higgs interactions. As already mentioned, $\Pg\Pg \to \PZ\PZ, \PW^+\PW^-$  featuring an off-shell Higgs boson, can provide complementary information and are sensitive to a wide range of new physics contributions coming from higher dimensional operators. The computation at NLO in QCD for such processes, relies on the knowledge of two-loop box amplitudes which are at the edge of the current loop technologies. A process-independent technology to obtain predictions at NNLO accuracy in QCD for final states featuring jets will also be needed for VBF. In all cases, fully differential predictions including EW effects matched to a parton shower at NLO accuracy will also be needed for all main production channels. 
\\
\item {\bf Precise and exclusive predictions for production and decay processes: dim=6 SM Lagrangian}\\
Assessing deviation from the SM only needs precise predictions from the dim=4 SM itself. Interpreting them, however, needs accurate and precise predictions in the context of an EFT. Inclusion of NLO in QCD for all processes and operators of interest in Higgs physics is certainly one of the main goals of this research program. Given the large number of operators and processes to cover, only an automatic approach will be able to deliver the predictions needed. Progress in this direction has started in the context of the Higgs~\cite{Artoisenet:2013puc,Maltoni:2013sma,Demartin:2014fia} and top-quark EFT.  It is also important to remember that attaining NLO in QCD accuracy is mandatory for processes for which no SM mechanism is present (or is highly suppressed). A glaring example are FCN interactions involving the Higgs and quarks for which new physics appear as  squared amplitudes (and not in the interference as usually is the case). All NLO predictions should be matched to a PS and available as event generators. 
Finally, it has to be foreseen that, were deviations found, even NNLO in QCD and EW corrections for some key observables could be needed.
\\
\item {\bf Global approach to the determination of the Wilson coefficients}\\
As already discussed above, UV priors should not be used  to constrain operators in an independent way (unless not enough data is available). This entails that all coefficients should be directly and only constrained by data via a global fit. Identifying the optimal set of key observables, their correlations in the measurements and the mapping of each observable to a given set of operators will be part of an important and non-trivial joint theory-experimental activity. Constraints will not only come from Higgs measurements proper, but also from  processes and final states that at first sight might not have any evident direct relation with Higgs physics. Well-known examples are $\PV\PV$ production cross sections, 
with $\PV=\PZ, \PW^\pm$  that test trilinear gauge couplings or top pair production cross section that constrains operators such as the chromo-magnetic/electric ones. 
This effort will need a coordinated action inside the groups interested in different final states/physics inside the experimental collaborations at the LHC, ATLAS and CMS, and outside with suitable working groups (such as the LHCEWWG and LEPEWWG) and theorists. 
\\
\item {\bf Advanced analysis techniques and boosted objects}\\
Among the foreseen experimental developments, the design, test, and deployment of advanced analysis techniques will be certainly one of the directions to invest in order to maximise the information that can be obtained from data. The contribution of new physics as parametrised by the EFT never shows up as bumps (no resonances are present) but mostly as changes in the distributions and typically as enhancements in the tails. Identifying such behaviours,  and quantifying them by connecting to a suitable set of operators, will need the development of dedicated tools and analyses. 
In this very same context, the tagging of boosted heavy objects, such as vector bosons, the top quark,  and also the Higgs itself, is expected to enhance the sensitivity to dim=6 operators considerably and will play an important role. 
\\
\item {\bf Beyond EFT: the connection with Dark Matter (DM) searches}\\
Evidence of physics beyond the SM comes from cosmological and astrophysical observations pointing to the existence of a form of matter, neutral and very weakly interacting. Current searches pose rather loose lower limits on its mass and simple estimates suggest a scale of the order of the EW interactions. Such states, could therefore have a mass of the order of the Higgs mass or even lighter, and more interestingly could couple to the Higgs field by the so called Higgs-portal, i.e., the $[\Phi^\dagger \Phi]$ term. The existence of such a state close to  the Higgs mass, would therefore invalidate the straightforward use of the EFT approach outlined above. On the other hand, generalisations to the case where only the DM candidate would lie at low scales, while possible mediators and new states would all be heavier could be rather easily treated in the same framework, even though not in a completely model-independent way, as the operators would depend on the properties of the DM candidate, such as its spin and gauge representations.
\end{itemize}

\hfill\eject
\section{The Higgs potential and the Electroweak vacuum\footnote{G.~Degrassi}}
The first run of the LHC has delivered two important messages: i) no
signal of physics beyond the SM (BSM) was discovered. ii) The Higgs
boson was found exactly in the mass range $110{-}160\UGeV$ predicted by the SM,
using the information from precision physics and that from the direct searches
at LEP and Tevatron before the turning on of the LHC. 

The fact that all the mass parameters of the SM have now been
experimentally determined constrains tightly  the model and possibly BSM
physics. New Physics (NP), if it exists,
should have a marginal effect on the SM electroweak fit in order not to 
spoil its very good agreement with the experimental results.  
This fact, together with 
the negative result of the run I of the LHC, indicates that BSM physics
is likely to be at a high scale, possibly out of  the reach of direct
LHC searches. 

The study of the stability of the SM Higgs potential, or 
if  the electroweak (EW)  minimum we live in is the true minimum of the 
SM  effective potential $ V^{\mathrm{eff}}$, is a general argument that can give 
us an indication of where the scale of  NP is, or if instead 
the validity of the SM can be extended up to the Planck  scale, $M_{ \rm Pl}$.

Below  $M_{ \rm Pl}$, the appearance in $ V^{\mathrm{eff}}$  of a second  minimum  
deeper than the EW minimum,  or the fact that  $ V^{\mathrm{eff}}$ at high scale 
is  not bounded from below, are signals of the need (with a caveat to be 
discussed below) of NP to rescue the stability of the EW vacuum.

We are interested in establishing if the EW vacuum is unstable 
and if NP is needed, more than
in pinning down the exact  value of the instability scale, $\Lambda_I$,
where $V^{\mathrm{eff}}$ becomes smaller than its
value at the EW minimum.
Therefore, it is sufficient to study the Renormalization Group (RG)
evolution of  the  Higgs quartic 
coupling, $\lambda(\mu)$, with the scale $\mu$. If  $\lambda(\mu)$ does not 
become negative
up to $M_{ \rm Pl}$ the stability of  $V^{\mathrm{eff}}$  is established.

Actually, $\lambda$ is the only SM coupling that is allowed to
change sign during its RG evolution because it is not multiplicatively 
renormalized. Its $\beta$ function, $\beta_\lambda$, contains two competing 
terms, one 
proportional to $\lambda$ itself, i.e.~the Higgs mass $\mh$,
and the other proportional to the fourth power 
of the top Yukawa coupling $y_{\PQt}$, i.e.~the top mass $\mt$,
which  drive the evolution of  $\lambda$ towards different
directions. For the present central values of  $\mt$ and $\mh$,  
(and the strong coupling $\alphas$ which affects $\beta_\lambda$ through its 
effect on the running of $y_{\PQt}$) a state-of-the-art computation,
based on  three-loop beta functions  for the evolution of the couplings
\cite{Tarasov:1980au,Larin:1993tp,Mihaila:2012fm,Chetyrkin:2012rz,Bednyakov:2012en}  and  two-loop matching conditions 
\cite{Bezrukov:2012sa,Degrassi:2012ry,Buttazzo:2013uya} for the extraction
of the couplings at the weak scale from the related experimental quantities,
shows that the term proportional to $\mt^4$
wins, driving the evolution of $\lambda$  towards smaller values and 
eventually going below zero   at a scale of about $10^{10}\UGeV$. A more
refined analysis \cite{Buttazzo:2013uya} shows that $\Lambda_I \sim 10^{11}\UGeV$
implying  that our EW minimum  is not  the true minimum of the Higgs potential
and that there is a tunnelling  probability between the EW false vacuum  and 
the  true vacuum at high field  values. In this situation,  we can be sure that
NP must appear below $\Lambda_I$ to cure  the instability of the SM potential 
only if the lifetime of EW vacuum is shorter than the observed age of the 
universe.

The rate of quantum tunnelling out of 
the EW vacuum is given by the probability $d\wp/dV\, dt$ of nucleating a
bubble of true vacuum within a space volume $dV$ and time interval
$dt$. The total probability $\wp$ for vacuum decay to have occurred during the 
history of the universe can be computed by integrating  $d\wp/dV\, dt$ over 
the space-time volume of our past light-cone, or
\bq
\wp \sim \tau_U^4  \Lambda_B^4\, e^{-S(\Lambda_B)}~~~~~~~~~~~~~~~~~~~
S(\Lambda_B)=\frac{8\pi^2}{3|\lambda(\Lambda_B)|},
\label{eq:3} 
\eq
where $\tau_U$ is the age 
of the universe and  $S(\Lambda_B)$ is the action of the bounce of size 
$R=\Lambda_B^{-1}$.
$\Lambda_B$ is determined as the scale at
which $\Lambda_B^4 e^{-S(\Lambda_B)}$ is maximized.  
In practice this
roughly amounts to minimizing $\lambda(\Lambda_B)$, which corresponds
to the condition $\beta_\lambda (\Lambda_B)=0$ which is fulfilled
for $\Lambda_B \sim M_{ \rm Pl}$.
By numerical inspection of
$\wp$ in \eqn{eq:3} one finds that the exponential suppression wins over
the large 4-volume factor if $|\lambda(\Lambda_B)|$ is less than $\sim 0.05$.

The fact that in \eqn{eq:3} the probability for the vacuum to decay is 
connected to  the scale $\Lambda_B$ close to $M_{ \rm Pl}$ 
and not to $\Lambda_I$  is a signal that Planck-scale physics
could affect the tunneling rate \cite{Branchina:2013jra}. 
It is conceivable that at scales close to $M_{ \rm Pl}$ the effective potential
could be sensitive to Planck-scale physics which  could dramatically modify
the tunneling rate. An explicit toy example of this possibility has been 
constructed\cite{Branchina:2013jra}. However, we do not know anything about 
Planck-scale  physics and therefore no conclusion can be drawn on
whether the tunneling rate is modified by Planck-scale effects or not.

For the sake of our argument of looking for an unambiguous motivation for NP,
possible Planck-scale effects are not relevant. Thus we 
discuss the lifetime of the EW vacuum  assuming that 
unknown Planck-scale physics  does not modify $\wp$ in \eqn{eq:3}. 
At the Planck scale one  finds for $\lambda$ \cite{Buttazzo:2013uya} 
\bqa
\lambda(M_{ \rm Pl}  ) &=& -0.0143 +  
0.0029\left( \frac{\mh }\UGeV-125.15  \right)
-0.0066 \left( \frac{\mt  }\UGeV -173.34 \right) 
\nonumber \\
&& +0.0018  \left( \frac{ \alphas(\mz) -0.1184}{\rm 0.0007} \right)
\eqa
that implies that our vacuum is metastable, i.e.  $\wp$
is extremely small (less than $10^{-100}$) or the lifetime of the EW vacuum is
extremely long, much larger than $\tau_U$. We must then conclude that the 
instability of the SM Higgs potential  cannot be taken as a motivation for NP. 
However, as shown in \fig{fig:4}, we would have reached a different conclusion 
if $\mh$ had been smaller, leading to a stronger instability of the Higgs 
potential
(the red region in \fig{fig:4}). Obviously, the vacuum stability analysis 
does not exclude BSM physics, which might have no impact on stability,
make it worse, or ameliorate it. Examples of all the three possibilities can be
easily found \cite{EliasMiro:2011aa,EliasMiro:2012ay}.
\begin{figure}[t]
$$\includegraphics[width=0.45\textwidth]{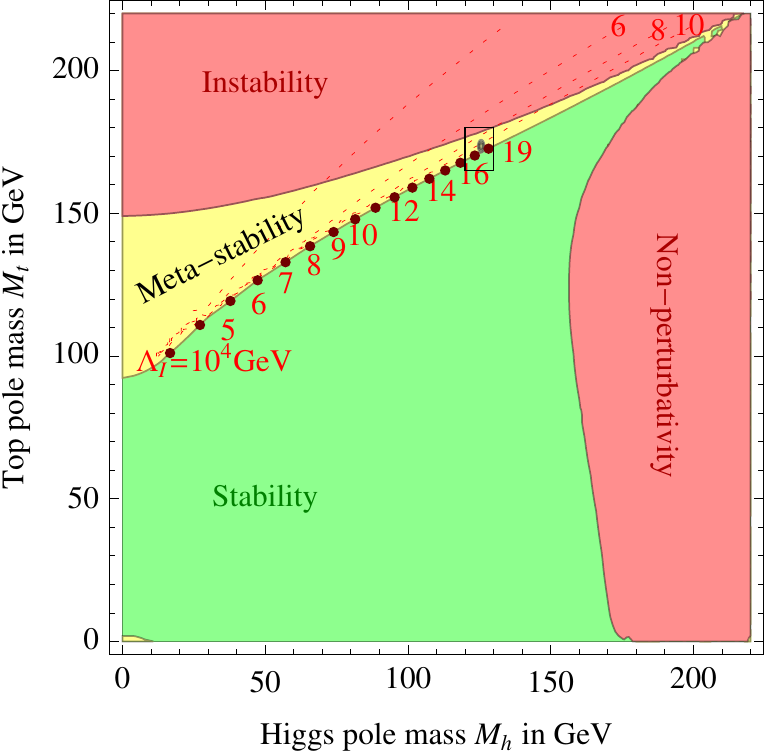}\qquad
  \includegraphics[width=0.46\textwidth]{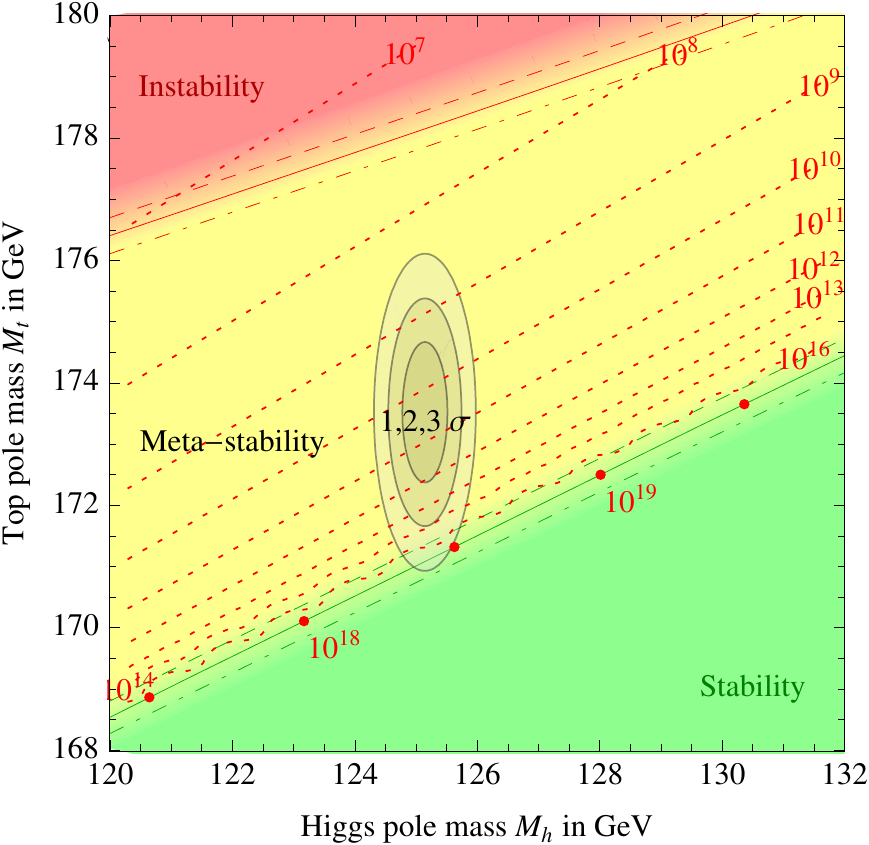}$$
\caption{\em {\bf Left}: SM phase diagram in terms of Higgs and top pole masses.
The plane is divided into regions of absolute stability, meta-stability, 
instability of the SM vacuum, and non-perturbativity of the Higgs quartic 
coupling.  The dotted contour-lines show the instability scale 
$\Lambda_I$ in GeV assuming $\alphas(\mz)=0.1184$.
{\bf Right}: Zoom in the region of the preferred experimental range of $\mh$ 
and $\mt$ (the grey areas denote the allowed region at 1, 2, and 3$\sigma$).
Plots taken from ref.\cite{Buttazzo:2013uya}.
\label{fig:4}}
\end{figure}

The regions of stability, metastability, and 
instability of the EW vacuum are shown in \Fref{fig:4} both for a broad range 
of $\mh$ and $\mt$, and after zooming into the region corresponding to the
measured values \cite{ATLAS:2014wva,Aad:2012tfa,Chatrchyan:2012ufa}. 
The latter  appear to be rather special, 
in the sense that 
the present central values of $\mh$ and $\mt$
place the SM vacuum  in a near-critical condition at the border
between stability and metastability.  The NNLO computation of the
stability bound , i.e. of the $\mh$ value that ensures 
a  stable potential up to $M_{ \rm Pl}$ (green region in \fig{fig:4}),
gives \cite{Buttazzo:2013uya}  
\bq M_{\PH} > 129.6\, \UGeV + 2.0 ( M_{\PQt} - 173.34\, \UGeV) -0.5
\UGeV  \frac{ \alphas(M_{\PZ}) -0.1184}{\rm 0.0007} \pm 0.3_{\mathrm{th}}\, 
\UGeV ~.
\label{stab} 
\eq
 \fig{fig:4} and \eqn{stab} show that to achieve 
   an EW vacuum stable up to $M_{ \rm Pl}$ the  value of top mass, identified 
   as the pole mass,  should be $\sim 2$ GeV lower than
   the present experimental central value, $\mt = 173.34$ GeV.
   In terms of $\mt$ the stability bound reads  
\bq
\mt < (171.53\pm 0.15\pm 0.23_{\alphas} \pm 0.15_{\mh})\, \UGeV~.
\label{mtsta}
\eq

The $\pm 0.3_{th}$ theoretical uncertainty in \eqn{stab} is an estimate 
of the unknown higher order corrections. It indicates that the factor that can  
discriminate  between a stable and a metastable EW vacuum  is the exact value 
of the top mass, rather  than a further refined computation.
Fig. \ref{fig:4}, as well as the bound (\ref{mtsta}), are obtained
using as renormalized mass for the top quark  the  pole mass,
$\mt^{\mathrm{pole}}$,  
and identifying it with the average of the Tevatron, CMS and ATLAS 
measurements, $\mt =173.34 \pm 0.76\UGeV$.
This identification can be debated in two aspects. i)
From a theoretical point of view the concept of pole mass for a quark is not
well defined as quarks are not free asymptotic states. Furthermore
the quark pole mass is plagued with an intrinsic non-perturbative ambiguity
of the order of $\Lambda_{QCD}$ due to the so-called infrared (IR) renormalon
effects. ii)  The top mass extracted by the experiments,  called 
Monte Carlo (MC) mass $\mt^{MC}$, is a parameter of a MC generator
determined via the comparison  between the kinematical reconstruction
of the top quark decay products and the  MC simulations of
the corresponding event.  $\mt^{MC}$ is  sensitive to the on-shell 
region of the top quark but it cannot be directly identified with 
$\mt^{\mathrm{pole}}$. 
The uncertianty quoted on $\mt$ by the experimental collaborations  refer to
$ \mt^{MC}$ and not to  $\mt^{\mathrm{pole}}$. We  do not know the exact relation 
between $ \mt^{MC}$ and   $\mt^{\mathrm{pole}}$. However, an ``educated guess'' is
to assume that $ \mt^{MC}$ can be interpreted as  $\mt^{\mathrm{pole}}$ 
within the ambiguity intrinsic in the definition of  
$\mt^{\mathrm{pole}}$, thus at the level of $\sim 250-500$ GeV.

 Alternative ways to get the top pole mass from the experimental determinations
 can be considered.
The MC mass  can be better related to a theoretically well defined 
short-distance 
mass, i.e. a mass defined in a renormalization scheme that avoids spurious 
higher-order  renormalon effects, taken at a low scale of 
the order of the top width. The  uncertainty in the translation
between  $\mt^{MC}$ and the short-distance mass is estimated to be
$\sim 1\UGeV$. The short-distance mass can be then 
converted to the pole  mass using the known relation up to 
${\cal O} (\alphas^3)$ with the conversion inducing a shift $\sim 600\UMeV$
\cite{Moch:2014tta}.

A further possibility is to extract a short-distance mass defined
in the $\overline{\mbox{\sc MS}}$ scheme, $\mt^{\overline{\mbox{\sc ms}}}$,
directly  from the total
production cross section for top quark pairs $\sigma(t \bar{t} + X)$.
A recent analysis reports  $\mt^{\overline{\mbox{\sc ms}}}(\mt) = 162.3 \pm 2.3\UGeV$
\cite{Moch:2014tta}, a value that translated 
in terms of  $\mt^{\mathrm{pole}}$ 
\bq
 \mt^{\overline{\mbox{\sc ms}}}(\mt) = 162.3 \pm 2.3 ~~\UGeV \rightarrow 
\mt^{\mathrm{pole}} = 171.2 \pm 2.4 ~~\UGeV
\eq
gives a central value compatible with the full stability of the Higgs potential.

As already discussed, the possibility of the full stability of the SM Higgs 
potential requires an $\mt^{\mathrm{pole}} \sim 171$ GeV.  The top pole mass  
is  the same object that enters  the EW fit and it can be predicted now  that
we know the Higgs mass quite accurately. A recent indirect determination
of $\mt^{\mathrm{pole}}$ from a global fit to EW precision data, i.e. without 
using in the fit the experimental information on $\mt$, reports
$\mt^{\mathrm{pole}} = 176.6 \pm 2.5\UGeV$ \cite{Ciuchini:2014dea}. This number 
shows the tendency of the EW fit to prefer high values of $\mt$, therefore not
supporting the possibility of an  EW vacuum stable up to $M_{ \rm Pl}$.

It is clear that the issue of the (meta)stability of the SM Higgs potential,
with its important  implications for the case of NP or cosmology (like the
possibility of vacuum decay during inflation), will not be fully clarified
until two conditions are realized: more precise measurements of $\mt$ and 
a better control of the uncertainty in the relation between the experimentally 
determined quantities and the corresponding theoretical parameters.  

\hfill\eject
\section{Jet physics}
Measurements of QCD processes are necessary to better control them in
their role as backgrounds to almost all the possible channels for discovering
new physics at LHC, and in order to refine our
understanding of the strong sector.  In the challenging task to
understand the mechanism of electroweak symmetry breaking~(EWSB) and to
explore the TeV scales, the energy and intensity of collisions have grown
in the past and will be further increased in future hadron colliders.  This
makes events at colliders a very busy hadronic environment.

The best reason for new physics to live anywhere near the weak scale is that
it is partially responsible for the generation of this scale.  New physics
that is related to EWSB will naturally couple most strongly to those
particles in the SM which feel EWSB most strongly, in particular the top
quark and the electroweak~(EW) bosons ($\PH$, $\PW$, and $\PZ$), and thus
will decay preferentially into these heavy particles rather than into light
quarks and leptons, which yield simpler final states.  Moreover, we have
compelling reasons to believe that the new particles or resonances will
naturally decay to boosted SM particles~\cite{Altheimer:2013yza}.  Even
before the LHC turned on, the lack of deviations from SM predictions for
flavor or precision electroweak observables already hinted that the
most-likely scale for new physics was not the vacuum expectation value
$v_{\rm EW}$, as naturalness might have suggested, but rather $\Lambda
\gtrsim $ few TeV.  Evidence for this ``little hierarchy'' problem has of
course only gotten stronger as the LHC has directly explored physics at the
TeV scales.  Thus many models which address the stabilization of the EW scale
will naturally give rise to final states rich in boosted tops, Higgses, $\PW$'s
and $\PZ$'s.  These particles will have an appreciable cross section to be
produced in a kinematic regime where they are boosted and give rise to
collimated decay products.  The
simple picture that one hard parton corresponds to one jet breaks down badly
in this scenario, and new tools are needed to separate out collimated decays
from standard QCD showers.

Even in the absence of a resonance or other mechanism to preferentially
populate boosted regions of phase space, looking for boosted signals can be
useful for improving the signal over background ratio. In fact, any change in
the reconstruction method affects both the signal and the backgrounds.
Background reduction comes in two forms:

\begin{enumerate}
\item In high-multiplicity final states, combinatorial background is often
  prohibitive.  When some or all of the final-state particles are boosted,
  the combinatorial background is greatly reduced.
\item
  In addition to this, it is also possible to use boosted selection
  techniques to identify regions where the background from other physics
  processes is intrinsically reduced.

\end{enumerate}
We will illustrate these features using $\Pt\tbar$ and $\PH\PV$ production as
example.

\subsubsection{$\Pt\tbar$ production}
To appreciate the need for new reconstruction techniques, consider the
production of a $\Pt\tbar$ pair at fixed center-of-mass energy.  If we set
the jet radius $R_0$ to 0.6, a typical value used in jet
reconstruction at the LHC, it is interesting to investigate the fraction of
top quarks that have all three, only two, or none of their decay products
($\Pb jj $) isolated from the others at that scale. This gives a rough
estimate of how well a jet algorithm with $R=R_0$ will be able to reconstruct
the three partonic top decay products as separate jets.  For a centre-of-mass
energy of 1.5~TeV, 20\% of the top quarks are reconstructed as three separate
jets, while 20\% appear as a single jet.  But at 2~TeV, only 10\% of the top
quarks are reconstructed as 3 separate jets, while 45\% appear as a single
jet~\cite{Shelton:2013an}.  Clearly, tops produced in the very interesting
high~TeV regime (>10~TeV) straddle the borderlines between several
different topologies.  For this reason it would be more desirable to have a
flexible reconstruction method that can handle semi-collimated tops in a
unified way.

\subsubsection{Boosted Higgs boson production}
Searching for the Higgs boson in its decay to $\Pb\bbar$ is very difficult at
the LHC, due to overwhelming QCD backgrounds.  Even in associated production
with a vector boson, $\Pp\Pp\to \PH \PZ$ or $\Pp\Pp\to \PH \PW$, the
background processes $\PZ \Pb\bbar$, $\PW\Pb\bbar$, and even $\Pt\tbar$ are
overwhelming.  Nonetheless, thanks to Ref.~\cite{Butterworth:2008iy}, $\Pp\Pp\to
\PH \PV$, $\PH\to \Pb\bbar$ is now an active search channel at the LHC.
If we consider, for example, $\PH \PZ$ production, with leptonic decay of the
$\PZ$ boson, the traditional approach was to look for final states with two
leptons, compatible with the $\PZ$ boson decay, and 2 $\Pb$-tagged jets, and
to reconstruct the invariant mass of the $\Pb$-jets, and look for a peak in
the distribution of $m_{\Pb\bbar}$.  The new approach, suggested by recent
developments in jet physics, is instead to focus on events where the Higgs
boson is produced with high transverse momentum, i.e.~the event is
characterized by $p_{\mathrm TV}$ > 200~GeV and cluster these events with a
large jet radius ($R= 1.2$), such that all of the Higgs decay products are
swept up in a single fat jet.  The signal is now a leptonic $\PZ$ + a fat
``Higgs-like'' jet, and the background to this signal is now $\PZ$ plus one
fat jet rather than $\PZ\Pb\bbar$.
For an unboosted search, the ultimate discriminator between signal and
background is the $\Pb \bbar$ invariant mass: the goal is to find a
resonance, a bump, in the $\Pb \bbar$ mass spectrum.  In the boosted regime,
the Higgs boson is collected into a single fat jet and the Higgs boson mass
should be reflected in the invariant mass of the fat jet itself.
The jet-substructure algorithms offer enough quantitative precision to
discriminate between a jet from Higgs boson decay and a QCD jet.  In fact, a
Higgs boson that decays perturbatively into a $\Pb \bbar$ pair tends to
generate two quarks that share in a more symmetric way the initial energy. On
the contrary, QCD splitting from shower is more asymmetric. In addition, the
Higgs boson decays into two (almost) massless quarks in one step, while QCD
splittings prefer to share their virtualities gradually.  Procedures like the
``mass-drop'' and ``filtering'' are conceived to resolve the fat jet and
distinguish QCD jets from Higgs-like ones.

In this way, the background is reduced by an extent that compensates the
acceptance price demanded by the high-$\pT$ cut.

\subsection{Grooming techniques}
Inspired by these new developments, a lively research field has emerged
in recent years, investigating how to best identify the characteristic
substructure that appears inside single ``fat'' jets from electroweak scale
objects (see e.g.~Refs.~\cite{Abdesselam:2010pt, Altheimer:2012mn,
  Plehn:2011tg} for a review).  Many ``grooming'' and ``tagging'' algorithms
have been developed and are now tested in experimental analyses (in
particular see Refs.~\cite{ATLAS:2012am, Aad:2012meb, Aad:2013gja,
  Chatrchyan:2013vbb} for studies on QCD jets).  An example of these of new
jet techniques are trimming~\cite{Krohn:2009th} and
pruning~\cite{Ellis:2009su,Ellis:2009me} algorithms.

All three grooming techniques (filtering, trimming, and pruning) increase the
signal to background ratio by both improving mass resolution for signal and
suppressing QCD background.  For example, QCD jets, whose jet masses are
generated by relatively softer and less symmetric emissions, are more likely
to have their masses shifted substantially downward by jet grooming than
collimated perturbatively decaying particles, thus depleting the
background to high-mass searches.

\subsection{Jet shapes}
Another field of investigation that has seen a rapid development and will
surely benefit from future investigation concerns the study of jet shapes. A
jet shape is typically a function $f$ defined on a jet $J$ that quantifies the properties
of the jet without the (explicit) use of any jet algorithm.  The approach is
conceptually similar to event shapes, which allow quantitative study of QCD,
without requiring specific characterization of an event in terms of jets, and
indeed many jet shapes are descendants of event shapes.
Again their aim is to target non-QCD-like substructures in jets from QCD
ones by studying the radial distribution of particles in the jet
(jet broadening, differential and integrated jet shape), the spread in
radiation in the plane perpendicular to the jet (planar
flow~\cite{Almeida:2008yp}), the existence of subjets
($N$-subjettiness~\cite{Thaler:2010tr}), colour structure of jets (jet
pull~\cite{Gallicchio:2010sw}, dipolarity~\cite{Hook:2011cq}), etc.
We recall in the following the definition and characteristics of a few of
them.

\subsubsection{Radial distribution of particles within a jet}
The probability of the splitting of a parton into two other partons
depends on the running coupling $\alphas$ evaluated at the $k_\perp$
scale of the splitting.  Jet shapes which measure the angular
distribution of particles in an event are therefore measuring both the
strength and the running of the strong coupling constant, and are
classic probes of QCD.  These jet shapes are also sensitive to the
colour charge of the parent parton: since $C_{\mathrm F} < C_{\mathrm A}$, an initial gluon
will radiate more, and at wider angles, than an initial quark.

\begin{itemize}

\item {\bf Jet broadening}: given a thrust axis $\hat n$, we can partition
  the particles into two hemispheres according to the sign of
  $(\vec{p}_i\cdot\hat n )$, where $\vec{p}_i$ is the three-momentum of the
  $i$-th particle. For example, for dijet-like events, this is equivalent to
  associating each particle to a jet.  Hemisphere broadening is then defined
  as the momentum-weighted transverse spread of the particles
\beq
B_H = \frac{1}{\sum_{i\in H} |\vec{p}_i|} \sum_{i\in H} |\vec{p}_i \times \hat n |
\eeq
where the sum runs over all particles $i$ in a hemisphere $H$.

\item {\bf Differential} $\rho(r)$ and {\bf integrated jet shapes} $\Psi(r)$
  characterize the radial distribution of radiation inside a jet.  Both of
  these shapes are defined on an ensemble of $N$ jets of radius $R$.
  Then for $r < R$, the integrated jet shape $\Psi(r)$ is the ensemble
  average of the fraction of a jet $\pT$ which is contained within a radius
  $r$ from the jet axis.  Defining $r_i$ as the distance of a constituent $i$
  from the jet axis
\begin{equation}
\Psi(r) = \frac{1}{N} \sum_{J} \sum_{i\in J} \frac{\pT(0 < r_i <r)}{p_{\mathrm T,J}}.
\end{equation}
Here the second sum runs over all constituents $i$ of a jet $J$.
The differential jet shape $\rho(r)$ is then given by
\begin{equation}
\rho(r) =\frac {1} {\delta r}\, \frac {1} {N}\sum_{J} \sum_{i\in J}
\frac{\pT(r <r_i < r+\delta r)}{p_{\mathrm T, J}} . 
\end{equation}
These variables are often included in the suite of QCD precision measurements
performed by experimental collaborations, and are useful for validating
parton shower models.

\end{itemize}

\subsubsection{Shape variables for boosted decay kinematics}
The radial distribution jet shapes discussed in the previous section are
geared toward probing the characteristic structure of QCD showers.  Here we
will recall a couple of examples of jet shapes that target evidence of
non-QCD-like substructure in jets.
\begin{itemize}

\item {\bf Planar flow}~\cite{Almeida:2008yp} considers the spread of the jet
  radiation in the plane transverse to the jet axis.  Since QCD coherence
  gives rise to angular-ordered showers, radiation subsequent to the first
  emission $P\to i j$ tends to be concentrated between the clusters of energy
  defined by $i$ and $j$, leading to a roughly linear distribution of energy
  in the jet. By contrast, boosted three-body decays, such as boosted tops,
  have a more planar distribution of energy.

Planar flow is defined in terms of an auxiliary tensor
\beq
I^{ab} = \frac{1}{m_J}\sum_{i\in J} \frac{p^a_{i,\perp} p^b_{i,\perp}}{E_i},
\eeq
where the indices $a,\,b$ span the plane perpendicular to the jet axis, and
$\vec{p}_{ i,\perp}$ denotes the projection of the momentum of the $i$-th
particle onto this plane.  Letting $\lambda_1,\,\lambda_2 $ be the
eigenvalues of $I^{ab}$, the planar flow of a jet is defined by
\beq
Pf_J=\frac{4\lambda_1\lambda_2}{ (\lambda_1+\lambda_2) ^ 2} =\frac {\det
  I}{({\rm tr} \,I)^2}.
\eeq
With this normalization, $Pf_J\in (0,1)$. Monte Carlo studies have
demonstrated that QCD events do indeed peak at low values of $Pf$,
while boosted top decays show a relatively flat distribution in $Pf$,
but preliminary results show some sensitivity to shower modeling
\cite{Thaler:2008ju} and the utility of this shape in data is so far
unclear. Further studies will be needed to clarify these issues.


\item {\bf ${\boldmath N}$-subjettiness}~\cite{Thaler:2010tr}: given $N$ axes
  $\hat n_k$, we define $N$-subjettiness as
\beq
\tau_N =\frac{\sum_{i\in J}p_{\mathrm T,i} \min (\Delta R_{i k} )
}{\sum_{i\in J} p_{\mathrm T,i} R_0} \,,
\eeq
where $R_0$ is the jet radius, and $\Delta R_{ik}$ is the distance
between particle $i$ and axis $\hat n_k$. The smaller $\tau_N$ is, the
more radiation is clustered around the chosen axes, or in other words,
smaller values of $\tau_N$ indicate a better characterization of the
jet $J$ as having $N$ (or fewer) subjets.  Conversely, if $\tau_N$ is
large, then a description in terms of $>N$ subjets is more desirable.

However, as QCD alone will easily make jets with subjets, to differentiate
boosted objects we need to probe not just the possible existence of subjets,
but their structure.  The real distinguishing power of $N$-subjettiness
occurs when looking at ratios. For instance, a two-prong boosted particle
such as a Higgs boson or a vector boson $\PV$ will have large $\tau_1$ and
small $\tau_2$.  QCD jets which have small $\tau_2$ will generically have
smaller $\tau_1$ than for signal, as the QCD jets are more
hierarchical. Conversely, QCD jets which have large $\tau_1$ are generally
diffuse, and will have larger $\tau_2$ as well than for signal.  Thus the
best single discriminating variable is $\tau_2/\tau_1$, or, more generally
\beq
r_{N} = \frac{\tau_N}{\tau_{N-1}}
\eeq
for a boosted $N$-prong particle.

\end{itemize}

\subsubsection{Colour-flow variables}
\begin{figure}
\begin{center}
\includegraphics[width=0.48\textwidth]{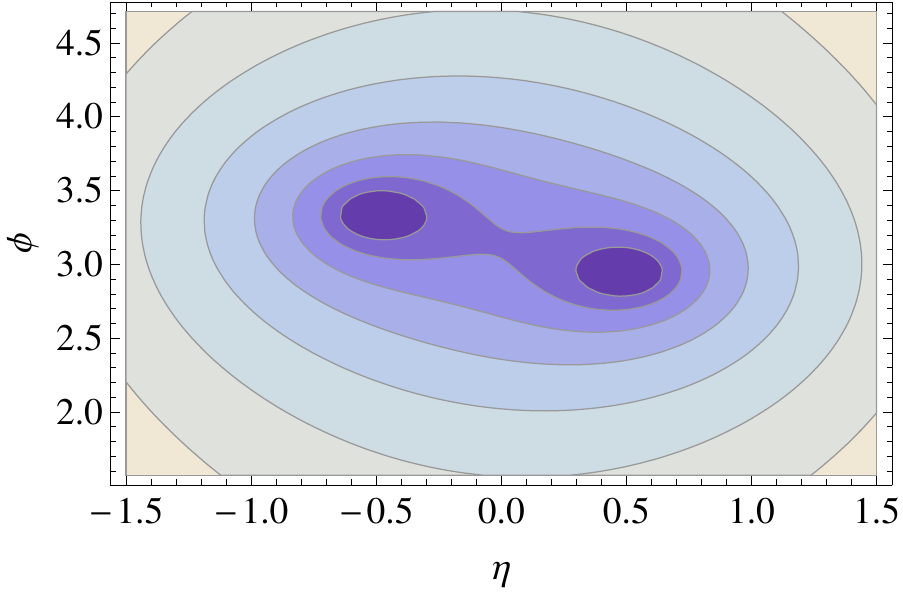}
\includegraphics[width=0.48\textwidth]{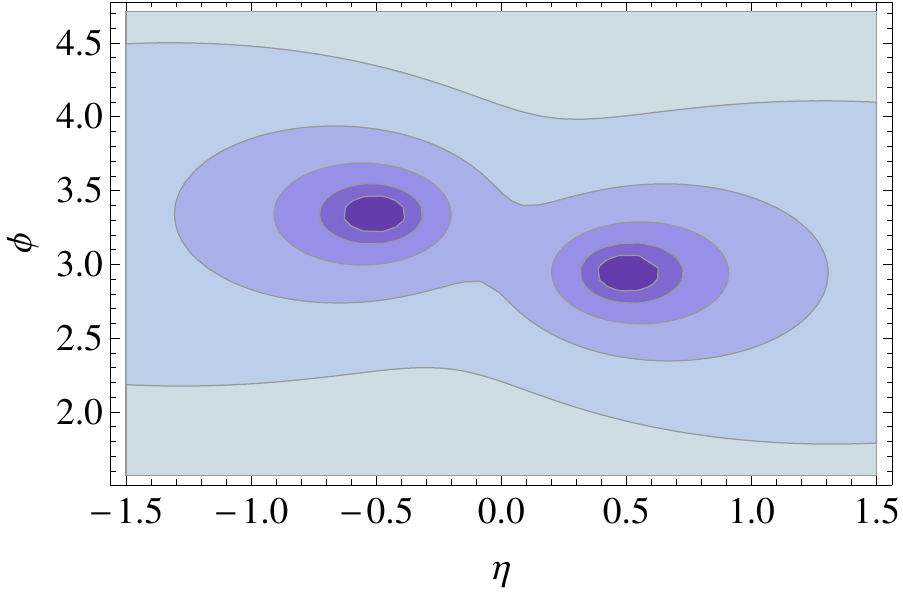}
\caption{Radiation patterns in the eikonal approximation for two
  triplet colour sources colour-connected to each other~(left) and to
  the beam~(right).  Contours are logarithmic, and the scales in the
  two figures are not the same. From Ref.~\cite{Shelton:2013an}.}
\label{fig:colorflow}
\end{center}
\end{figure}
Beyond kinematics, boosted perturbative decays can also differ from
QCD backgrounds in their colour structure.  Consider a colour singlet
such as a $\PH$ or $\PV$ boson decaying to a quark-antiquark pair.  The
decay quark jets form a colour dipole: they are colour-connected to
each other, but not to the rest of the event. Meanwhile, the
backgrounds to these processes come from QCD dijets, which necessarily
have different colour connections, as shown in
Fig.~\ref{fig:colorflow}~\cite{Shelton:2013an}, where the radiation patterns for a
colour-singlet signal are plotted on the left and for a typical
background on the right, as computed in the eikonal (soft)
approximation. This observation has motivated work on variables which
can add colour flow to the suite of variables which can discriminate
signal from background.

\subsection{Conclusions}
Until very recently, nearly all theoretical studies of jet substructure have
been performed using Monte Carlo parton shower programs (see for instance
Ref.~\cite{Altheimer:2012mn} and references therein), with tools such as
Herwig and Pythia. While these are powerful general purpose tools, their
numerical nature masks insight into the dependence on tagger and jet
algorithm parameters, which should ideally be optimised for the purposes of
detecting new physics. Such a detailed level of understanding, manifested as
accurate analytic QCD predictions, is a key ingredient for substructure
analyses to reach their full potential. However it is far from obvious that,
given their inherent complexity, jet-substructure observables can be
understood to a high level of accuracy analytically.

Future progress on analytic calculations, along the lines of what has been done
in Refs.~\cite{Dasgupta:2013ihk,Dasgupta:2013via} (see for instance
Ref.~\cite{Altheimer:2013yza} for a review), and on the
merging of high-precision fixed-order calculations with parton shower
algorithms, as in POWHEG and MC@NLO, will doubtless shed more light in
regards to jet substructure in the near future.

\hfill\eject
\section{Higher order QCD corrections and resummation\footnote{L.~Magnea, G.Ferrera}}

\subsection{Introduction\footnote{L.~Magnea}}
\label{intro}

It is clear that precision QCD calculations are a necessary ingredient
for future discoveries. The argument, which has been implicit in most
of the previous discussion in this document, can be summarized as follows.

\begin{itemize}

\item The first run of the LHC successfully discovered a scalar boson which closely matches
the properties of the Standard Model Higgs boson. This was largely expected, although by 
no means guaranteed. On the other hand, while operating at an energy four times higher 
than previously achieved, the LHC failed to uncover any signs of ``new physics'': in fact, all
results to date are in impressive agreement with Standard Model (SM) predictions. This is 
true not only for the experiments probing the high-energy end of the spectrum (roughly `top 
and Higgs' at ATLAS and CMS), but also for intensity/precision experiments (witness the 
spectacular measurements~\cite{Chatrchyan:2013bka,Aaij:2013aka,CMS:2014xfa} of the 
branching ratio of the decay $B_s \rightarrow \PGmp \PGmm$ by CMS and LHCb).

\item In the next run, the LHC will extend its energy reach by a factor of roughly $1.6$. 
Spectacular discoveries in this new energy range (typically new resonances directly produced 
in the $s$ channel) are possible, and should indeed be hoped for. One must however
be realistic: the impressive agreement of all existing data with the SM, and the relatively 
modest increase in the available energy, make such discoveries unlikely, in the following 
limited sense: we have at this point no compelling reason to expect new physics to become 
directly accessible between $8$ and $13\UGeV$. 

\item Such a situation is not new nor exceptional. From Kepler's Laws to Bohr's atomic 
model, disruptive physics discoveries have more often come from increased precision in 
the measurements of existing phenomena than from the opening of new energy ranges.
It is likely that we will find ourselves, in the next several years, once again in a situation 
in which our best available option for discovery will be increasing the accuracy and precision
of experimental measurements, and of theoretical predictions based on existing theories.

\item More specifically, in collider physics language, even if we don't have the energy to 
directly access new very massive states via `$s$-channel' production, we can still (hopefully)
measure their contributions to low-energy observables via virtual exchanges. These can be
for example $t$-channel exchanges, which would induce deviations from SM predictions
in the tails of energy distributions, or loop-level exchanges affecting SM parameters such
as couplings or mixing angles. Such small deviations from SM predictions can only be 
reliably observed if the SM-based theoretical prediction is sufficiently precise and accurate. 
In this limited sense, it is quite possible that future discoveries in high-energy physics may 
hinge on the degree of accuracy that our calculations can reach.

\end{itemize}

With these general premises, one hardly needs to emphasize the relevance of advanced
QCD calculations for future experiments. The LHC is a hadron collider, and any precision 
measurement in such a collider requires, one way or another, a detailed understanding 
of the underlying QCD phenomenology. Future linear (or circular) $\Pep\Pem$ colliders
may focus part of their program on the production of electroweak final states ($\PZ, \PZ\PZ, \PH, 
\PH\PH \dots$), however many crucial inputs and outputs from such searches will be driven
by QCD (top production, jet studies, $\alphas$ measurements, to name a few).

The focus of this section will be on QCD resummations: the question is then 
how resummations, given the context described above, can be 
useful for selected phenomenological applications. A related question is what kind of 
resummation technology we may expect to be available on the time scale 
of several years. We will first summarize some basic facts
about the classic formalism of soft-gluon resummation, and then turn
to more recent developments, by tackling the above two questions in 
reverse order: first tentatively outlining the likely future developments in QCD resummations,
and then making a few observations on possible phenomenological applications.

\section{Higher order QCD corrections and resummation\footnote{L.~Magnea, G.Ferrera}}

\subsection{Sudakov resummation~\footnote{G.~Ferrera}}

The all-order summation of the logarithmically-enhanced perturbative corrections
produced by soft-gluon radiation, also known as Sudakov resummation~\cite{Sterman:1986aj,
Catani:1989ne,Contopanagos:1996nh,Forte:2002ni},
is a very important topic 
for physics studies within (and beyond) the Standard Model, at present and future 
accelerator energies.

The origin of the Sudakov logarithms is well known. 
In particular kinematical regions, where the contributions of real and virtual parton emissions are
highly unbalanced, the reliability of the standard  perturbative expansion 
(\ie \ order-by-order in powers of the QCD coupling $\alphas$)
is spoiled by the presence of large double-logarithmic terms,
which are finite residual effects of the cancellation of infrared (soft and collinear) 
singularities 
in IR-safe 
QCD cross sections. 
The predictivity of the perturbative expansion can be restored by
summing these logarithmically-enhanced contributions to all order in $\alphas$.

Sudakov resummation can be performed with analytical techniques 
by exploiting dynamics and kinematics factorizations.
While dynamical factorization
is a general property of multi-gluon QCD amplitudes in the soft limit,
phase-space factorization strongly depends
on the observable under consideration.
If the phase-space 
in the soft limit factorizes, 
multi-gluon emissions can be written in the form of a generalized
exponentiation of the single-gluon emission probability.
In such cases it is possible 
to perform an improved perturbative expansion 
that systematically resums, to all orders in $\alphas$, the 
leading (LL), next-to-leading (NLL), next-to-next-to-leading (NNLL)  (and so forth)
logarithmic contributions. 

In the following we briefly discuss 
Sudakov resummation in the case of two of the most important hard-scattering 
observables in hadronic collisions:
inclusive cross sections in the threshold region 
and transverse-momentum ($q_{\mathrm T}$) distributions at low $q_{\mathrm T}$.

\subsubsection*{Threshold resummation}

Threshold logarithms appear in the perturbative expansion of inclusive cross sections 
when the observed high mass ($M^2$) system  
is forced to carry 
a very large fraction $x$ of
the available (partonic) centre-of-mass energy $\sqrt s$. 
The kinematical variable $x=M^2/s$ parametrises the distance from the partonic 
threshold $x=1$.
In the kinematical region close to the partonic threshold ($x\to 1$) 
the emission of real radiation at higher orders is strongly suppressed.
As a result, large 
logarithms of the type $L = \ln(1-x)$ 
appear order-by-order in the perturbative expansion, in the form
\begin{equation}
c_{nm} \alphas^n L^{m}\,,~~\mbox{with}~~1\leq m\leq 2n\,.
\end{equation}
In order to get reliable perturbative predictions, these  
logarithmic corrections (which diverge in the  $x \to 1$ limit) 
have to be resummed
to all orders~\cite{Sterman:1986aj, Catani:1989ne, Catani:1990rp}.
In the case of inclusive cross sections near threshold, phase-space factorization is obtained by working in the conjugated
Mellin ($N$-moment) space where
soft-gluon resummation  can be systematically performed (see, however, \cite{Becher:2006nr} for an alternative viewpoint).

Due to finite experimental acceptances,  
theoretical predictions in Mellin space cannot be compared directly with data, 
and the inversion to the physical $x$-space has to be performed. Resummed expressions,
however, diverge at very large $N$, where the Landau singularity in the QCD running coupling signals the onset of non-perturbative
phenomena. The Mellin inversion can be performed only after the introduction of
a prescription which regularizes the Landau singularity~\cite{Catani:1996yz,Magnea:2000ss,Forte:2006mi, Abbate:2007qv}. 

The formalism to perform threshold resummation was first developed 
in the case of processes involving two QCD partons at the Born 
level~\cite{Sterman:1986aj, Catani:1989ne, Catani:1990rp,Contopanagos:1996nh,Forte:2002ni}
and successively
extended to the more general case of inclusive cross sections
in multiparton processes~\cite{Kidonakis:1996aq, Bonciani:1998vc, Kidonakis:1998nf, Laenen:1998qw, Catani:1998tm,Bonciani:2003nt}.
More recently threshold resummation techniques based on effective theories have been developed
\cite{Manohar:2003vb, Idilbi:2006dg, Becher:2007ty, Ahrens:2008nc, Beneke:2009rj, Becher:2009th, 
Bonvini:2012az}.

Some processes in hadronic collision where 
threshold resummation is particularly important 
are the 
production of vector and Higgs bosons 
\cite{Vogt:2000ci, Forte:2002ni, Catani:2003zt, Laenen:2005uz,Moch:2005ky,Ravindran:2005vv, Ravindran:2006bu, Bonvini:2010tp, Ahmed:2014uya, Bonvini:2014joa},
prompt photons 
\cite{Laenen:1998qw, Catani:1998tm, Sterman:2000pt, Bolzoni:2005xn, Becher:2009th},
heavy quarks
\cite{Kidonakis:1996aq, Bonciani:1998vc, Laenen:1998qw, Almeida:2008ug, 
Beneke:2009rj, Czakon:2009zw, Ahrens:2010zv}, and
jet 
and  single-hadron inclusive production 
\cite{Catani:1996yz, Kidonakis:1998bk, deFlorian:2007fv, deFlorian:2005yj, Catani:2013vaa}.

\subsubsection*{Transverse-momentum resummation}

Transverse-momentum logarithms
occur in the transverse-momentum ($q_{\mathrm T}$)
distribution  of high invariant mass systems  ($M$)
in the region of small $q_{\mathrm T}$ ($q_{\mathrm T} < \hspace{-5pt} < M$).
Also in this case the suppression of real emissions 
gives rise
to large double-logarithms of the type $L = \ln M^2/q_{\mathrm T}^2$, order-by-order in perturbation theory.

Transverse-momentum resummation 
for the 
hadroproduction of an arbitrary system of colourless particles,
first developed in the series of 
papers~\cite{Dokshitzer:1978yd, Dokshitzer:1978hw, Parisi:1979se, 
Curci:1979bg, Collins:1981uk, Kodaira:1981nh, Davies:1984hs,
Collins:1984kg, Catani:1988vd}, 
is nowadays well understood~\cite{deFlorian:2000pr, Catani:2000vq, Catani:2010pd, 
Catani:2013tia}. 
Some examples of such systems are DY lepton pairs~\cite{Bozzi:2010xn, Guzzi:2013aja}, 
Higgs boson~\cite{Bozzi:2005wk, deFlorian:2011xf, Wang:2012xs} and diboson 
production~\cite{Balazs:2006cc, Cieri:2015rqa, Grazzini:2015wpa}.

On the
contrary, in the case of hadroproduction 
of systems that involve coloured QCD partons,
the structure of colour correlations and coherence effects lead to theoretical complications
which have still prevented a fully general extension of the resummation formalism.
Nonetheless the phenomenological importance of multiparton 
scattering processes
together with the high 
precision experimental data, strongly demand generalizations of the transverse-momentum 
resummation formalism for 
such processes.
Recent theoretical progress in this direction  was obtained in Refs.~\cite{Zhu:2012ts, Catani:2014qha}
by considering the specific case of the hadroproduction of a heavy-quark pair ($Q\bar{Q}$) with a small $q_{\mathrm T}\ll m_Q$.

Transverse-momentum resummation has also been reformulated
in the framework of effective theories~\cite{Gao:2005iu, Idilbi:2005er, Mantry:2009qz, 
Becher:2010tm, GarciaEchevarria:2011rb, Becher:2012yn, D'Alesio:2014vja}
and it can also be performed by using approaches
beyond the customary QCD framework of collinear factorization, 
that use transverse-momentum dependent (TMD) factorization
and introduce transverse-momentum dependent parton distributions functions (TMD PDFs)
(see \Bref{Collins:2011zzd} and references therein).


\subsubsection*{Universality of Sudakov resummation}

An important aspect of Sudakov resummation is related to its universality 
(\ie \ 
process independence).
Resummed cross sections 
can be expressed in a factorized form which involves 
a {\itshape process-independent form factor}
which resums to all orders the corrections
due to soft and collinear parton emissions,
and a {\itshape process-dependent hard factor} which takes into account 
hard-virtual contributions. 
The all-order resummation of the logarithmic corrections
is controlled by these
factors which are expressed in terms of perturbative functions
with coefficients computable order-by-order in perturbation theory. 

In the case of hadroproduction of colourless particles, it has been shown that
the hard factor, despite its intrinsic process dependence,  has 
an  all-order  universal structure~\cite{Catani:2013tia}. 
The  process-dependent information encoded in the hard factor 
can be entirely extracted by the scattering amplitude of the
Born-level partonic subprocess and its virtual radiative corrections~\cite{Catani:2013tia}.
The hard resummation factor is directly determined
by a universal (process-independent) all-order factorization formula,
that originates from the factorization properties of soft and collinear parton radiation,
and by the knowledge of the corresponding scattering amplitude. 
This factorization formula has been explicitly evaluated,
in the case of hadroproduction of an arbitrary system  of colourless particles,
up to the next-to-next-to-leading order (NNLO)  in the case of $q_{\mathrm T}$-resummation~\cite{Catani:2013tia} and 
N$^3$LO in the case of threshold 
resummation~\cite{Catani:2014uta}.
Results in the case of the production of coloured objects have been obtained in 
Refs.~\cite{Catani:2013vaa,Catani:2014qha}.


\subsubsection*{Other aspects}

Sudakov resummation techniques have nowadays reached a high level of accuracy
and 
resummed calculations up to NNLL order are available for various observables in many different processes. 
This increasing
precision is  of fundamental importance to fully exploit the discovery potential provided by
the high quality of the collected and forthcoming collider data.
For instance the successful accomplishment of the LHC physics program will depend on the 
ability to provide precise theoretical predictions.
Many experimental results are indeed sensitive to soft-gluon effects and resummed calculations
(consistently matched to standard fixed-order results) 
allow us to enlarge the applicability of precise perturbative 
QCD predictions.

Let us finally stress that analytic techniques to perform all-order Sudakov resummation
are also important for other aspects of perturbative QCD.
The parton shower algorithms which are implemented in Monte Carlo event generators 
resum to all-order leading-logarithmic corrections due to collinear and soft emissions.
Analytic resummation techniques can thus be used to improve parton showers  beyond 
their present logarithmic accuracy.
Another important aspect is related to fixed-order computations. 
This is the case, for instance, 
of the {\em subtraction formalism} of \Bref{Catani:2007vq} 
which permits to perform fully-exclusive NNLO calculations 
using the knowledge of the transverse-momentum distributions in the small $q_{\mathrm T}$ 
region.


\subsubsection*{Conclusions}

Sudakov resummation techniques have nowadays reached a high level of accuracy
and 
resummed calculations up to NNLL order are available for various observables in many different processes. 
This increasing
precision is  of fundamental importance to fully exploit the discovery potential provided by
the high quality of the collected and forthcoming accelerator data.
For instance the successful accomplishment of the LHC physics program will depend on the 
ability to provide precise theoretical predictions.
Many experimental results are indeed sensitive to soft-gluon effects and resummed calculations
(consistently matched to standard fixed-order results) 
allow us to enlarge the applicability of precise perturbative 
QCD predictions.

Let us finally stress that analytic techniques to perform all-order Sudakov resummation
are also important for other aspects of perturbative QCD.
The parton shower algorithms which are implemented in Monte Carlo event generators 
resum to all-order leading-logarithmic corrections due to collinear and soft emissions.
Analytic resummation techniques can thus be used to improve parton showers  beyond 
their present logarithmic accuracy.
Another important aspect is related to fixed-order computations. 
This is the case, for instance, 
of the {\em subtraction formalism} of \Bref{Catani:2007vq} 
which permits to perform fully-exclusive NNLO calculations 
using the knowledge of the transverse-momentum distributions in the small $q_{\mathrm T}$ 
region.


\subsection{Resummations: future developments\footnote{L.~Magnea}}
\label{FutRes}

The development of resummation technology proceeds mainly in two ways. On the one 
hand, there are well-established theorems (see, for example,~\Bref{Sterman:1995fz}) stating 
that, for certain inclusive cross sections, {\it all} logarithms associated with soft and collinear 
emissions exponentiate. For such cross sections progress comes in the form of increased 
accuracy: new finite-order calculations provide the values for the relevant anomalous 
dimensions, and the contributions of more towers of logarithms become explicitly known. 
On the other hand, resummation theorems can be extended in various directions, for 
example to less inclusive or more complicated observables, or to new classes of logarithms. 
Let us tackle these two lines of progress in turn.

\subsubsection{Towards greater logarithmic accuracy}
\label{Accu}

Recent years have seen remarkable progress in high-order perturbative calculations 
(see for example~\cite{Henn:2014yza}). Further progress is to be expected in the next
several years, as new techniques are brought to fruition. While it is very difficult to predict
developments on a time scale of five to ten years, it is perhaps useful to attempt to list what 
might happen. The following is a tentative list of finite order perturbative calculations, relevant 
for resummations, which might be completed within the time frame we are considering.

\begin{itemize}

\item Fully inclusive electroweak final state process. The three-loop (${\rm N}^3$LO) 
corrections to the inclusive cross sections for the Drell-Yan process, for W and Z production,
and for Higgs production via gluon fusion, are being computed~\cite{Anastasiou:2014vaa,
Anastasiou:2014lda} (time scale: 1 year\footnote{The calculation of the Higgs production 
cross section in the gluon fusion channel at N$^3$LO in QCD has recently been made 
available as a power series in the threshold expansion in~\Bref{Anastasiou:2015ema}}). 
The contributions at this order which are relevant for threshold resummation are already 
known~\cite{Moch:2005ky,Laenen:2005uz,Anastasiou:2013mca} and being put to use by 
several groups~\cite{Ahmed:2014cla,Ahmed:2014uya,Bonvini:2014joa,Catani:2014uta,
Li:2014bfa,Li:2014afw}.

\item It is to be expected that simple distributions for these processes ($p_{\mathrm T}$, rapidity)
at the same accuracy will become known in the medium term (time scale: three years),
since the required techniques are known and the increase in complexity is incremental.

\item The fully subtracted ${\rm N}^2$LO cross section for two-jet production in QCD is being 
computed~\cite{Currie:2013dwa} (time scale: 2 years). This is only marginally useful for 
resummation since all relevant anomalous dimensions are known at this accuracy, but virtual 
corrections (which have been known for some time) provide necessary matching conditions 
for possible NNLL resummations.

\item A fully exclusive analysis of three-jet production at NNLO may require significant 
refinements of current subtraction techniques (witness the time scale of the two-jet 
calculation). It is however likely that virtual two-loop corrections, necessary for matching
conditions, will become known on a time scale of three to five years.

\item The techniques for the calculation of DGLAP splitting functions at four loops 
(${\rm N}^3$LO), and in fact of the complete DIS structure functions at ${\rm N}^4$LO 
are in principle available, and the calculation could be performed on a time scale of several
years.

\item In the meantime, techniques are becoming available to compute the anomalous 
dimensions relevant for resummations directly, without resorting to fitting these values from 
finite order calculations of specific processes. The three-loop soft anomalous dimension
matrix for generic multi-parton scattering processes in QCD is being computed (time scale
of one year for the massless case\footnote{After the completion of this manuscript, the 
missing ingredient for the computation of the three-loop massless multi-parton soft anomalous 
dimension matrix, a quadrupole contribution, was presented in~\cite{Almelid:2015jia}}, and 
two to three years for the massive case, see for example 
Refs.~\cite{Gardi:2013saa,Falcioni:2014pka,Gardi:2014kpa,Magnea:2014vha}
for a review of recent progress).

\end{itemize}

\noindent This (potential) wealth of new finite order results would almost automatically lead 
to a considerable refinement of existing resummation techniques. Here's a list of what
could become available within the stated timescale.

\begin{itemize}

\item The inclusive cross sections and simple inclusive distributions (such as $p_{\mathrm T}$ 
and rapidity) for electroweak annihilation processes will be available with ${\rm N}^3$LL 
accuracy, fully matched to exact ${\rm N}^3$LO calculations. The processes include Drell-Yan, 
$\PW$ and $\PZ$ production, Higgs production in gluon fusion, but also for example di-boson 
production (two $\PZ$'s, two Higgses, ...) where however the matching to ${\rm N}^3$LO will 
remain incomplete for some time.

\item Sufficiently inclusive jet distributions (single inclusive jet $p_{\mathrm T}$, dijet mass, ...)
will be known to ${\rm N}^2$LO, with ${\rm N}^2$LL threshold resummation. This involves
some subtle issues of non-universality {\it w.r.t.} jet algorithms, and the existence of non-Sudakov
logarithms, which however are likely to have been tackled within our stated time frame.

\item The calculation of four-loop DIS structure functions (supplemented with a value for 
the five-loop light-like cusp anomalous dimension) would lead to a fairly stunning resummed 
prediction at ${\rm N}^4$LL. Given the rather detailed knowledge of power-suppressed 
corrections near threshold in this case~\cite{Gardi:2002bk}, structure functions would stand 
to remain the best predicted quantities in perturbative QCD for quite some time, with likely 
effects on the accuracy of PDF fits.

\item If the need were to arise, typically if a Linear Collider or CLIC is built, much of the
knowledge of space-like QCD processes will not be too difficult to transfer to time-like 
kinematics, and we can expect more detailed calculations of event shape distributions, 
resummed with ${\rm N}^3$LL (and perhaps at some point ${\rm N}^4$LL) accuracy, 
matched to NNLO (and perhaps in future at ${\rm N}^3$LO), and with detailed QCD-motivated
models of power corrections.

\end{itemize}

\subsubsection{Theoretical developments}
\label{Theo}

The second line of development in resummations is the extension of existing
techniques to new observables or new classes of logarithms. This is of course
much more difficult to predict, since it involves fundamental theoretical progress. 
Here are some examples of what can be expected to happen.

\begin{itemize}

\item Studies are under way to extend threshold resummations to logarithms suppressed
by a power of the threshold variable, or `next-to-leading-power' (NLP) threshold logarithms,
see for example Refs.~\cite{Laenen:2008ux,Laenen:2008gt,Laenen:2010uz,Almasy:2010wn,
Bonocore:2014wua,Presti:2014lqa,Bonocore:2015esa}). 
Partial resummed formulas already exist for some inclusive cross sections and a systematic 
treatment is likely to be available within a few years. The phenomenological impact of these 
logarithms is not yet clear~\cite{Kramer:1996iq,Ball:2013bra,deFlorian:2014vta,Anastasiou:2014lda}, 
but experience suggests that further reductions in scale uncertainties are a likely effect.

\item Anomalous dimensions required for ${\rm N}^2$LL resummations for multi-leg
processes have been available for some time, and those needed at ${\rm N}^3$LL
will become available within a few years. Here the issues are: the selection of appropriate
observables, involving only a limited number of scales and not affected (or affected in 
a controlled way) by non-Sudakov logarithms, and the availability of matching conditions
to preserve an adequate finite-order accuracy.

\item Jet cross sections, and in general less inclusive cross sections, are affected by 
new classes of potentially large logarithms arising from phase space cuts and constraints.
Examples are non-global logarithms~\cite{Dasgupta:2001sh} and clustering 
logarithms~\cite{Delenda:2006nf,Kelley:2012kj}. These logarithms typically
enter at NLL level in the threshold counting, they can be numerically important and contain
interesting physical information. Several groups are engaged in studying the resummation
of these logarithms, or the optimization of observables in order to minimize their effects
(see, for example, \cite{Caron-Huot:2015bja}).

\item In view of the complexity of typical LHC observables, and also of the flexibility 
required to consider many possible cross sections, a very important development
in resummation techniques is going to be the extension of existing numerical codes
(such as {\tt Caesar}~\cite{Banfi:2004yd}) to ${\rm N}^2$LL accuracy~\cite{Banfi:2014sua}. 
This is non-trivial, since the logarithms involved are to some extent non universal, and one 
should ultimately include non-Sudakov logarithms as well, which are not known at this accuracy. 
Solving these problems would however provide a tool applicable to a vast array of processes.

\item In a similar vein, an important development, which would be to a
large extent numerical, is the matching/merging of the analytic resummation
techniques, so far applied to highly inclusive cross sections at high accuracies, with
the parton shower language, which is much more flexible but not easy to extend to
higher logarithmic accuracy. Work in this direction is in progress by several groups, the 
difficulty to a large extent being the very different languages spoken by the two communities.

\end{itemize}

\subsection{Resummations: applications\footnote{L.~Magnea}}

Aside from generic claims that greater theoretical accuracy is important, and resummations
are likely to help to achieve it, one should ask what specific processes/quantities are
most likely to be relevant for new physics searches, and are also affected by 
resummations. A negative example would be the Higgs mass, which is of course very 
important, but can be precisely determined from very clean processes such as
$\PH \to \PZ \PZ \to 4 \mu$ where QCD corrections and extra QCD radiation are not relevant. 

Let us consider briefly two situations where resummations can be important. On 
the one hand, the precise determination of (certain) Standard Model parameters 
and input data, which are of great relevance because they enter almost all theoretical 
predictions\footnote{For an in-depth and up-to-date discussion of these issues, 
see~\Bref{Moch:2014tta}, and references therein.}; on the other hand, a few classes 
of processes where resummed predictions are likely to make a significant impact.

\subsubsection{Resummations for precision in SM parameters}
\label{resprec}

\begin{itemize}

\item Although the strong coupling has been quoted with an uncertainty of the order 
of $0.7\%$~\cite{Moch:2014tta}, there is evidence of unsolved theoretical problems 
that might lead to an upwards revision of the stated uncertainty. 
On the one hand, there are tensions between determinations from different process: 
for example, early values extracted at the LHC tend  to be significantly lower than the 
world average~\cite{Moch:2014tta}. On the other hand, some of the best controlled 
predictions, which involve event shapes at $\Pep \Pem$ colliders, give different 
results depending on the detailed treatment of resummation effects and power 
corrections~\cite{Becher:2008cf,Abbate:2012jh,Hoang:2015hka}. An improved 
prediction for jet cross sections at the LHC, involving NNLO and NNLL contributions, 
as well as a treatment of non-Sudakov logarithms and an improved understanding 
of power corrections, is very likely to make an impact on the determination of 
$\alphas$ in hadron-hadron collisions. Further studies of resummation and power 
correction effects are also probably needed to shed light on the tensions between 
different methods to determine $\alphas$ at lepton colliders.

\item Somewhat similarly, the top quark mass has an official uncertainty well below 
$1\%$~\cite{Moch:2014tta}, which is almost certainly underestimated. In fact, when the 
stated uncertainty goes below $1\UGeV$, it becomes inevitable to deal with the theoretical 
details of the definition of the mass parameter~\cite{Moch:2014tta,Hoang:2014oea}. 
The most precise determination of the top mass is likely to come ultimately from a lepton collider, and 
(as shown by existing studies), improvements in resummation technology have been 
and will be important ingredients in achieving the impressive goal of an uncertainty 
in the per mil range (here both threshold and Coulomb enhancements need to 
be addressed~\cite{Beneke:2009rj,Beneke:2015kwa}).

\item Another ubiquitous ingredient for precision LHC predictions is the determination
of Parton Distribution Functions (PDFs). These are currently determined by means
of global fits  to DIS and collider data by several collaborations, and the standard is 
NNLO accuracy. It was observed already some years ago~\cite{Corcella:2005us} that 
the technology exists to determine PDFs with resummed NLL+NNLO accuracy, which 
could soon be extended to NNLL+${\rm N}^3$LO. Since, for several cross sections of 
interest at the LHC, PDFs are now, or could become, a dominant source of uncertainty, 
such improvements are likely to play an important role. In some cases (for example
the Higgs production cross section) the uncertainty on matrix elements is accidentally
still large enough to compete with the one associated with PDFs. Matrix element 
uncertainties are however going to decrease with time as new calculations become 
available, and the general need for more precise PDFs is likely to increase. One must
finally consider the fact that PDFs determined at finite orders are routinely being used
in conjunction with resummed matrix elements, an inconsistency which could be 
physically relevant in some cases, and which can be corrected with existing 
techniques: indeed, first steps towards a global PDF fit including resummation effects
were taken in~\cite{Bonvini:2015ira}, where this issue is discussed in greater detail.

\end{itemize}

\subsubsection{Resummations for new physics}
\label{resprecnp}

\begin{itemize}

\item Threshold resummations are playing a role in setting limits for the masses of
heavy new physics states, both colorless~\cite{Broggio:2011bd,Fuks:2013vua,Fuks:2013lya} 
and colored~\cite{Beenakker:2011dk,Beenakker:2011sf,Beneke:2013opa,Beenakker:2014sma}, 
which could be produced at hadron colliders. The reason is simple: if the states are heavy 
as compared to the available center-of-mass energy, they must be produced (if at all) near 
threshold. In that case, typically threshold logarithms are large and enhance the production 
cross section. If this enhancement is known, it leads to sharper mass limits in the
case of non-observation. This technique has already been applied to a selection 
of supersymmetric models, for the production of sleptons, gauginos, and colored SUSY
partners such as gluinos and squarks. Of course, if in due course some of these 
states are observed, resummed calculations will help a precise determination
of their quantum numbers and interactions.

\item A fashionable topic of investigation is the subtle relation between the SM 
parameters (most notably the top mass and the Higgs mass) and the stability of 
the electroweak broken-symmetry vacuum, also discussed here in section 5. 
Renormalization-group arguments suggest that current experimental values of 
$m_{\PQt}$ and $m_{\PH}$ place the universe close to the edge between stability 
and metastability~\cite{Buttazzo:2013uya}, and various possibilities are being explored 
as to why it should be so. The universality of this conclusion has recently been 
challenged~\cite{Branchina:2013jra,Branchina:2014usa}, with the argument 
that new physics effects even at the Planck scale are likely to drastically alter the 
scenario. These studies however remain a strong motivation for a precision determination 
of $m_{\PQt}$ and $m_{\PH}$: even if the location of the stability boundary is not 
universal, for a given new physics model it can in principle be determined.  A precise 
knowledge of the parameters can then be used to sharpen the limits on the new 
physics arising from the requirements of stability. This provides extra motivation for 
accurate determinations of $m_{\PQt}$, which, as discussed, crucially involve 
resummation techniques.

\item  A broad and quickly developing field of investigation is the study of jet shapes
(see, for example, Refs.~\cite{Banfi:2010xy,Stewart:2010tn}), with special emphasis 
on the detection of heavy, but also heavily boosted, objects, which are produced and 
then decay inside a jet cone (for a review of recent developments, see~\Bref{Adams:2015hiv}). 
These techniques are already being used to explore Higgs and top properties in channels 
previously thought to be inaccessible due to large backgrounds~\cite{Butterworth:2008iy}. 
Most of the currently available techniques are numerical, and make use of showering 
algorithms. Work is however starting on analytic techniques~\cite{Dasgupta:2013via,
Dasgupta:2013ihk}, which involve the resummation of logarithms of scales arising from 
the jet substructure (for example the ratio of a jet radius to the radius of a selected 
subjet). Analytic control on such resummations is likely to improve our understanding
also of power-suppressed corrections linked to hadronization, which are known to
become large for small jet radii~\cite{Korchemsky:1994is,Dasgupta:2007wa}. Furthermore, 
this field sees parallel developments in numerical, shower-driven techniques, and resummations: 
it would be an interesting area for cross-fertilization between the two methods.

\end{itemize}

\hfill\eject
\section{Monte Carlo tools\footnote{P.~Nason}}
\label{sec:montecarlo}

The development of high energy physics experiments carried out at
colliders with increasing centre-of-mass energy, has seen a parallel
development in the tools for the calculation and simulation of hard
processes.  In the 1980's, calculation of collider processes were
typically performed at tree level, and full simulation of the events
relied upon the Leading-Log (LL) shower approximation. Next-to-leading order
calculations were only available for a handful of processes.

In the last fifteen years, prompted by the perspective of the LHC runs, a
remarkable progress has taken place in several areas. Fully automated
techniques have been developed for the calculation of Next-to-Leading
Order (NLO) cross sections, by several collaborating and competing
groups.  Techniques for combining fixed order calculations with parton
shower generators have appeared, and have been widely applied to
collider processes. Intensive work on Next-to-Next-to-Leading Order
(NNLO) calculations has been carried out by several groups, with
several new NNLO results having appeared since
a little more than a year. Methods for interfacing NNLO
calculations to shower Monte Carlo generators have also appeared for relatively
simple processes.

This section summarizes what is available at present, and
illustrates what can be considered to be frontier research in this field.
Although it is impossible to predict what will be available ten years from
now, it may be safely assumed that current frontier research
will have turned into commonly used tools by that time.

\subsection{Presently available results}
Parton Shower Monte Carlo generators (PS) fully simulate hadronic
production processes by merging together a QCD component (the Shower itself)
and a model for hadron formation. The QCD component is typically given in the
collinear approximation. When applied to infrared finite observables, 
PS generators are accurate only in the collinear and soft regions, failing
to predict hard, large angle emissions even at leading order.
In Ref.~\cite{Catani:2001cc} a procedure was developed for matching matrix element
calculations with PS generators (ME+PS), such that the production of hard, widely
separated jets could be improved to LO accuracy. This prompted the application
of ME+PS techniques to various ME generation tools,
like, for example in \ALPGEN{} with the MLM matching procedure
(for a list of available ME+PS generators see \Bref{Alwall:2007fs}).

In the past 10 years, considerable effort has gone in building
NLO-improved PS generators (NLO+PS).
Methods like \MCatNLO{} \cite{Frixione:2002ik} and
\POWHEG{}~\cite{Nason:2004rx,Frixione:2007vw} allow to interface fixed order NLO
calculations to parton shower generators like
{\sc PYTHIA}~\cite{Sjostrand:2003wg,Sjostrand:2007gs} and
{\sc HERWIG}~\cite{Corcella:2000bw,Bahr:2008pv}.
In essence, for a given process, these techniques extend the precision
of the generator to NLO QCD accuracy for inclusive processes, and to tree level
for the given process in association with one jet. For example, an NLO+PS
generator for Higgs production (a process of order $\alphas^3$
at the Born level) will yield distributions accurate up to
order $\alphas^4$. That amounts to NLO accuracy for inclusive
quantities (i.e. quantities that do not depend upon the emission of
associated jets, like the rapidity distribution of the Higgs, and already
receive contributions at order $\alphas^2$),
and to LO accuracy for processes involving the emission of an
associated jet that start at order $\alphas^3$.
Recently, these techniques have seen
considerable progress, due to the appearance of computer
frameworks that automatize some or all aspects of the calculation:
the virtual
contributions, the implementation of a subtraction framework for the
real corrections, and the interface to a PS. In the {\sc
  MadGraph5\_aMC@NLO} framework~\cite{Alwall:2014hca}, all aspects of
an NLO calculation are automatized, starting from the generation of
the LO and NLO matrix elements, down to the event generation interfaced
to a PS program. The {\sc GoSam}~\cite{Cullen:2011ac},
{\sc Recola}~\cite{Actis:2012qn} and {\sc Open Loops}~\cite{Cascioli:2011va}
frameworks deal with the automatic generation of general-purpose virtual
amplitudes. The BlackHat~\cite{Berger:2008sj} generator provides
virtual corrections for selected processes (vector boson production in association with
jets) and is capable to deal with fairly high jet multiplicities.
In fact it was recently used to compute $W$ production with five
associated jets at NLO~\cite{Bern:2013gka}.
The {\sc Sherpa} generator~\cite{Gleisberg:2008ta}
implements a framework for NLO calculations and
for NLO+PS generation based upon a variant of the \MCatNLO{} method.
The so called MatchBox framework~\cite{Platzer:2011bc} implements
NLO+PS generators within the \Herwigpp{}~\cite{Bahr:2008pv} PS generator.
The \POWHEGBOX{} framework automatizes all aspects of the NLO calculation
interfaced to a PS generator, except for the computation of the matrix elements.
For these it relies upon other programs, like \MadGraph{}, for the real matrix elements,
and \GoSam{} for the virtual corrections.

Electroweak corrections are not presently included in any publicly available
automatic NLO calculators. It is however clear that the same techniques that
have been applied for automated NLO QCD can be extended to the full Standard 
Model, as well as to any renormalizable model.
Interfacing calculations including Electro-Weak corrections to Shower
Monte Carlo requires the ability to handle together QED and QCD collinear showers,
but it does not present new conceptual problems with respect to QCD
corrections alone. In fact, in few simple cases NLO calculation matched
with Shower generators have appeared in the
literature~\cite{Barze':2013yca,Barze:2012tt}.

\subsection{NNLO calculations}

Next-to-next-to-Leading Order calculations (NNLO) for collider
processes have first appeared in 1990 for the Drell-Yan
process~\cite{Hamberg:1990np}, followed more than ten years later by
the NNLO computation of the total Higgs cross section in gluon
fusion~\cite{Harlander:2002wh,Anastasiou:2002yz,Ravindran:2003um}, and
of Higgs differential distributions~\cite{Anastasiou:2004xq,Catani:2007vq}.
We have witnessed since then
a steady increase in the complexity of the processes for which NNLO
calculations have become available: 3 jet cross sections in $\Pep\Pem$
annihilation~\cite{GehrmannDeRidder:2007hr}, $\PW\PH$ and $\PZ\PH$
production~\cite{Brein:2003wg,Ferrera:2011bk}, $\PGg\PGg$
production~\cite{Catani:2011qz}. In a little more than a year from now,
several new results for complex $2 \to 2$ processes have become available:
Higgs production in association with a jet~\cite{Boughezal:2013uia},
$\PQt\PAQt$ production~\cite{Czakon:2013goa}, a partial result on
inclusive jets production~\cite{Currie:2013dwa},
$\PZ/\PW+\PGg$ production~\cite{Grazzini:2013bna},
$\PZ\PZ$ production~\cite{Cascioli:2014yka},
$\PW^+\PW^-$ production~\cite{Gehrmann:2014fva}
and $t$-channel single
top production~\cite{Brucherseifer:2014ama}.
Important results have also been obtained for decay
processes~\cite{Brucherseifer:2013iv,Gao:2012ja,Gao:2014nva}.

There are several components that make up a NNLO calculation, besides
the two loop corrections. One must also supply the square of 1-loop
contribution (double virtual), the virtual correction to one real emission
(real-virtual) and the two-real-emission contributions. Each contribution
contains soft and collinear divergences, that must cancel in the
sum. This also constitutes a challenging aspect of
NNLO calculations.
There are several techniques currently developed for implementing these
cancellations. The $q_{\mathrm T}$ subtraction method~\cite{Catani:2007vq} has
been used for Higgs, Drell-Yan, $\PGg\PGg$, $\PW\PH$, $\PZ\PH$
and $\PZ\PZ$ production processes. It is particularly useful for processes where the
final state is a colour neutral system. The Antenna subtraction
method~\cite{GehrmannDeRidder:2005cm} has been used for the computation
of $\Pep\Pem \to 3\,{\rm jets}$ and for dijets, and is presently
also used in an effort to compute fully differential $\PQt\PAQt$
production at NNLO~\cite{Abelof:2014fza} (now including only the $q\bar{q}$
initial state). The so-called
STRIPPER method
(Sector Improved Phase sPaCe for real
Radiation)~\cite{Czakon:2010td,Boughezal:2011jf} has been used for
$\PQt\PAQt$, $\PH+j$ and $t$-channel single top production. Another
method being developed is described in a series of publications
(see~\cite{DelDuca:2013kw} and references therein).

The computation of the double virtual
contribution is very demanding. Recent progress with integrals
including massive
particles~\cite{Gehrmann:2014bfa,Caola:2014lpa,Caola:2014iua} have
opened the possibility of computing NNLO corrections to pairs of
massive vector bosons. In general, it seems that today
two-loop virtual corrections to generic $2\to 2$ processes are
feasible. A recent groundbreaking technique introduced
by Henn~\cite{Henn:2013wfa}
is among the developments that have made this possible.
$\PW +\,$jet is now know at NNLO~\cite{Boughezal:2015dva} and
there is a phenomenologically complete calculation of $\PH +\,$jet through
NNLO in~\Bref{Boughezal:2015dra}. 

\subsection{Current developments: NLO+PS merging and NNLO+PS generators}
NLO+PS merging deals with the merging of NLO+PS generators of
different associated jet multiplicity. Consider for example
Higgs production in gluon fusion, a process of order $\alphas^2$ at the
Born level. Let us call \Hgg{}, \HggJ{} and \HggJJ{} the NLO+PS
generators for the production of a Higgs, of a Higgs in association
with a jet, and of a Higgs in association with two jets
respectively. The \Hgg{} generator will yield $\alphas^3$ accuracy; that is to say
NLO accuracy for observable that are inclusive in the emission of
associated jets, like the Higgs rapidity
distribution, that include terms of order $\alphas^2$ (LO
terms) plus terms of order $\alphas^3$ (NLO terms), and LO accuracy
for observables requiring an associated jet, that are given at
the lowest order by terms of order $\alphas^3$. Observables requiring more
than two associated jets will be generated by the shower Monte Carlo
in the collinear approximation. The \HggJ{} generator is capable
of yielding NLO accuracy (\ie{} $\alphas^4$ accuracy) for observables
involving the Higgs plus one jet, that are inclusive in the emission
of further jets, and LO accuracy for those requiring two jets.
It would be however unpredictive for fully inclusive observables.
A merged \Hgg{}-\HggJ{} generator would have, in addition, NLO (i.e. $\alphas^3$)
accuracy for fully inclusive observables. In general one may ask to merge
even more NLO+PS generators, for example \Hgg{}+\HggJ{}+\HggJJ{}, in order to have
NLO accuracy (i.e. $\alphas^5$ accuracy) also for observables involving
two associated jets, and thus LO accuracy for those involving three
associated jets.

Notice that NLO+PS merging can be seen as an intermediate step
in the construction of NNLO+PS generators. Thus, for example,
if we have an \Hgg{}+\HggJ{} merged generator, we know that it is
already accurate at the $\alphas^4$ level for all observables, except for those
that are totally inclusive in the emission of associated partons,
where the accuracy is instead $\alphas^3$. If we could reach $\alphas^4$ accuracy
for inclusive observabes, we would have full NNLO accuracy.

Several methods have been proposed for NLO+PS merging, although the
accuracy that they really achieve is still a debated
matter~\cite{Hoeche:2012yf,Frederix:2012ps,Platzer:2012bs,Lonnblad:2012ix,%
Alioli:2012fc,Hartgring:2013jma}. In particular,
in the calculations of Refs. \Brefs{Hoeche:2012yf,Frederix:2012ps},
carried out in the frameworks of the \Sherpa{} and \MCatNLO{} collaborations respectively,
merging is performed using a merging scale. One clusters the event using some jet
clustering procedure, characterized by a merging scale $Q_0$, and uses the
generator with the appropriate number of jets. 
In~\cite{Frederix:2012ps}, stability under variations of the merging scale is
interpreted as an indication of accuracy. In \Bref{Lonnblad:2012ix},
NLO accuracy is adjusted by forcing the inclusive distribution to agree
with the NLO one. This is achieved by subtracting appropriate terms, with a procedure dubbed
UNLOPS (standing for ``Unitary'' NLOPS).
In Ref.~\cite{Alioli:2012fc}, within the so called GENEVA framework,
the merging scale is defined in such a way that
resummation can be carried out up to the NNLL level.
In Refs.~\cite{Hamilton:2012np} a method (called MiNLO) was
proposed to improve the accuracy of generators involving the production
of associated jets, in such a way that it
becomes reliable also after integrating out the associated jets.
In particular, in \Bref{Hamilton:2012rf} it was shown that in certain simple
cases the MiNLO method applied to generators for a boson (Higgs, $Z$
or $W$) plus one jet, can be refined in such a way that observables
that are inclusive in the associated jet (i.e. such that the associated
jet is integrated ou) becomes NLO accurate.

In Ref.~\cite{Hamilton:2013fea}
a first NNLO+PS accurate generator for Higgs production in gluon fusion was presented, based upon the MiNLO procedure of \Bref{Hamilton:2012rf}.
The same method discussed above was also applied recently to the
Drell-Yan process~\cite{Karlberg:2014qua}.
In \Brefs{Hoeche:2014aia,Hoche:2014dla} NNLOPS generators were built for
the Drell-Yan process and for Higgs production respectively.

In \Bref{Alioli:2013hqa}, a general strategy for NNLO+PS
generators based upon the GENEVA framework was outlined. No complete
application of this method to physical processes has been published,
although preliminary results on the Drell-Yan process
have been presented at conferences~\cite{AlioliMunster}.

\section{Conclusions}
At present generators for NLO calculations matched
to parton shower are obtainable with a certain ease for processes with up to
four particles in the final state. It is conceivable to imagine that
automated generators for electroweak corrections for generic processes
may become available soon.
While generators for merged mutltijet samples (i.e. for processes
with an arbitrary number of associated jets), with LO accuracy,
have been available for quite some time, NLO-accurate merged generators
are now beginning to appear.
NNLO calculation for processes with up to two particles in the
final state have recently appeared for a considerable number
of processes, and NNLO calculation matched to shower generators have
appeared only for Higgs production in gluon fusion and Drell-Yan
processes. It is concievable that within the next decade NNLO calculations
matched to shower will become generally available, and that the problem
of merging for NLO generators will be solved.

\hfill\eject
\section{Tools for precision electroweak physics\footnote{C.M. Carloni Calame, M. Chiesa, G. Montagna, O. Nicrosini, F. Piccinini}}

In this section we give a brief overview of 
the state of the art of the tools for precision 
electroweak physics, in view of the forthcoming 
experiments at the LHC run-II and the prospects of 
developments for future experiments 
at very high energy colliders, 
like the FCC-hh and FCC-ee. Some emphasis 
will be put on codes for hadronic collisions, 
while for $\Pep \Pem$ colliders we will refer to the 
state of the art at the end of the LEP data analysis, 
discussing some issues and prospects relevant for 
future high luminosity/energy machines. 

\subsection{Hadron colliders}
As already noted in section~\ref{sec:montecarlo}, the experimental 
precision foreseen for LHC run-II will require the inclusion 
of the complete SM, both the QCD and the electroweak part, in the evaluation 
of quantum corrections for accurate simulations. 
The processes that have been most accurately measured, 
where the inclusion of electroweak radiative 
corrections is already mandatory, are charged and neutral Drell-Yan, 
in addition to Higgs channels for the precise determination 
of its properties. In the past, 
i.e. at Tevatron and LHC run-I, the simulations and analyses 
have been performed by exploiting the dominance 
of QED LL photonic emission from external leptons 
and the relative suppression of QED radiation from quarks with 
respect to gluon radiation. In practice this was achieved by describing 
final state leptonic QED radiation by means of process-independent 
codes such as PHOTOS~\cite{Golonka:2005pn} or internal algorithms 
provided by the shower MC itself, as for instance in 
HERWIG++~\cite{Bellm:2013lba}, 
PYTHIA(8)~\cite{Sjostrand:2007gs} and 
SHERPA~\cite{Gleisberg:2003xi,Gleisberg:2008ta}. 
With the ultimate precision reached at Tevatron measurements, in particular 
the combined CDF and D0 $\PW$-boson mass measurement~\cite{Aaltonen:2013iut}, 
a more precise theoretical description of Drell-Yan processes became necessary, 
at least for the estimate of the systematic uncertainties 
induced by the approximate factorized QCD$\otimes$QED$+$PS approach 
of the simulation tools. In fact, several complete fully-differential 
electroweak NLO calculations are available in the literature and 
implemented in corresponding simulation codes, 
such as {\sc HORACE}~\cite{CarloniCalame:2006zq,CarloniCalame:2007cd}, 
{\sc RADY}~\cite{Dittmaier:2001ay,Brensing:2007qm,Dittmaier:2009cr}, 
{\sc SANC}~\cite{Arbuzov:2005dd,Arbuzov:2007db}, {\sc WGRAD}~\cite{Baur:2004ig}, 
{\sc WINHAC}~\cite{Placzek:2003zg,Bardin:2008fn}, and {\sc ZGRAD}~\cite{Baur:2001ze}. 
These codes share the common feature of LO QCD and NLO electroweak accuracy. 
Several detailed comparisons exist in the 
literature~\cite{CarloniCalame:2004qw,Gerber:2007xk,Buttar:2008jx,Buttar:2006zd}, 
which allow to 
understand the level of technical as well as physical precision 
reached on the electroweak side of the calculations. Among the 
fixed order codes, it is worth mentioning that SANC can calculate the 
NLO contributions of ${\mathcal O}(\alphas)$ and ${\mathcal O}(\alpha)$, 
while the code {\sc FEWZ}~\cite{Li:2012wna} adds up the EW NLO corrections 
to the QCD NNLO corrections for the neutral Drell-Yan process. The HORACE 
generator includes also the effect of all order photonic effects, 
consistently matched to the NLO calculation without double counting, 
in analogy with the QCD NLOPS codes such as MC@NLO and POWHEG. 
Only recently a consistent merging of NLO EW and NLO QCD corrections 
within a single event generator, matched with higher order QED and 
QCD emissions has been achieved within the {\sc POWHEG} 
framework~\cite{Barze:2013yca,Barze:2012tt}. An independent 
implementation has been presented in \Bref{Bernaciak:2012hj}, 
where the higher order shower corrections are given by the 
QCD shower only. In this way also terms of 
${\mathcal O}(\alpha \alphas)$ are included. In particular, terms of 
${\mathcal O}(\alpha)$ dressed with soft/collinear QCD radiation and 
terms of ${\mathcal O}(\alphas)$ dressed with soft/collinear QED radiation 
are correctly accounted for. The remaining ${\mathcal O}(\alpha \alphas)$ 
terms are a source of theoretical uncertainty which can be assessed 
by comparison with a complete two-loop ${\mathcal O}(\alpha \alphas)$ 
calculation. At present such a calculation has been carried out 
in the pole approximation for the charged and neutral Drell-Yan 
processes~\cite{Dittmaier:2014qza}. A solid estimate of these 
and NNLO EW perturbative contributions will be crucial for 
future precision measurements of the $\PW$-boson mass 
at the LHC (see section \ref{sec:WZ}). The complete NNLO calculation, 
beyond the pole approximation, will be a challenge for future 
theoretical advances. 

Besides the Drell-Yan processes, exact NLO EW calculations exist 
for a limited number of final states, such as dijets, 
$\PV + 1$~jet, $\PQt \PAQt$, single-top, $\PV(=\PW,\PZ,\PGg) + 3$~jets, 
$\PH + \PV$, $\PH + 1$~jet, and $\PH + 2$~jets. 
The recent progress in the automation of NLO QCD calculations, described 
in section~\ref{sec:montecarlo} is being extended to include also the 
calculation of NLO EW corrections. There are in principle no obstacles 
to this, even if the EW corrections are more involved due to the presence 
of different mass scales circulating in the loops, together with 
the presence of unstable particles and chiral interactions. 
Several groups are working in this direction: 
{\sc GoSaM}~\cite{Cullen:2011ac,Cullen:2014yla}, 
{\sc HELAC-NLO}~\cite{Bevilacqua:2011xh}, 
{\sc MadLoop}~\cite{Hirschi:2011pa,Alwall:2014hca}, 
{\sc OpenLoops}~\cite{Cascioli:2011va} and {\sc Recola}~\cite{Actis:2012qn}. 
First complete results obtained with automated tools appeared 
in Refs.~{\cite{Denner:2014ina,Frixione:2014qaa,Frixione:2015zaa,Kallweit:2014xda}}. 
During the 2013 edition of the Les Houches Workshop on Physics at 
TeV Colliders a ``High Precision Wish List'' 
has been proposed~\cite{Butterworth:2014efa}, which can be considered 
as the goal of the high precision calculations 
for the coming years. By inspection of Tables [1] through [3] of 
\Bref{Butterworth:2014efa}, we can see that the NLO EW corrections, 
consistently added to the (N)NLO QCD ones and matched with 
higher order QCD/QED PS contributions are required for all the 
processes in the tables. This list of processes will allow to fully exploit 
the LHC run-II data in understanding the Standard Model. 
It is worth noticing that, in addition to the already discussed Drell-Yan processes, 
the consistent matching of NLO EW corrections with higher order QED PS 
is only available for Higgs decay to four leptons~\cite{Boselli:2015aha}. 

Usually the size of the ``genuine'' EW corrections (i.e. excluding the 
leading terms of electromagnetic origin) is moderate, at the 
few percent level. However, when the scales involved in the considered 
scattering process become large with respect to $M_{\PW}$, 
the NLO EW corrections can be particularly enhanced, 
because of the presence of logarithmic terms of the form 
$\alpha \ln^2(Q^2/M_{\PW}^2)$ and $\alpha \ln(Q^2/M_{\PW}^2)$, where 
$Q^2$ is a typical energy scale of the process. These terms are known 
as Sudakov logarithms and correspond to the soft and collinear singularities 
of QCD and QED, induced by the presence of massless particles. 
In the case of the EW corrections, however, 
the $\PW$ boson mass acts as a physical cutoff so that 
the virtual corrections can be considered separately 
from the real contributions~\footnote{The real corrections produce different 
final states, which usually in the experimental analysis are considered as 
different processes with respect to the one under consideration.}, 
giving rise to large negative corrections in the phase space regions where 
$Q^2 \gg M_{\PW}^2$. Moreover, on pure theoretical grounds, 
the cancellation of Sudakov logarithms in the EW sector can only be partial, 
due to the incomplete summation of the contribution of $SU(2)$ doublets 
in the initial state. The Sudakov logarithmic 
structure of the electroweak corrections has been studied in detail in the 
literature~\cite{Ciafaloni:1998xg,Beccaria:1999fk,
Ciafaloni:2000df,Ciafaloni:2000rp,Ciafaloni:2001vt,
Fadin:1999bq,Stirling:2012ak} and a general algorithm able to extract, 
in a process-independent way, the coefficients of the double and single 
logarithms has been presented in Refs.~\cite{Denner:2000jv,Denner:2001gw}. 
Such an algorithm has been recently implemented in the ALPGEN event 
generator, with first phenomenological results for 
$\PZ/\PGg +$~jets production~\cite{Chiesa:2013yma}, a particularly important 
background for the search of new physics in the kinematic 
regime at the LHC. Further studies at the energies of $33\UTeV$ and $100\UTeV$, 
typical reference energies for future hadronic colliders, 
have been carried out within the 2013 Snowmass Community Summer 
Study~\cite{Campbell:2013qaa}. 
For example, for a few selected processes, such as dijet production, 
inclusive vector boson production, $\PV$ + jets, and vector boson pair production, 
it has been shown that, in the extreme regions probed at the LHC with 
$\sqrt{s} = 8\UTeV$, the electroweak effects on the tails of 
some distributions become of the same order of magnitude of the 
experimental accuracy. This means that with the future run-II of the LHC 
we will enter the Sudakov zone, where the EW corrections are relevant 
for data analysis and will be even more important for higher energies, 
as shown in Table~\ref{tab:sudakov}, where the size of the corrections 
can reach several tens of percent~\cite{Mishra:2013una}. 
With such large effects also the 
issue of the resummation of EW corrections should be addressed, as 
suggested in Refs.~\cite{Fadin:1999bq,Ciafaloni:2001mu,Ciafaloni:2005fm,
Chiu:2007yn,Chiu:2007dg}. 
\begin{table}[htbp]
\begin{center}
\begin{tabular}{l c  c  c} \hline \hline
Process & $\sqrt{s}=8\UTeV$ &  $\sqrt{s}=14\UTeV$  & $\sqrt{s}=33, 100\UTeV$ \\
\hline
Inclusive jet, dijet &  Yes  & Yes  & Yes  \\  \hline
Inclusive $\PW/\PZ$ tail   &  $\sim$ Yes  & Yes  & Yes \\
$\PW\PGg$, $\PZ\PGg$ tail ($\ell\PGn\PGg, \ell\ell\PGg$) & No &  $\sim$ Yes &  Yes \\  
$\PW/\PZ$ + jets tail & $\sim$ Yes & Yes  & Yes  \\  \hline
$\PW\PW$ leptonic  & Close & $\sim$ Yes  & Yes  \\  
$\PW\PZ$, $\PZ\PZ$ leptonic  & No & No  & Yes  \\ 
$\PW\PW, \PW\PZ, \PZ\PZ$ semileptonic & $\sim$ Yes & Yes  & Yes  \\  \hline
\end{tabular}
\end{center}
\caption{\small Are we in the electroweak Sudakov zone yet? Taken from \Bref{Mishra:2013una}.}
\label{tab:sudakov}
\end{table}

\subsection{Lepton colliders}
The simulation tools for lepton colliders can be grouped in two 
different classes, according to the physics purpose: generators 
for the precise luminosity determination on the one side 
and programs for the analysis of the large angle data. 
These two kinds of theoretical tools allow for the completion 
of a high precision physics program of an $\Pep \Pem$ collider.

\subsubsection{Event generators for luminosity}
The luminosity can be determined through a counting measurement 
of a process which has a large cross section and is 
calculable to a high accuracy, such as the small angle Bhabha scattering. 
This process is in fact largely dominated by QED $t$-channel photon 
exchange and its cross section can be calculated 
perturbatively with a high level of accuracy. During the LEP1 and LEP2 
eras the reference generator for small angle Bhabha scattering 
was BHLUMI~\cite{Jadach:1991by,Jadach:1996is}, which was based 
on QED NLO corrections to $t$-channel scattering, supplemented with 
higher-order corrections in the Yennie-Frautschi-Suura exponentiation 
approach. The physical precision of BHLUMI was scrutinized by means of 
independent calculations, such as for instance SABSPV~\cite{Cacciari:1995fq}, 
mainly based on QED NLO precision plus 
higher-orders photonic corrections in the QED structure function approach. 
The final theoretical accuracy on Bhabha scattering at LEP1 was 
at the level of $0.05\%$. 

The experience gained at LEP has been fruitful for the 
development of Monte Carlo tools for the luminosity 
determination at the low-energy 
flavour factories by means of large angle Bhabha scattering, 
cross-checked with $\Pep \Pem \to \PGg\PGg$ measurements. 
In this context the first QED parton shower matched to the 
NLO fixed order calculation for the QED processes 
$\Pep \Pem \to \Pep \Pem$, $\Pep \Pem \to \PGmp \PGmm$ and 
$\Pep \Pem \to \PGg\PGg$, BABAYAGA@NLO, has been 
realized~\cite{CarloniCalame:2000pz,Balossini:2006wc,Balossini:2008xr}. 
In parallel, an impressive effort has been devoted to the calculation
of the exact NNLO QED corrections to Bhabha scattering, see for
example \Bref{Actis:2010gg} and references therein. A
future consistent inclusion of these results into Monte
Carlo generators could push the accuracy to the level of a few
$0.01\%$, at least as far as QED corrections are concerned.

A source of theoretical uncertainty (driven by experimental uncertainties)
is the hadronic contribution to the running of the QED coupling constant 
$\Delta\alpha_{\mathrm{had}}(q^2)$, which is derived from low
energy data through dispersion relations. In this context, the present
measurements at low energy machines are extremely important to reduce
the dominant uncertainties at LEP. 

It is worth mentioning that an alternative process to Bhabha
scattering for luminometry is $\Pep\Pem\to\PGg\PGg$, which 
is not affected, at least up to NNLO order, by the error on
$\Delta\alpha_{\mathrm{had}}$ and thus, in principle, it could be calculated 
with higher theoretical precision.

\subsubsection{Simulation tools for $\PZ$ and $\PW$ bosons at FCC-ee}
Given the available statistics at LEP1, a $0.1\%$ precision level 
was reached for most of the observables. With such a level of precision, 
the necessary ingredients for the simulation tools 
(event generators and seminalitical programs, such as for instance  
{\sc KORALZ}~\cite{Jadach:1993yv,Jadach:1999vf}, 
{\sc TOPAZ0}~\cite{Montagna:1993py,Montagna:1993ai,Montagna:1995ja,Montagna:1998kp} 
and {\sc ZFITTER}~\cite{Bardin:1989di,Bardin:1989tq,Bardin:1992jc,Bardin:1999yd}) 
were the exact NLO EW corrections to the $\Pep \Pem \to \Pf\PAf$ hard 
scattering, convoluted with QED final and initial state radiation. 
Since around the $\PZ$ resonance the latter contribution is very large, 
of the order of $30\%$, higher order effects were included through the 
Yennie-Frautschi-Suura formalism or the QED structure function approach.
In order to match the target accuracy, 
also higher order effects of weak and QCD origin, contributing 
for instance to the $\rho$ and $\Delta r$ parameters, had to be included 
in the computational tools. It is clear that a future GigaZ run 
of an $\Pep \Pem$ collider will require complete EW NNLO calculations, 
supplemented with 
improved higher order QED corrections. While the theoretical framework of 
Standard Model two-loop renormalization has been set in 
Refs.~\cite{Actis:2006ra,Actis:2006rb,Actis:2006rc}, 
the calculation of observables at NNLO accuracy 
for the processes $\Pep \Pem \to \Pf\PAf$ is still a challenge for the 
future. 

The high luminosity run of a future $\Pep \Pem$ collider at energies 
close and above the $\PW\PW$, $\PZ\PZ$ and $\PZ\PH$ thresholds will be very 
challenging for the development of Monte Carlo codes 
able to provide precise theoretical predictions. In fact, at LEP2 
most of the tree-level predictions for four-fermion final states were based 
on tree-level matrix element, supplemented with convolution with 
initial state radiation effects and leading electroweak corrections 
in the form of running couplings, together with a scheme for the 
treatment of the unstable virtual bosons  
(see \Bref{Beenakker:1996kt,Bardin:1997gc,Carena:1996bj} 
for a review). Complete NLO predictions for $\Pep \Pem \to 4$~fermions final 
states  appeared only after the end of LEP2 
operations~\cite{Denner:1999kn,Denner:2002cg,Denner:2005es,Denner:2005fg}. 
Most probably the required theoretical accuracy at FCC-ee will be NNLO 
EW corrections to $\Pep \Pem \to 4$~fermion final states, interfaced with 
algorithms for the treatment of QED higher order initial state radiation, 
a very challenging task for the presently available theoretical knowledge. 


\newpage
\section{Conclusions \footnote{S.~Forte, G.~Passarino}}
The SM is always the background to all of our experimental explorations. 
The discovery of the SM-like Higgs boson is a milestone in particle 
physics. Direct study of this boson will hopefully shed light on the mysteries
surrounding the origin of the electroweak scale, and possibly 
provide insight into observations that remain unexplained by the SM.

In this report we have taken the viewpoint that in the next several
years an important 
window to explore the theory space of physics beyond the Standard
Model - perhaps the only window - will be provided  by precision
physics. This expectation is based on the twin observations that
effective field theory provides the general framework for
consistent calculation of higher orders in studying deviations from
the standard model, and that ongoing 
and near future experiments can achieve an estimated per mille
accuracy on precision Higgs and EW observables.

Effective field theory is superior to a generic parametrization of
higher-dimensional operators (such as the so-called $\upkappa\,$-framework of
\Bref{LHCHiggsCrossSectionWorkingGroup:2012nn}) in that it
automatically implements gauge symmetry and unitarity, and, as
discussed in the introduction and then in Sect.~\ref{sec:EFT}, it may
point to the ultraviolet completion which provides hints for the
underlying theory. 
However,  EFT itself is subject to assumptions and limitations that
one should be aware. Firstly, in principle EFT is defined in a
Wilsonian approach, in which heavy degrees of freedom are integrated
above a cutoff. In practice, however, computations beyond
leading order are performed in a continuum (cutoff-independent) EFT,
in which heavy degrees of freedom are not integrated out, but rather
compensated for through an appropriate matching
calculation~\cite{Georgi:1994qn}.  This implies that decoupling of
heavy degrees of freedom is assumed.
Furthermore, while being the only approach that can be systematically improvable 
by including higher dimension operators and higher-order corrections in QCD and EW, 
in practice the EFT will be compared to data at a given accuracy. For example, the impact 
of ${\mathrm{dim}} = 8$ operators in some key observables will need to be evaluated as 
well as possibly the effect of NLO EW corrections. 
Finally, the most common EFT parametrisations are based on a linear realisation of 
the gauge symmetry. Work on non-linear realisation can be found in \Bref{Buchalla:2013rka}.

This then raises the question of whether results from LHC should be
cast in a language which is as much as possible independent of our
current conceptual framework. Theoretical and phenomenological
developments are currently making this increasingly possible at the
level of data analysis and of comparison between data and theory. For
instance, it is now increasingly clear that cross-sections should be
published as differential as possible, at the fiducial level, 
without the subtraction of electroweak corrections, and so on. 
Old hadron collider data are often  obsolete because, say,
they were analyzed using outdated parton distributions and
leading-order theory, or infrared-unsafe QCD definitions, and this
should surely be avoided. 

However, this it is not enough: LEP results, which  were
free of these problems, could be stored in the form of 
Pseudo-Observables (PO), see \Brefs{Bardin:1999gt,Bardin:1998nm} and
\Bref{Z-Pole}, thereby allowing experimentalists and 
theorists to meet half way, without theorists having to run full simulation and 
reconstruction and experimentalists not having to fully unfold to model-dependent
parameter spaces. The situation at the LHC is harder not only because
it is a hadron collider, with the corresponding aforementioned
problem (so that at the LHC fiducial cross sections should always be reported), but also
because 
$4\Pf$ decays are $40\%$ 
of $2\Pf$ decays, so most of the time we face off-shell unstable
particles, even at the $\PH$ peak cross-section, and signal and
background are then inextricably tangled and interfering.

It is thus important to build a simple platform to bridge
between realistic observables and theory parameters working in the space of signals 
but having in mind the space of theories. Realistic proposals will necessarily involve 
a combination of fiducial observables, and 
pseudo-observables~\cite{Gonzalez-Alonso:2014eva,Gonzalez-Alonso:2015bha,Bordone:2015nqa}, 
linked through the language of effective Lagrangians~\cite{POtalk,Ghezzi:2015vva,David:2015waa}. 


%
\clearpage
\section{Acknowledgements}

The work of G.~Passarino and L.~Magnea is supported by MIUR under contract
$2001023713\_006$, by UniTo - Compagnia di San Paolo under contract 
ORTO11TPXK and by FP7-PEOPLE-$2012$-ITN HiggsTools 
PITN-GA-$2012{-}316704$. 

The work of S.~Forte is supported by an  Italian PRIN2010 grant, by a 
European Investment Bank EIBURS grant, and by the European
Commission through the HiggsTools Initial Training Network
PITN-GA-2012-316704.

The activity of C~.M.~Carloni Calame is supported by MIUR under the 
PRIN project $2010$YJ$2$NYW. 
This work was supported in part by the Research Executive Agency (REA) of 
the European Union under the Grant Agreement number 
PITN-GA-$2010$-$264564$ (LHCPhenoNet) and by MIUR under the PRIN project 
$2010$YJ$2$NYW. 

The work of F.~Maltoni is partly supported by the ERC grant $291377$
``LHCtheory: Theoretical predictions and analyses of LHC physics:
advancing the precision frontier'', by the FP7 Marie Curie Initial 
Training  Network MCnetITN (PITN-GA-$2012$-$315877$), the Belgian Federal 
Science Policy Office through the Interuniversity Attraction Pole P7/37 
and the FNRS via the IISN $4.4511.10$ and $4.4517.08$ conventions.

The work of F.~Riva is supported by the Swiss National Science Foundation, under
the Ambizione grant PZ$00$P$2136932$.

The work of A.~Vicini was supported by the Munich Institute for Astro- and 
Particle Physics (MIAPP) of the DFG cluster of excellence "Origin and 
Structure of the Universe".

\newpage
\bibliographystyle{atlasnote}
\bibliography{WNSM_revp}
\providecommand{\href}[2]{#2}\begingroup\raggedright\endgroup


\end{document}